\documentclass[a4paper,11pt,usenames,dvipsnames]{article}
\usepackage{jheppub}
\usepackage[utf8]{inputenc}
\usepackage{todonotes}
\usepackage{subcaption}
\usepackage{placeins}
\usepackage{pdflscape}
\usepackage{afterpage}
\usepackage{paralist}

\newcommand{\cO}{\mathcal{O}}
\newcommand{\cN}{\mathcal{N}}

\newcommand{\cM}{\mathcal{M}}
\newcommand{\bC}{\mathbb{C}}

\newcommand{\cE}{\mathcal{E}}
\newcommand{\cH}{\mathcal{H}}
\newcommand{\cL}{\mathcal{L}}
\newcommand{\e}{\epsilon}

\newcommand{\bZ}{\mathbb{Z}}
\newcommand{\bR}{\mathbb{R}}

\newcommand{\bP}{\mathbb{P}}

\newcommand{\blambda}{\boldsymbol\lambda}
\newcommand{\fq}{\mathfrak{q}}

\newcommand{\GW}{\textrm{GW}}
\newcommand{\DT}{\textrm{DT}}
\newcommand{\GV}{\textrm{GV}}
\newcommand{\tX}{\widetilde{X}}

\newcommand{\NEbar}{\overline{\mathrm{NE}}}

\usepackage{tikz}
\usetikzlibrary{positioning}
\usetikzlibrary{arrows}
\usetikzlibrary{decorations.pathreplacing} 
\usetikzlibrary{decorations.pathmorphing} 
\usetikzlibrary{decorations.markings} 
\usetikzlibrary{shapes.geometric}
\usetikzlibrary{arrows.meta}
\usetikzlibrary{calc}
\usetikzlibrary {math}
\tikzset{
  mid arrow/.style={
    postaction={decorate},
    decoration={markings, mark=at position 0.5 with {\arrow{Latex}}}
  }
}
\tikzset{bd/.style={circle, draw=black, inner sep=0pt, fill=black, minimum size=2mm}}
\tikzset{wd/.style={circle, draw=black, inner sep=0pt, fill=white, minimum size=2mm}}
\tikzstyle{ligne}=[draw, very thick] 
\tikzstyle{brane}=[draw] 
\tikzset{sevenb/.style={circle, draw,fill=white,inner sep=2.5pt}}
\tikzstyle{gridline}=[draw, black!20!] 
\tikzset{hasse/.style={circle, fill,inner sep=2pt}}   

\preprint{}

\title{BPS Invariants for Generalized Toric Calabi-Yau Threefolds}

\author[1]{Antoine Bourget,}
\author[2]{Peng Cheng,}
\author[1]{and Quentin Lamouret}
\affiliation[1]{Institut de physique théorique, CEA, CNRS, Université Paris-Saclay, 91191 Gif-sur-Yvette, France}
\affiliation[2]{Johannes Gutenberg-Universit\"at, Staudinger Weg 7, 55128 Mainz, Germany}

\abstract{We apply topological vertex techniques to Calabi-Yau threefolds dual to brane webs where several 5-branes can end on the same 7-brane. In this context, we determine how topological string partition functions transform under Hanany-Witten transitions and flops, which allows us to track curves and the associated invariants under such transitions. The contributions of parallel external branes form a universal sector invisible to the 5d SCFT; once it is removed, invariants can be transported between different geometries engineering the same theory. This yields an efficient technique to compute Gopakumar-Vafa invariants at any degree and genus. We illustrate this with local Hirzebruch and del Pezzo surfaces, including $dP_4$, whose invariants we obtain at high degree for the first time.}

\begin{document} 

\maketitle
  
\section{Introduction}
Geometric engineering of supersymmetric field theories in M-theory and string theory is a wonderful laboratory where geometric insights turn into physical ones, and conversely physics can be used to learn about geometry. In particular, 5d $\mathcal{N}=1$ QFTs can be realized as M-theory placed on local Calabi-Yau threefolds. If $X$ is a singular such threefold and $\tX \rightarrow X$ is a resolution, M-theory associates in general a strongly-coupled SCFT to $X$ and relevant deformations thereof to $\tX$. 

The topological string partition function $Z_{\mathrm{top}}(\tX ; q , Q)$ on $\tX$ contains a wealth of information about the geometry of $\tX$. It depends on the string coupling constant $g_s$ through the combination $q = e^{i g_s}$, a notation we keep throughout the paper (this $q$ should not be confused with the instanton counting parameter which we denote $\fq$). It also depends on the exponentiated and complexified volumes $Q^{\beta}$ of the two-cycle class $\beta \in H_2^{\mathrm{cpt}}(\tX , \mathbb{Z})$. The dependence of $Z_{\mathrm{top}}$ in $q$ and $Q$ can be expressed in terms of Gopakumar-Vafa (GV) \cite{Gopakumar:1998ii,gopakumar1998m} coefficients $\GV_{r , \beta} (\tX) \in \mathbb{Z}$ which depend on the genus $r \geq 0$ and the homology class $\beta$, see \eqref{eq:GVPlethystic}. These are related to Gromov--Witten and Donaldson--Thomas \cite{maulik2004gromovwittentheorydonaldsonthomastheory} invariants as briefly recalled in Appendix \ref{app:notations}, which summarizes the definitions, conventions and useful formulas. In this paper we concentrate on unrefined GV invariants, although many results in this paper can be easily generalized to refined GV invariants as well. Roughly speaking, they count certain massive BPS states in the 5d theory, corresponding to M2 branes wrapping compact two-cycles $\beta$. In the SCFT limit, corresponding to the map $\tX \rightarrow X$, all these two-cycles shrink to zero size, and the particles become massless. One subtlety is that not all zero size cycles contribute to the 5d SCFT per se. This is true if the singularity in $X$ is isolated. If it is not, one might need to decouple higher-dimensional degrees of freedom. Formally, we write \cite{Iqbal:2003ix,Eguchi:2003sj,Taki:2007dh}
\begin{equation}
    Z_{5d} (q , Q) = \frac{Z_{\mathrm{top}}(\tX ; q , Q)}{Z_{\mathrm{decoupled}}(\tX ; q , Q)} \, . 
\end{equation}
This can be seen as the definition of $Z_{\mathrm{decoupled}}$, provided we know how to compute the other two terms. The supersymmetric index $Z_{5d} (q , Q)$ counting BPS particles of the theory on $\mathbb{R}^{4} \times S^1$ can be evaluated in certain cases, in particular if the theory admits a gauge theory phase (i.e. if there is a relevant deformation of the SCFT which is a 5d $\mathcal{N}=1$ gauge theory). When this is the case, one can express the gauge theory parameters, i.e. the gauge couplings and masses of hypermultiplets, which we call collectively $\mu (Q)$, in terms of $Q$ \cite{Katz:1996fh,Leung:1997tw}. In general, finding this explicit map is hard; for the GTP geometries to be considered below, it is relatively straightforward.  Placing the theory on\footnote{We consider only the symmetric $\Omega$-deformed $\mathbb{C}_{\e_1 , \e_2}$ background, $\e_1 + \e_2 = 0$. In the general case $\e_1 + \e_2 \neq 0$, one needs to use the refined topological vertex and it is not clear if $Z_{\mathrm{top}}$ remains purely topological, i.e. invariant under the complex structure deformations implied by GTPs.  } $\Omega$-deformed $\mathbb{C}^2_{\epsilon , -\epsilon} \times S^1_R$ with $R \epsilon = i g_s = \log q$, one can combine the perturbative sector and Nekrasov instanton counting \cite{Nekrasov:2003rj,Nekrasov:2004vw,Nekrasov:2009rc,Nekrasov:2015wsu,Nekrasov:2016qym,Nekrasov:2017gzb,Nekrasov:2016ydq,Braverman:2004vv,Nakajima:2003uh,Closset:2022vjj} and write \cite{nekrasov2003seiberg,Nekrasov:2003rj}
\begin{equation}
    Z_{5d} (q , Q) = Z_{\mathrm{Pert}} (q,\mu(Q)) \; Z_{\mathrm{Inst}} (q,\mu(Q)) \, . 
\end{equation}
We refer to Appendix \ref{app:instantoncounting} for some aspects of these computations. 

We now review an incomplete set of techniques to compute $Z_{\mathrm{top}}$ with focus on non compact $CY3$:
\begin{itemize}
    \item Topological vertex \cite{Aganagic:2002qg,Aganagic:2003db} for toric threefolds. This technique is based on the open (Chern--Simons theory) / closed topological string duality \cite{Gopakumar:1998ki}. A smooth \textit{toric} threefold $\tX$ is defined by a three-dimensional fan $\Sigma \subset \mathbb{R}^3_{x,y,z}$ whose generators of one-dimensional cones lie in the plane $\{z=1\}$. The intersection $P = \Sigma \cap \{z=1\}$ is the \textit{toric diagram}, in which 
\begin{itemize}
    \item Dots represent 1-dimensional cones in $\Sigma$, hence divisors in $X$. The divisors are compact for internal dots and non-compact for external dots. 
    \item Edges represent 2-dimensional cones in $\Sigma$, hence curves in $X$. Internal edges are compact $\mathbb{P}^1$ curves, edges on the boundary are non-compact curves.  
    \item Faces represent 3-dimensional cones in $\Sigma$, hence toric fixed points in $X$. 
\end{itemize}
The dual graph can be read as a web of 5-branes in type IIB string theory \cite{Aharony:1997bh}. These are two dual ways of engineering the same 5d theory: M-theory reduction and brane web world-volume. Examples of this correspondence are depicted below:
\begin{subequations}
\renewcommand{\theequation}{\theparentequation.\alph{equation}}
\begin{align}
    \raisebox{-.5\height}{ \scalebox{.8}{  
    \begin{tikzpicture}[x=.5cm,y=.5cm] 
    \draw[gridline] (-1,-1.5)--(-1,1.5); 
    \draw[gridline] (0,-1.5)--(0,1.5); 
    \draw[gridline] (1,-1.5)--(1,1.5); 
    \draw[gridline] (-1.5,-1)--(1.5,-1); 
    \draw[gridline] (-1.5,0)--(1.5,0); 
    \draw[gridline] (-1.5,1)--(1.5,1); 
    \draw[ligne] (0,-1)--(1,0); 
    \draw[ligne] (1,0)--(0,1); 
    \draw[ligne] (0,1)--(-1,0); 
    \draw[ligne] (-1,0)--(0,-1);  
    \node[bd] at (1,0) {}; 
    \node[bd] at (0,1) {}; 
    \node[bd] at (-1,0) {}; 
    \node[bd] at (0,-1) {};  
    \end{tikzpicture}}}  
    \qquad &\longleftrightarrow \qquad  
    \raisebox{-.5\height}{\scalebox{.8}{\begin{tikzpicture}[scale=.8]  
        \draw (-1,-1) -- (1,1);      
        \draw (-1,1) -- (1,-1);    
    \end{tikzpicture}}} \\
    \raisebox{-.5\height}{ \scalebox{.8}{  
    \begin{tikzpicture}[x=.5cm,y=.5cm] 
    \draw[gridline] (-1,-1.5)--(-1,1.5); 
    \draw[gridline] (0,-1.5)--(0,1.5); 
    \draw[gridline] (1,-1.5)--(1,1.5); 
    \draw[gridline] (-1.5,-1)--(1.5,-1); 
    \draw[gridline] (-1.5,0)--(1.5,0); 
    \draw[gridline] (-1.5,1)--(1.5,1); 
    \draw[ligne] (0,-1)--(1,0); 
    \draw[ligne] (1,0)--(0,1); 
    \draw[ligne] (0,1)--(-1,0); 
    \draw[ligne] (-1,0)--(0,-1); 
    \draw[ligne] (-1,0)--(1,0); 
    \draw[ligne] (0,-1)--(0,1); 
    \node[bd] at (1,0) {}; 
    \node[bd] at (0,1) {}; 
    \node[bd] at (-1,0) {}; 
    \node[bd] at (0,-1) {}; 
    \node[bd] at (0,0) {}; 
    \end{tikzpicture}}}  
    \qquad &\longleftrightarrow \qquad  
    \raisebox{-.5\height}{\scalebox{.7}{\begin{tikzpicture}[scale=.8]
        \coordinate (A) at (0,0);
        \coordinate (B) at (2,0);
        \coordinate (C) at (2,1);
        \coordinate (D) at (0,1);
        \draw (A) -- (B);
        \draw (B) -- (C);
        \draw (C) -- (D) node[midway, above] {$Q_F$};
        \draw (D) -- (A) node[midway, left] {$Q_B$};
        \draw (D) -- (-1,2);      
        \draw (C) -- (3,2);    
        \draw (B) -- (3,-1);      
        \draw (A) -- (-1,-1);        
    \end{tikzpicture}}}  
\end{align}
\end{subequations}
The left-hand side depicts $P$, which is a (subdivided) lattice convex polygon. In the first line, the polygon is not subdivided, and the associated threefold $X$ is affine and singular. In the second line, the polygon is maximally subdivided: the associated threefold $\tX$ is smooth, and $\tX \rightarrow X$ is a resolution of singularities.  The right-hand side shows the dual web of 5-branes. A $(p,q)$ 5-brane extending to infinity can be thought of as ending on a 7-brane of the same $(p,q)$ type. 
The topological vertex computation is superficially similar to the computation of Feynman diagrams, see Table \ref{tab:TV}: starting from the brane web, each finite-size 5-brane carries an integer partition and contributes a "propagator", each vertex contributes an "interaction", and the partition function is obtained by summing over all the internal partitions. This is reviewed in Section \ref{sec:topologicalVertex}.  
    \item Mirror symmetry and holomorphic anomaly equations from string worldsheet perspective \cite{Bershadsky:1993cx,Bershadsky:1993ta,Witten:1991zz,Candelas:1990rm}. This technique is based on duality of a pair of $(2,2)$ non linear sigma model (NLSM)\footnote{Here we assume both theories are $(2,2)$ NLSM with target space $CY3$ manifolds. Although this technique applies in more general cases.} with target space a mirror pair of $CY3$. This technique can be applied to compact $CY3$ and gives higher genus topological string partition function, up to a finite number of ambiguities at every genus, which makes it hard in practice in general cases. The local version has been developed and applied to compute topological string partition function in \cite{chiang1999localmirrorsymmetrycalculations} for $K_{dP_{5\leq n \leq8}}$.  
    \item Supersymmetric localization \cite{Witten:1993yc,Morrison:2016bps,Jockers:2012dk,Closset:2015rna,Benini:2016qnm}. When the $(2,2)$ NLSM admits a simple gauge linear sigma model (GLSM) realization in the UV. Then the topological string partition function can be contained in the GLSM partition function, which can be calculated using supersymmetric localization techniques.
    \item Non critical (E,M) string techniques \cite{Haghighat:2014pva,Haghighat:2014vxa,Gu:2018gmy,Gu:2019dan,Gu:2019pqj,Gu:2020fem,DelZotto:2017mee,DelZotto:2018tcj,DelZotto:2022ras,DelZotto:2023rct,DelZotto:2023ryf,Huang:2010kf,Huang:2013yta,Huang:2015sta,Minahan_1997,Minahan:1997ct,Minahan:1998vr}. Via a chain of dualities in F/M/type-II theories, the elliptic genus of $2d$ supersymmetric theory on non critical string worldsheet is related to the (subset of) topological string partition function of corresponding $CY3$. For example, the elliptic genus of (degree-$k$, massless) $E$ string gives the topological string partition function of $K_{dP_9}$ associated to curve classes $k\bP^1 +nF, n\geq 0$, where $F$ is the elliptic fiber of $dP_9$. By taking suitable massive limit, degree $k$ E-string partition function gives (subset of) topological string partition function of$K_{dP_{n\leq 8}}$.  There is also a holomorphic anomaly equation relates E-strings with different degree $k$. Although in each degree the $E$-string partition function contains all genus topological string partition function, in practice it is hard to go to higher degree.
\end{itemize}

For reasons grounded both in SCFT physics and in brane dynamics, it is necessary to go beyond this toric realm, and introduce boundary conditions on the brane web. An example is 
\begin{equation}
       \raisebox{-.5\height}{ \scalebox{.8}{  
    \begin{tikzpicture}[x=.5cm,y=.5cm] 
    \draw[gridline] (0,-0.5)--(0,3.5); 
    \draw[gridline] (1,-0.5)--(1,3.5); 
    \draw[gridline] (2,-0.5)--(2,3.5); 
    \draw[gridline] (-0.5,0)--(2.5,0); 
    \draw[gridline] (-0.5,1)--(2.5,1); 
    \draw[gridline] (-0.5,2)--(2.5,2); 
    \draw[gridline] (-0.5,3)--(2.5,3); 
    \draw[ligne] (0,2)--(1,0); 
    \draw[ligne] (1,0)--(2,3); 
    \draw[ligne] (2,3)--(0,3); 
    \draw[ligne] (0,3)--(0,2); 
    \draw[ligne] (1,0)--(1,2)--(2,3); 
    \draw[ligne] (0,2)--(1,1)--(2,3); 
    \draw[ligne] (0,2)--(1,2); 
    \node[bd] at (1,0) {}; 
    \node[bd] at (2,3) {}; 
    \node[bd] at (0,3) {}; 
    \node[bd] at (0,2) {}; 
    \node[bd] at (1,1) {}; 
    \node[bd] at (1,2) {}; 
    \node[wd] at (1,3) {}; 
    \end{tikzpicture}}} 
    \qquad  \longleftrightarrow \qquad   
    \raisebox{-.5\height}{\scalebox{.7}{\begin{tikzpicture}
        \tikzmath{\x = .07;}
         \draw (-1,0) -- (0,1) node[midway, above left] {$Q_1$};
        \draw (0,1) -- (1,1) node[midway, below] {$ Q_2$};
        \draw (1,1) -- (3,0) node[midway, above right] {$Q_1$};
        \draw (3,0) -- (-1,0) node[midway,below] {$Q_1^3 Q_2$};
        \draw (0,1) -- (0,2-2*\x) node[midway, left] {$ Q_2$};
        \draw (0,2- 2*\x) arc (270:90:\x);
        \draw (0,2.0) -- (0,3);
        \draw (-1,0) -- (-2,-.5);
        \draw (1,1) -- (\x,2-\x) node[midway, above right] {$Q_2$};
        \draw (\x,2-\x) -- (\x,3);
        \draw (3,0) -- (3.6,-.2);
        \draw (\x,2-\x) -- (-1,2-\x);
    \end{tikzpicture}}}
    \label{eq:ex3}
\end{equation} 
One reason is provided by Hanany-Witten moves: when a 7-brane is dragged across the brane web, 5-branes are created or annihilated, and one can end up with several 5-branes of the same type $(p,q)$ ending on a single $(p,q)$ 7-brane. If this were the only reason, it would not matter much: one could always undo such moves and return to a toric configuration. But it turns out that allowing several 5-branes to end on the same 7-brane is required in order to account for certain SCFTs, which cannot be described otherwise -- simple examples include the $E_7$ and $E_8$ rank-1 SCFTs. In fact, this appears to be the case for the majority of 5d SCFTs, as suggested by surveys such as \cite{Bourget:2021csg}.
Concretely, for a stack of $N$ parallel 5-branes, non-trivial boundary conditions can be specified, in one-to-one correspondence with integer partitions $\lambda$ of $N = \lambda_1 + \dots + \lambda_r$: one introduces $r$ 7-branes and ends $\lambda_i$ 5-branes on the $i$-th one. These partitions are encoded as a pattern of white dots in the dual toric polygon, which is then called a \emph{generalized toric polygon} (GTP) \cite{Benini:2009gi}. These are the diagrams which appear on the left hand side in \eqref{eq:ex3}. They have subsequently been used to
study Higgs branches, magnetic quivers and RG flows
\cite{Cabrera:2018jxt,Bourget:2019rtl,Bourget:2020mez,vanBeest:2020kou,VanBeest:2020kxw}, their relation to T-branes \cite{Bourget:2023wlb,Arias-Tamargo:2024fjt}, and their behavior under Hanany--Witten transitions and polytope mutations \cite{Franco:2023flw,Franco:2023mkw,CarrenoBolla:2024fxy,CarrenoBolla:2025rkv,Kho:2026zwc}.
The threefold dual to a GTP remained obscure for several decades, but is now much better understood \cite{Bourget:2023wlb,alexeev2025non,BCL}; we refer to such constructions as \emph{generalized toric geometry}. For a (possibly subdivided) generalized toric polygon $P$ (this includes the toric case), we still call $X[P]$ the associated threefold, and $\mathcal{T}_{X[P]}$ the 5d SCFT. We note that GTPs are the smallest set of polygons such that the class of threefolds constructed from them is closed under Hanany-Witten transitions.  On the computational side, the topological vertex formalism has been extended to some (but not all) GTP geometries \cite{Hayashi:2015xla}; we review this in Section \ref{sec:topologicalVertex}. It allows for instance to compute $Z_{\mathrm{top}}(\tX ; q , Q)$ for $\tX$ the generalized toric threefolds defined by the diagrams in \eqref{eq:ex3}, by adding one simple rule, see the last row in Table \ref{tab:TV}.

\paragraph{Summary of results}
\begin{itemize}
    \item We explain how to compute GV invariants for $X[P]$, where $P$ is any GTP tiled with $\mathrm{SL}(2,\mathbb{Z})$ transforms of \eqref{eq:simpleCrossing}. 
    \item We demonstrate how GV invariants are mixed under Hanany-Witten transitions. We specify explicitly how cycles are transformed, in particular how one curve class maps to another. As conjectured in the field theory community, \emph{although two threefolds connected by a Hanany-Witten move are geometrically distinct, the SCFTs engineered on them are the same}. As an illustration, we study in detail the rank 1 SCFTs. Already at rank 0, the subtleties arise, see equations \eqref{eq:relationF0F2}-\eqref{eq:equalityF1F3GTP}.
    \item When M-theory on a threefold $X$ engineers a $5d $ $\cN=1$ theory $\mathcal{T}_{X}$, it does not imply $Z_{\textrm{top}}[X] = Z_{5d} [\mathcal{T}_{X}] $, as the full $M$ theory on $X$ may contain extra decoupled degrees of freedom. In the topological string partition function $Z_{\textrm{top}}$, we identify the SCFT contribution and the decoupled contribution, which for GTP geometries is entirely accounted for by external parallel branes, confirming earlier investigations including \cite{Bergman:2013aca,Bao:2013pwa,Taki:2014pba,Hayashi:2015xla}. In equation, this reads 
    \begin{equation}
    \label{eq:main}
        Z_{5d} [\mathcal{T}_{X}] = \frac{Z_{\textrm{top}} [X]}{Z_{\parallel}[X]} \, 
    \end{equation} 
    when $X$ is a GTP. 
    The fact that only the parallel branes contribution $Z_{\parallel}[X]$ needs to be removed from $Z_{\textrm{top}}$ is checked explicitly in the rank 1 case using local mirror symmetry. Alternatively, observe that when gluing $N$ external parallel legs to a second toric diagram carrying the same $N$ parallel legs, the compact curves between the legs give rise to the $SU(N)$ $W$-bosons. Without this gluing, the compact curves between the parallel legs account for only half of the $SU(N)$ $W$-boson spectrum -- by CPT, this cannot be the spectrum of a consistent, interacting $5d$ SCFT.
     \item We use \eqref{eq:main} to compute new invariants as follows. Let $X$ and $Y$ be two CY threefolds. Assume that 
    \begin{compactenum}[(i)]
        \item They realize the same 5d SCFT, $\mathcal{T} = \mathcal{T}_{X} = \mathcal{T}_{Y}$ ; 
        \item We can identify the decoupled sectors in both $X$ and $Y$. 
    \end{compactenum}
    Then we can relate their invariants using 
    \begin{equation}\label{eq:main2}
        \frac{Z_{\textrm{top}}[X]}{Z_{\textrm{decoupled}}[X]} = Z_{\textrm{5d SCFT}} [\mathcal{T} ] = \frac{Z_{\textrm{top}}[Y]}{Z_{\textrm{decoupled}}[Y]} \, . 
    \end{equation}
    We apply this to the case where $X = K_{dP_n}$, which has $Z_{\textrm{decoupled}}[X] = 1$ because $dP_n$ is Fano, and $Y$ is a generalized del Pezzo surface or pure GTP geometry, with $Z_{\textrm{decoupled}}[Y] = Z_{\parallel}[Y]$ from \eqref{eq:main}.
Identifying $Z_{\textrm{decoupled}}$ case by case is not a bookkeeping technicality: it resolves a genuine puzzle about how one and the same $5d$ SCFT can be engineered on geometrically different surfaces. The $E_{n\leq 8}$ SCFTs \cite{Seiberg:1996bd,Morrison:1996xf} are related to one another by mass deformations, which in geometric engineering correspond to blowing up or down points on the surface $S$ defining the local threefold $K_S$. From the $5d$ perspective, which points are blown up should be immaterial, as long as an exceptional divisor is introduced; yet this choice visibly affects the geometry of $S$ -- for instance, whether $S$ remains toric. The resolution is that M-theory on $K_S$ depends on the \emph{full} geometry, so blowing up different points changes the full M-theory spectrum, and the $5d$ SCFT is only the part of that spectrum insensitive to the location of the blown-up points.  For instance, $K_{dP_4}$ is non-toric, and unlike $K_{dP_{5\leq n\leq 8}}$, its NLSM doesn't admit a convenient GLSM realization nor a convenient Picard Fuchs system\footnote{ The Picard Fuchs equation for $4d$ $KK$ theory from $5d$ $E_4$ SCFT can be found in \cite{Closset:2021lhd}. It would be interesting to understand it from enumerative geometry perspective.}. The E-string method could in principle give the topological string partition function of $K_{dP_{4}}$, but it seems hard to go to high degree. However, $dP_4$ and the toric $GdP_4$ engineer the same $5d$ $E_4$ SCFT, and we can compute $Z_{\textrm{decoupled}}[GdP_4] = Z_{\parallel}[GdP_4]$ while $Z_{\textrm{decoupled}}[dP_4] = 1$: the additional sector of M-theory present on $GdP_4$ is entirely accounted for by the parallel branes, while $dP_4$, being Fano, carries no additional sector at all. The local $dP_4$ case illustrates the power of the method, the GV invariants being explicitly presented here for the first time, see Table \ref{tab:GVdP4}. At low degree, we checked our results match with E-string results.
\end{itemize} 

\paragraph{Outline.}
In Section \ref{sec:topologicalVertex}, we review the topological vertex formalism, including its extension to certain GTP geometries, and we determine how the topological string partition function transforms under flops and Hanany-Witten moves.  In Section \ref{sec:HWMori}, we spell out the geometric counterpart of the Hanany-Witten move at the level of curve classes and Mori cones, using the pairs $(\mathbb{F}_0, \mathbb{F}_2)$ and $(\mathbb{F}_1, \mathbb{F}_3)$ as detailed case studies. In Section \ref{sec:rank1}, we apply our strategy to the rank-1 $E_n$ SCFTs and their realizations on local del Pezzo and generalized del Pezzo surfaces.

\section{Topological Vertex and Hanany-Witten transitions}
\label{sec:topologicalVertex}

In this section, we review how to compute the topological string partition function $Z_{\text{top}}$ using the topological vertex, for some (generalized) toric geometries. Then we examine how $Z_{\text{top}}$ changes under flops and Hanany-Witten moves. 

\subsection{Formal prescription}

Recall that the partition function of topological strings on a toric Calabi-Yau threefold can be computed from the defining brane web using the topological vertex. In essence, one associates an integer partition to all 5-brane finite segments (and the trivial partition $\emptyset$ to the semi-infinite 5-branes), then multiplies contributions from vertices and 5-brane finite segments according to Table \ref{tab:TV}, and sums over all integer partitions. Our conventions for partition and Schur functions are gathered in Appendix \ref{app:notations}. 

\begin{table}[t]
    \centering
    \begin{tabular}{|c|c|}
    \hline 
    Brane web element & Topological Vertex Contribution \\ \hline 
       \raisebox{-.5\height}{\scalebox{.8}{\begin{tikzpicture}[>=stealth, thick]
  \draw[->] (-1,0) -- (-2,1);
  \draw (-2,-1) -- (-1,0);
  \draw (1,0) -- (2,1);
  \draw[->] (1,0) -- (2,-1);
  \draw[->] (-1,0) -- (1,0) node[midway, above] {$Q , \lambda$};
  \node[left] at (-2,1) {$(p_1, q_1)$};
  \node[right] at (2,-1) {$(p_2, q_2)$};
\end{tikzpicture}}}  &  $(-Q)^{|\lambda|} \left[ (-1)^{|\lambda|} q^{\frac{1}{2} \kappa_\lambda} \right]^{p_2 q_1 - p_1 q_2}$\\ \hline 
     \raisebox{-.5\height}{\scalebox{.8}{\begin{tikzpicture}
      \draw[->] (0,0) -- (.8,0);
      \draw[->] (0,0) -- (-.5,.75);
      \draw[->] (0,0) -- (-.5,-.75);
      \node at (.8,.3) {$\lambda$};
      \node at (-.8,.75) {$\mu$};
      \node at (-.8,-.75) {$\nu$};
 \end{tikzpicture}}}      &  $C_{\nu \mu \lambda} = q^{\frac{1}{2} (\kappa_{\mu} + \kappa_{\lambda})} s_{\lambda} (q^{-\rho}) \sum\limits_{\sigma} s_{\nu^T / \sigma} (q^{-\lambda - \rho}) s_{\mu / \sigma} (q^{-\lambda^T - \rho})$ \\ \hline \hline 
 \scalebox{.8}{\raisebox{-.4\height}{     \begin{tikzpicture}
    \tikzmath{\x = .1;}
    \node at (\x/2,3) {$\bullet$};
    \draw (0,1) -- (0,2-2*\x) node[midway,left]{$\mu$};
    \draw (0,2- 2*\x) arc (270:90:\x);
    \draw (0,2.0) -- (0,3);
    \draw (1,1) -- (\x,2-\x) node[midway,right]{$\nu$};
    \draw (\x,2-\x) -- (\x,3) node[midway,right] {$\emptyset$};
    \draw (\x,2-\x) -- (-1,2-\x)node[midway,above]{$\lambda$};
    \node at (0,.9) {};
\end{tikzpicture}}} & $ \delta_{\mu\emptyset} C_{\lambda\emptyset\nu}(q)$ \\ \hline 
    \end{tabular}
    \caption{Rules for computing $Z_{\mathrm{top}}$ from the topological vertex. Here $\lambda, \mu , \nu , \sigma$ are integer partitions, $Q$ is an exponentiated complexified K\"ahler parameter, and $q = e^{i g_s}$. See Appendix \ref{app:notations} for details. Note that $C_{\emptyset \mu \lambda^T} = q^{\frac{1}{2} \kappa_{\mu} } s_{\lambda} (q^{-\rho}) s_{\mu } (q^{-\lambda  - \rho})$. The last row tells how to compute in the presence of a 7-brane (represented by a black disk). }
    \label{tab:TV}
\end{table}

\paragraph{Examples. }
For the web associated with the conifold, we have only one internal edge. We associate the empty partition to the external edges. Applying the topological vertex rules, one finds 
\begin{equation}
    Z_{\mathrm{top}} \left[ \scalebox{.7}{\raisebox{-.5\height}{\begin{tikzpicture}
  \draw[mid arrow] (0,0)--(1,1) node[midway, above left]{$Q,\lambda$};
  \draw[mid arrow] (1,1)--(2,1);
  \draw[mid arrow] (1,1)--(1,2);
  \draw[mid arrow] (0,-1)--(0,0);
  \draw[mid arrow] (-1,0)--(0,0);
\end{tikzpicture}}} \right] =  \sum\limits_{\lambda} (-Q)^{|\lambda|} s_{\lambda} (q^{-\rho}) s_{\lambda^T} (q^{-\rho}) = \mathcal{R}_{00} (Q,q) = \operatorname{PExp} \left( - \frac{Q q}{(1-q)^2} \right)  \, .  \label{eq:conifold}
\end{equation}
using \eqref{eq:CauchyIdentitiesR} and \eqref{eq:R00}. 
Similarly one can compute, using \eqref{eq:CauchyIdentitiesR} repeatedly \cite{Iqbal:2004ne}  
\begin{equation} 
Z \left[     \raisebox{-.5\height}{\scalebox{.7}{\begin{tikzpicture}
  \draw[mid arrow] (-2,-1.5)--(-1,-1);
  \draw[mid arrow] (-1,-1)--(0,0) node[midway, above left]{$Q_1 , \nu_1$};
  \draw[mid arrow] (0,0)--(0,1) node[midway, right]{$\dots$};
  \draw[mid arrow] (0,1)--(-1,2) node[midway, left]{$Q_{N-1}, \nu_{N-1}$};
  \draw[mid arrow] (-1,2)--(-2,2.5);
  \draw[mid arrow] (-1,-1)--(2,-1) node[midway, above]{$\lambda_0$};
  \draw[mid arrow] (0,0)--(2,0) node[midway, above]{$\lambda_1$};
  \draw[mid arrow] (0,1)--(2,1) node[midway, above]{$\lambda_{N-2}$};
  \draw[mid arrow] (-1,2)--(2,2) node[midway, above]{$\lambda_{N-1}$};
\end{tikzpicture}}} \right] = \prod\limits_{j=0}^{N-1}   s_{\lambda_j^T} (q^{-\rho}) \, \, \prod\limits_{0 \leq i < j \leq N-1} \frac{1}{\mathcal{R}_{\lambda_i^T \lambda_j} (Q_{i+1} \cdots Q_j , q)}   \, .  \label{eq:strip}
\end{equation} 
These expressions can be computed exactly in closed form because there is no closed loop in the brane web. An example with a rank-1 theory is\footnote{The sums over $\mu$ and $\sigma$ is performed using \eqref{eq:strip}, leaving the sums over $\lambda$ and $\nu$. One then uses \eqref{eq:CauchyIdentitiesR} and finally \eqref{eq:RtoN} to introduce the factor $\mathcal{N}$. } 
\begin{align} 
    Z_{\text{top}} \left[ \scalebox{.5}{\raisebox{-.5\height}{\begin{tikzpicture}[xscale=1.2]
  \draw[mid arrow] (0,0)--(1,1) node[midway, left]{$Q_1 , \mu$};
  \draw[mid arrow] (1,1)--(2,1) node[midway, above]{$Q_2 , \lambda$};
  \draw[mid arrow] (2,1)--(2,0) node[midway, right]{$Q_1 , \sigma$};
  \draw[mid arrow] (2,0)--(0,0) node[midway, below]{$Q_1^m Q_2, \nu$};
  \draw[mid arrow] (0,0)--(-1,-.5) node[midway, left]{$(-2,-1) \quad$};
  \draw[mid arrow] (1,1)--(1,2) node[midway, left]{$(0,1)$};
  \draw[mid arrow] (2,1)--(3,2) node[midway, right]{$(2-m,1)$};
  \draw[mid arrow] (2,0)--(3,-1) node[midway, right]{$(m,-1)$};
\end{tikzpicture}}} \right] =& \operatorname{PExp} \left[  2 Q_1\frac{q}{(1-q)^2} \right]  \times \\ & \sum\limits_{\lambda , \nu} ((-1)^m Q_2)^{|\lambda| + |\nu|} Q_1^{m |\nu|} q^{\frac{m}{2} (\kappa_\lambda -\kappa_\nu)} \left( \frac{s_{\lambda} (q^{-\rho}) s_{\nu} (q^{-\rho})}{ \mathcal{N}_{\lambda^T \nu} (Q_1 ; q) }\right)^2 \, .  \label{eq:FmZtop}
\end{align} 
Here the result is expressed as a sum and can not be simplified further. However, in Section \ref{sec:rank1} we rewrite this expression in the form of a Nekrasov instanton counting partition function. 

In \cite{Hayashi:2015xla}, the authors propose an adapted topological vertex which can be applied to the geometries associated with some GTPs. The GTPs in question are the ones with a tessellation using only lattice triangles of area 1/2 (as in the toric case) and trapezoids which are $SL(2,\mathbb{Z})$-equivalent to  
\begin{equation} \label{eq:simpleCrossing}
\raisebox{-.5\height}{ \scalebox{1}{  
\begin{tikzpicture}[x=.5cm,y=.5cm] 
\draw[gridline] (0,-0.5)--(0,1.5); 
\draw[gridline] (1,-0.5)--(1,1.5); 
\draw[gridline] (2,-0.5)--(2,1.5); 
\draw[gridline] (3,-0.5)--(3,1.5); 
\draw[gridline] (4,-0.5)--(4,1.5); 
\draw[gridline] (-0.5,0)--(4.5,0); 
\draw[gridline] (-0.5,1)--(4.5,1); 
\draw[ligne] (0,1)--(0,0); 
\draw[ligne] (0,0)--(3,0); 
\draw[ligne] (3,0)--(4,1); 
\draw[ligne] (4,1)--(0,1); 
\node[bd] at (0,0) {}; 
\node[bd] at (3,0) {}; 
\node[bd] at (4,1) {}; 
\node[bd] at (0,1) {}; 
\node[wd] at (1,0) {}; 
\node[wd] at (2,0) {}; 
\node[wd] at (3,1) {}; 
\node[wd] at (2,1) {}; 
\node[wd] at (1,1) {}; 
\draw [decorate,decoration={brace,amplitude=8pt},yshift=2pt]  (3.2,-.4) -- (-.2,-.4) node [black,midway,yshift=-14pt] {$n$};
\end{tikzpicture}}} \qquad \longleftrightarrow \qquad   
\raisebox{-.5\height}{\begin{tikzpicture}
    \tikzmath{\x = .1;}
    \tikzmath{\t = .1;}
    \foreach \i in {-2,-1,0} {
        \draw (\i*\t, 1) -- (\i*\t, 2-2*\x);
        \draw (\i*\t, 2-2*\x) arc (270:90:\x);
        \draw (\i*\t, 2.0) -- (\i*\t, 3);     }
    \draw (1,1) -- (\x,2-\x);
    \draw (\x,2-\x) -- (\x,3);
    \draw (\x,2-\x) -- (-1,2-\x);
\end{tikzpicture}}
\end{equation}
corresponding to the shown "simple crossings". In the language of \cite{CarrenoBolla:2024fxy}, this corresponds to trivial T-cones only. 
In the present paper, we will only need the simplest case, namely \eqref{eq:simpleCrossing} where the two coincident branes are external : they do not meet any other 5-branes and end on the same 7-brane. In that case, we simply use the entry in the third line of Table \ref{tab:TV}. 

\subsection{Decoupled sectors for GTPs}\label{sec:Zdec_GTP}

For GTP geometries, dual to a brane web, the decoupled sector is fully accounted for by semi-infinite parallel branes, which support higher dimensional non-abelian excitations: 
\begin{equation}
Z_{\rm{decoupled}}[X[P]] = Z_{\parallel}[P]
\end{equation}

Geometrically, in the toric case, these give rise to lines of $\bC^2/\mathbb Z_n$ singularities, as we discuss in details in Appendix \ref{app:GVcomplexStructureVariation}.

The contribution to $ Z_{\parallel}$ of a set of $N$ parallel 5-branes (without 7-branes) is given by equation \eqref{eq:strip}, with trivial external partitions $\lambda_i =\emptyset$. This gives :
\begin{equation}\label{eq:Zpar_strip}
    Z_{\parallel}[N \text{ parallel branes}] = \operatorname{PExp}\left(\frac{q}{(1-q)^2}\sum_{0\leq i<j\leq N-1} Q_{i+1} Q_{i+2} \cdots Q_j\right)
\end{equation}

Here, $Q_i$ is the Kähler moduli of the $(0,-2)$ rational curve extended between the $i-1$ and $i$-th branes. In general, it may be given by a sequence of edges in the brane web. For example, for the brane web in Figure \ref{fig:strip_web} below, there are two sets of two parallel branes, giving :
\begin{equation}
    Z_{\parallel} \left[\raisebox{-.5\height}{\scalebox{.8}{\begin{tikzpicture}[x=.7cm,y=.7cm,every node/.style={inner sep=2pt}]
        \draw (-1,0) -- (0,0) -- (0,-1);
        \draw (0,0) -- (1,1) node[midway, below right,inner sep=0pt] {$Q_1$};
        \draw (1,1) -- (2.5,1);
        \draw (1,1) -- (1,2) node[midway,left] {$Q_2$};
        \draw (1,2) -- (-.5,2);
        \draw (1,2) -- (2,3) node[midway,below right,inner sep=0pt] {$Q_3$};
        \draw (3,3) -- (2,3) -- (2,4);
    \end{tikzpicture}}}\right] = \operatorname{PExp}\left[\frac{q}{(1-q)^2}\left(Q_1 Q_2 + Q_2Q_3\right)\right]
\end{equation}
In total there are $\frac{N(N-1)}{2}$ terms in the argument of the plethystic exponential.

For a GTP with a set of $N$ parallel branes ending on $7$-branes according to a partition $\lambda \vdash N$, the situation can be described as follows : with $\mu = \lambda^T$, there are $\mu_1$ $7$-branes, each with a set of $5$-branes ending on it. In each of these sets, one $5$-brane ends on a first strip, whose contribution is given by \eqref{eq:Zpar_strip} with $N=\mu_1$ and the suitable Kähler parameters. The other branes go through this strip, following the last rule in  Table \ref{tab:TV}. For the $\mu_2$ 7-branes with two or more $5$-branes, another $5$-branes terminates on a second strip, giving again a factor of \eqref{eq:Zpar_strip} (this time with $N=\mu_2$), and so on. In total, we get the plethystic exponential of a sum containing $\frac{1}{2}\sum_i \mu_i (\mu_i-1)$ terms.

For example, for the brane web shown in figure \ref{fig:web_SU3_wd} below, we have one set of $4$ parallel branes ending on $7$-branes according to $\lambda = (2^2)$. Following the prescription described above, we get: 
\begin{equation}
Z_{\parallel}\left[
    \raisebox{-.5\height}{\scalebox{.75}{
        \begin{tikzpicture}[every node/.style={inner sep=2pt},x=.8cm,y=.8cm]
        \tikzmath{\ee = .1;};
            \draw (\ee,0) -- (\ee,1-\ee) --node[midway,below left,inner sep=0pt] {$Q_2^2$} (-2,3) -- (-3,3.5);
            \draw (-2,3) --node[midway,above] {$Q_2$} (-1,3) --node[midway,above right,inner sep=0pt] {$Q_2$} (0,2) -- node[midway,below] {$Q_1$}(2,2) --node[midway,above left,inner sep=0pt] {$Q_2$} (3,3) --node[midway,above] {$Q_2$} (4,3) -- (5,3.5);
            \draw (-1,3) --node[midway,right] {$Q_2$} (-1,4) --node[midway,above] {$Q_1Q_2^2$} (3,4) --node[midway,left] {$Q_2$} (3,3);
            \draw (-1,4) -- (-2,5);
            \draw (3,4) -- (4,5);
            \draw (2-\ee,0) -- (2-\ee,1-\ee) --node[midway,below right,inner sep=0pt] {$Q_2^2$} (4,3);
            \draw  (\ee,1-\ee)--node[midway,below] {$Q_1$} (2-\ee,1-\ee);
            \draw (0,2) -- (0,1+\ee) arc(90:270:\ee) -- (0,0);
            \draw (2,2) -- (2,1+\ee) arc(90:-90:\ee) -- (2,0);
            \end{tikzpicture}}}
    \right] = Z_{\rm{top}}\left[
    \raisebox{-.5\height}{\scalebox{.75}{
    \begin{tikzpicture}[every node/.style={inner sep=2pt},x=.8cm,y=.8cm]
        \tikzmath{\ee = .1;};
        \draw (\ee,0) -- (\ee,1-\ee) -- (-1,2);
        \draw  (-1,3) -- (0,2) -- node[midway,below] {$Q_1$}(2,2)  -- (3,3);
        \draw (2-\ee,0) -- (2-\ee,1-\ee) -- (3,2);
        \draw  (\ee,1-\ee)--node[midway,below] {$Q_1$} (2-\ee,1-\ee);
        \draw (0,2) -- (0,1+\ee) arc(90:270:\ee) -- (0,0);
        \draw (2,2) -- (2,1+\ee) arc(90:-90:\ee) -- (2,0);
    \end{tikzpicture}
    }}
    \right] = \operatorname{PExp} \left[\frac{2Q_1q}{(1-q)^2} \right] \, . 
\end{equation}

\subsection{\texorpdfstring{$\mathbb{P}^1$}{P1} curves}

From the topological string partition function $Z_{\text{top}}$, one can extract the Gopakumar-Vafa invariants using equation \eqref{eq:GVPlethystic}.  
We first consider the case of a local $\mathbb P^1$ curve. Embedded in a Calabi-Yau threefold, its normal bundle is of the form $\mathcal{O}(n-1) \oplus \mathcal{O}(-n-1)$ for some integer $n \geq 0$. For $n =0$, the total space of this bundle is the resolved conifold. For $n=1$, the total space is the product of $\mathbb C$ with a resolved $A_1$ singularity. In these two cases, the curves are shrinkable and give rise to rank $0$ SCFTs. For $n=2$, the curve is no longer shrinkable. It can nevertheless be deformed to give rise to an SCFT. We will come back to this point later in Section \ref{sec:HWMori}.

The topological string partition function is :
\begin{equation}
    Z_{\text{top}} = \sum_{\lambda} Q^{|\lambda|} (-1)^{(n+1)|\lambda|}q^{n\kappa_\lambda/2}s_\lambda(q^{-\rho}) s_{\lambda^T}(q^{-\rho})
\end{equation}
For $n=0,1$, the sum over $\lambda$ simplifies nicely, and only one GV-invariant is non-zero :
\begin{subequations}
\begin{align}
Z_{\text{top}}^{n=0} &= \operatorname{PExp}\left(-\frac{q Q}{(1-q)^2}\right)\\
Z_{\text{top}}^{n=1} &= \operatorname{PExp}\left(\frac{q Q}{(1-q)^2}\right)
\end{align}
\end{subequations}
The non-trivial invariants are, for $n=0$, $\GV_{0,1} = 1$, and for $n=1$, $\GV_{0,1} = -1$. 
For $n=2$, no such simplification occurs and there are infinitely many non-trivial GV-invariants. Their values in low degree are shown in Table \ref{tab:GV(1,-3)curve}.

\begin{table}[t]
    \centering
\begin{tabular}{c|ccccccccccc}
 $d \backslash g$&  0 & 1 & 2 & 3 & 4 & 5 & 6 & 7 & 8 & 9 & 10 \\
\hline
0 &  &   &   &   &   &   &   &  &  & &   \\
1 &   1 &   &   &   &   &   &   &  &  & &   \\
2 &  -1 &   &   &   &   &   &   &  &  & &   \\
3 &   2 &   -1 &   &   &   &   &   &  &  & &   \\
4 &  -7 &   11 &   -6 &    1 &   &   &   &  &  & &   \\
5 &  31 & -106 &  161 & -130 &   56 &  -12 &    1 &  &  &   &   \\
6 & -156 & 975 & -2956 & 5404 & -6361 & 4927 & -2516 & 834 & -172 & 20 & -1
\end{tabular}

    \caption{GV invariants of the local $(1,-3)$-curve in low degree.}
    \label{tab:GV(1,-3)curve}
\end{table}

\subsection{Hanany-Witten transitions}
\label{sec:HWtopologicalVertex}

In the brane web picture, a $7$-brane can be brought to finite distance, with one or more $5$-brane ending on it. As the $7$-brane is moved through the brane web, it causes Hanany-Witten transitions, in  which $5$-branes are created or destroyed. Once on the other side, the $7$-brane can be sent to infinity. We say that the Calabi-Yau threefold dual to the brane webs before and after this process are related by a Hanany-Witten move. Crucially, the low energy effective theory does not change during this process. Therefore, using \eqref{eq:main}-\eqref{eq:main2} we can relate the topological string partition functions of geometries related by a Hanany-Witten move. We now describe some of the simplest examples of this phenomenon.

\subsubsection{Closed topological vertex and quotient of conifold}

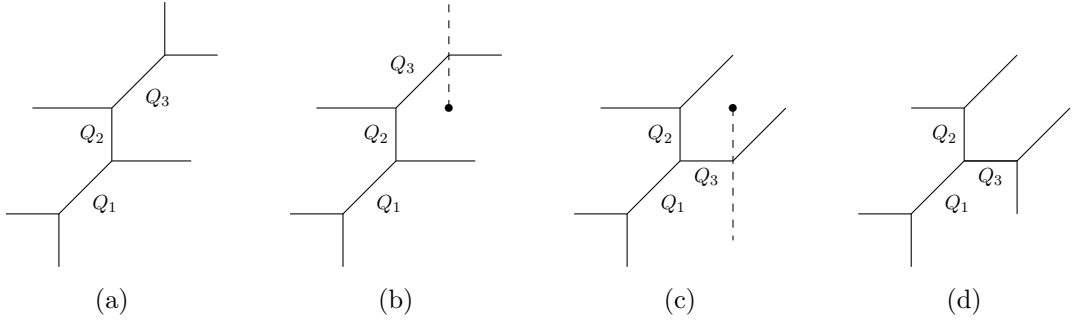
\begin{figure}[t]
\centering
\begingroup
\tikzset{every picture/.append style={scale=0.7, transform shape}}
\begin{subfigure}[t]{0.24\textwidth}
    \centering
    \begin{tikzpicture}
        \draw (-1,0) -- (0,0) -- (0,-1);
        \draw (0,0) -- (1,1) node[midway, below right] {$Q_1$};
        \draw (1,1) -- (2.5,1);
        \draw (1,1) -- (1,2) node[midway,left] {$Q_2$};
        \draw (1,2) -- (-.5,2);
        \draw (1,2) -- (2,3) node[midway,below right] {$Q_3$};
        \draw (3,3) -- (2,3) -- (2,4);
    \end{tikzpicture}
    \caption{}\label{fig:strip_web}
\end{subfigure}
\begin{subfigure}[t]{0.24\textwidth}
    \centering
    \begin{tikzpicture}
        \draw (-1,0) -- (0,0) -- (0,-1);
        \draw (0,0) -- (1,1) node[midway, below right] {$Q_1$};
        \draw (1,1) -- (2.5,1);
        \draw (1,1) -- (1,2) node[midway,left] {$Q_2$};
        \draw (1,2) -- (-.5,2);
        \node at (2,2) {$\bullet$};
        \draw[dashed] (2,2) -- (2,4);
        \draw (1,2) -- (2,3) node[midway,above left] {$Q_3$};
        \draw (2,3) -- (3,3);
    \end{tikzpicture}
    \caption{}
\end{subfigure}
\begin{subfigure}[t]{0.24\textwidth}
    \centering
    \begin{tikzpicture}
        \draw (-1,0) -- (0,0) -- (0,-1);
        \draw (0,0) -- (1,1) node[midway, below right] {$Q_1$};
        \draw (1,1) -- (2,1) node[midway,below ] {$Q_3$};;
        \draw (2,1) -- (3,2);
        \draw (1,1) -- (1,2) node[midway,left] {$Q_2$};
        \draw (1,2) -- (-.5,2);
        \node at (2,2) {$\bullet$};
        \draw[dashed] (2,2) -- (2,-.5);
        \draw (1,2) -- (2,3) ;
    \end{tikzpicture}
    \caption{}
\end{subfigure}
\begin{subfigure}[t]{0.24\textwidth}
    \centering
  \begin{tikzpicture}
        \draw (-1,0) -- (0,0) -- (0,-1);
        \draw (0,0) -- (1,1) node[midway, below right] {$Q_1$};
        \draw (1,1) -- (2,1);
        \draw (1,1) -- (1,2) node[midway,left] {$Q_2$};
        \draw (2,3)--(1,2) -- (0,2);
        \draw (1,1) -- (2,1) node[midway,below ] {$Q_3$};
        \draw (2,0) -- (2,1) -- (3,2);
    \end{tikzpicture}
    \caption{}
\end{subfigure}
\endgroup
\caption{ Hanany-Witten transition between a strip (a) and the closed topological vertex (d). The black disk is a 7-brane of type $(0,1)$, with monodromy cut represented by a dashed line, which can be rotated freely as done between (b) and (c).  }
\label{fig:HW_closedTopVertex}
\end{figure}

The simplest example of Hanany-Witten move relates two toric geometries whose topological string partition function admit closed form expressions : the closed topological vertex \cite{Karp_2006} and a quotient of a conifold by a $\mathbb Z/2$ action, which is an example of a strip geometry \cite{Iqbal:2004ne}. The two brane webs and the transition between them are shown in Figure \ref{fig:HW_closedTopVertex}. The topological string partition functions are :
\begin{gather}
Z_{\text{strip}} = \operatorname{PExp}\left[ \frac{q}{(1-q)^2} (-Q_1 - Q_2 - Q_3 + Q_1 Q_2 + Q_2 Q_3 - Q_1 Q_2 Q_3)\right] \\
Z_{\text{closed top. vert.}} = \operatorname{PExp}\left[ \frac{q}{(1-q)^2} (-Q_1 - Q_2 - Q_3 + Q_1 Q_2 + Q_2 Q_3 + Q_3 Q_1 - Q_1 Q_2 Q_3)\right]
\end{gather}
Both geometries contain parallel branes. After removing these contributions, we see that the partition function of the two rank $0$ SCFTs match :
\begin{align*}
    Z_{SCFT} 
    &= \frac{\operatorname{PExp}\left[ \frac{q}{(1-q)^2} (-Q_1 - Q_2 - Q_3 + Q_1 Q_2 + Q_2 Q_3 - Q_1 Q_2 Q_3)\right]}{\operatorname{PExp}\left[ \frac{q}{(1-q)^2} (Q_1 Q_2 + Q_2 Q_3)\right]}\\
    &= \operatorname{PExp}\left[ \frac{q}{(1-q)^2} (-Q_1 - Q_2 - Q_3 - Q_1 Q_2 Q_3)\right] \, , 
\end{align*}
and 
\begin{align*}
    Z_{SCFT} 
    &= \frac{\operatorname{PExp}\left[ \frac{q}{(1-q)^2} (-Q_1 - Q_2 - Q_3 + Q_1 Q_2 + Q_2 Q_3 + Q_3 Q_1 - Q_1 Q_2 Q_3)\right]}{\operatorname{PExp}\left[ \frac{q}{(1-q)^2} (Q_1 Q_2 + Q_2 Q_3 + Q_3 Q_1)\right]}\\
    &= \operatorname{PExp}\left[ \frac{q}{(1-q)^2} (-Q_1 - Q_2 - Q_3 - Q_1 Q_2 Q_3)\right] \, . 
\end{align*}

\subsubsection{Local \texorpdfstring{$\mathbb F_0$}{F0} and \texorpdfstring{$\mathbb F_2$}{F2}}

Our next example relates the two toric local Hirzebruch surfaces $\mathbb F_0$ and $\mathbb F_2$ (see Figure \ref{fig:HW_F0F2F1F3}). These engineer the $E_1$ SCFT which admits a mass deformation flowing to pure $\mathcal N=1$ $\operatorname{SU}(2)$ super-Yang-Mills. As $\mathbb F_0$ is Fano, there are no decoupled contributions to remove. On the other hand $\mathbb F_2$ is only weak Fano, so there is a contribution from two parallel branes and we get :

\begin{equation}\label{eq:HW_F0F2}
\frac{Z \left[ \scalebox{.5}{\raisebox{-.5\height}{      \begin{tikzpicture}
        \draw (0,0) -- (0,1);
        \draw (0,0) -- (1,0) node[midway,above] {$Q_2$};
        \draw (1,0) -- (1,1);
        \draw (1,0) -- (2,-1) node[midway,above right] {$Q_1$};
        \draw (0,0) -- (-1,-1) node[midway,above left] {$Q_1$};
        \draw (-1,-1) -- (2,-1) node[midway,below] {$Q_2 Q_1^2$};
        \draw (-1,-1) -- (-2,-1.5);
        \draw (2,-1) -- (3,-1.5);
    \end{tikzpicture}}} \right] }{\operatorname{PExp}\left( \frac{q Q_2}{(1-q)^2}\right) }
=   Z \left[ \scalebox{.5}{\raisebox{-.5\height}{ 
    \begin{tikzpicture}
        \draw (0,0) -- (0,1);
        \draw (0,0) -- (2,0) node[midway,above] {$ Q_1Q_2$};
        \draw (1,-1) -- (1,-2);
        \draw (2,0) -- (1,-1) node[midway,below right] {$Q_1$};
        \draw (0,0) -- (-1,-1) node[midway,above left] {$Q_1$};
        \draw (-1,-1) -- (1,-1) node[midway,below] {$Q_2 Q_1$};
        \draw (-1,-1) -- (-2,-1.5);
        \draw (2,0) -- (3,.5);
    \end{tikzpicture}}} \right]  
\end{equation}
The left hand side is divided by the contribution $\operatorname{PExp}\left( \frac{q Q_2}{(1-q)^2}\right)$ of the two parallel external branes, an inverse conifold factor in the curve class of the base $b_2$ (with K\"ahler parameter $Q_2$). The right hand side is the topological string partition function of the local $\mathbb F_0$ geometry, with the K\"ahler parameters matched through the Hanany-Witten move as indicated on the brane webs: the fiber classes are identified, while the base class of $\mathbb F_0$ maps to the sum of the base and fiber classes of $\mathbb F_2$. The equality states that, once the decoupled parallel-brane sector is removed, the two geometries carry the same BPS spectrum, as expected since both engineer the $E_1$ SCFT. We have verified this identity order by order in a series expansion in the K\"ahler parameters.

\subsubsection{Local \texorpdfstring{$\mathbb F_1$}{F1} and a GTP geometry} 

\begin{figure}[t]
\centering
\begingroup
\tikzset{every picture/.append style={scale=0.65, transform shape}}
\makebox[\textwidth][c]{\makebox[1.18\textwidth][c]{%
\begin{subfigure}[t]{0.28\textwidth}
    \centering
      \begin{tikzpicture}
        \draw (0,0) -- (0,1);
        \draw (0,0) -- (1,0) node[midway,above] {$Q_2$};
        \draw (1,0) -- (1,1);
        \draw (1,0) -- (2,-1) node[midway,above right] {$Q_1$};
        \draw (0,0) -- (-1,-1) node[midway,above left] {$Q_1$};
        \draw (-1,-1) -- (2,-1) node[midway,below] {$Q_2 Q_1^2$};
        \draw (-1,-1) -- (-2,-1.5);
        \draw (2,-1) -- (3,-1.5);
    \end{tikzpicture}
    \caption{}
\end{subfigure}\hfill
\begin{subfigure}[t]{0.28\textwidth}
    \centering
    \begin{tikzpicture}
        \draw (0,0) -- (0,1);
        \draw (0,0) -- (1,0) node[midway,above] {$Q_2$};
        \node at (1,-.5) {$\bullet$};
        \draw[dashed] (1,-.5) -- (1,1);
        \draw (1,0) -- (2,-1) node[midway,above right] {$Q_1$};
        \draw (0,0) -- (-1,-1) node[midway,above left] {$Q_1$};
        \draw (-1,-1) -- (2,-1) node[midway,below] {$Q_2 Q_1^2$};
        \draw (-1,-1) -- (-2,-1.5);
        \draw (2,-1) -- (3,-1.5);
    \end{tikzpicture}
    \caption{}
\end{subfigure}\hfill
\begin{subfigure}[t]{0.28\textwidth}
    \centering
    \begin{tikzpicture}
        \draw (0,0) -- (0,1);
        \draw (0,0) -- (2,0) node[midway,above] {$ Q_1Q_2$};
        \node at (1,-.5) {$\bullet$};
        \draw[dashed] (1,-.5) -- (1,-2);
        \draw (2,0) -- (1,-1) node[midway,below right] {$Q_1$};
        \draw (0,0) -- (-1,-1) node[midway,above left] {$Q_1$};
        \draw (-1,-1) -- (1,-1) node[midway,below] {$Q_2 Q_1$};
        \draw (-1,-1) -- (-2,-1.5);
        \draw (2,0) -- (3,.5);
    \end{tikzpicture}
    \caption{}
\end{subfigure}\hfill
\begin{subfigure}[t]{0.28\textwidth}
    \centering
    \begin{tikzpicture}
        \draw (0,0) -- (0,1);
        \draw (0,0) -- (2,0) node[midway,above] {$ Q_1Q_2$};
        \draw (1,-1) -- (1,-2);
        \draw (2,0) -- (1,-1) node[midway,below right] {$Q_1$};
        \draw (0,0) -- (-1,-1) node[midway,above left] {$Q_1$};
        \draw (-1,-1) -- (1,-1) node[midway,below] {$Q_2 Q_1$};
        \draw (-1,-1) -- (-2,-1.5);
        \draw (2,0) -- (3,.5);
    \end{tikzpicture}
    \caption{}
\end{subfigure} 
}}

\makebox[\textwidth][c]{\makebox[1.18\textwidth][c]{%
        \begin{subfigure}[t]{0.28\textwidth}
            \centering
            \begin{tikzpicture}
                \draw (0,0) -- (-1,1) node[midway, left] {$Q_1$};
                \draw (-1,1) -- (1,1) node[midway, above] {$Q_1 Q_2$};
                \draw (1,1) -- (3,0) node[midway, above right] {$Q_1$};
                \draw (3,0) -- (0,0) node[midway,below] {$Q_1^2 Q_2$};
                \draw (0,0) -- (0,-1);
                \draw (-1,1) -- (-2,1.5);
                \draw (1,1) -- (0,2);
                \draw (3,0) -- (3.6,-.2);
            \end{tikzpicture}%
            \caption{}
        \end{subfigure}\hfill
        \begin{subfigure}[t]{0.28\textwidth}
            \centering
            \begin{tikzpicture}
                \draw (0,0) -- (-1,1) node[midway, left] {$Q_1$};
                \draw (-1,1) -- (1,1) node[midway, above] {$Q_1 Q_2$};
                \draw (1,1) -- (3,0) node[midway, above right] {$Q_1$};
                \draw (3,0) -- (0,0) node[midway,below] {$Q_1^2 Q_2$};
                \node at (0,.5) {$\bullet$};
                \draw[dashed] (0,.5) -- (0,-1);
                \draw (-1,1) -- (-2,1.5);
                \draw (1,1) -- (0,2);
                \draw (3,0) -- (3.6,-.2);
            \end{tikzpicture}%
            \caption{}
        \end{subfigure}\hfill
        \begin{subfigure}[t]{0.28\textwidth}
            \centering
            \begin{tikzpicture}
                \draw (-1,0) -- (0,1) node[midway, above left] {$Q_1$};
                \draw (0,1) -- (1,1) node[midway, below] {$ Q_2$};
                \draw (1,1) -- (3,0) node[midway, above right] {$Q_1$};
                \draw (3,0) -- (-1,0) node[midway,below] {$Q_1^3 Q_2$};
                \node at (0,.5) {$\bullet$};
                \draw[dashed] (0,.5) -- (0,3);
                \draw (-1,0) -- (-2,-.5);
                \draw (1,1) -- (0,2)node[midway, above right] {$Q_2$};
                \draw (3,0) -- (3.6,-.2);
                \draw (0,2) -- (-1,2);
            \end{tikzpicture}%
            \caption{}
        \end{subfigure}\hfill
        \begin{subfigure}[t]{0.28\textwidth}
            \centering
            
            \begin{tikzpicture}
                \tikzmath{\x = .07;}
                 \draw (-1,0) -- (0,1) node[midway, above left] {$Q_1$};
                \draw (0,1) -- (1,1) node[midway, below] {$ Q_2$};
                \draw (1,1) -- (3,0) node[midway, above right] {$Q_1$};
                \draw (3,0) -- (-1,0) node[midway,below] {$Q_1^3 Q_2$};
                \draw (0,1) -- (0,2-2*\x) node[midway, left] {$ Q_2$};
                \draw (0,2- 2*\x) arc (270:90:\x);
                \draw (0,2.0) -- (0,3);
                \draw (-1,0) -- (-2,-.5);
                \draw (1,1) -- (\x,2-\x) node[midway, above right] {$Q_2$};
                \draw (\x,2-\x) -- (\x,3);
                \draw (3,0) -- (3.6,-.2);
                \draw (\x,2-\x) -- (-1,2-\x);
            \end{tikzpicture}
            \caption{}
        \end{subfigure}%
    }} 
\endgroup
\caption{\textit{Top:} Hanany-Witten transition between a local $\mathbb F_2$ surface (a) and a local $\mathbb F_0$ surface (d). \textit{Bottom:} Hanany-Witten transition between a toric local $\mathbb F_1$ surface (e) and a generalized toric CY3 (h). }
\label{fig:HW_F0F2F1F3}
\end{figure}
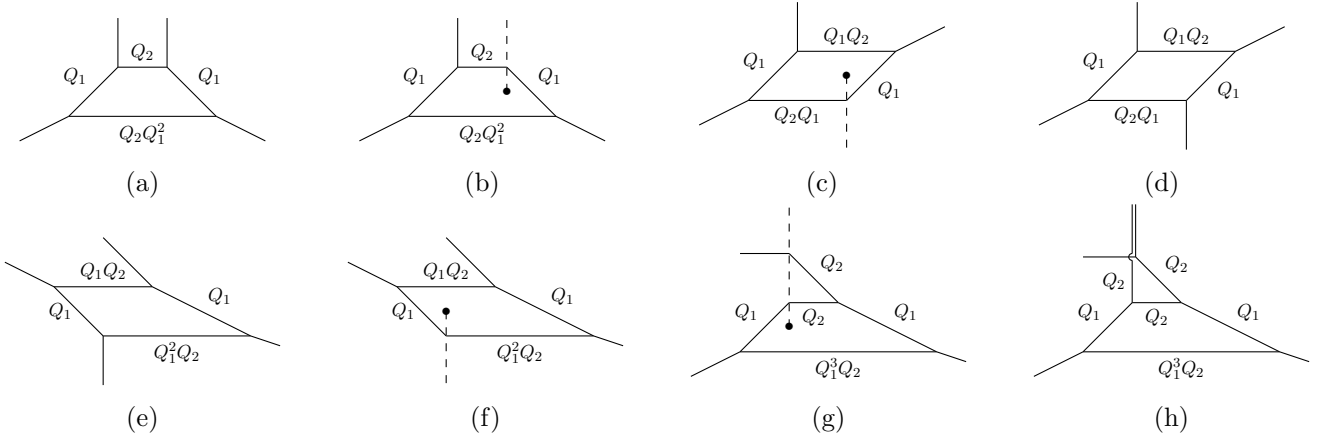

If we perform a Hanany-Witten move starting with the local Hirzebruch surface $\mathbb F_1$ (see Figure \ref{fig:HW_F0F2F1F3}), the $7$-brane comes out with two $5$-branes ending on it. The associated geometry is therefore no longer toric : it is a GTP geometry. We explain how it differs from the local $\mathbb F_3$ threefold in Section \ref{sec:F3vsGTP}. Nonetheless, we can use the last row of Table \ref{tab:TV} to express the topological string partition function on this generalized toric CY3 as 
\begin{equation}
\begin{split}
    &Z \left[ \scalebox{.5}{\raisebox{-.5\height}{\begin{tikzpicture}
    \tikzmath{\x = .1;}
     \draw (-1,0) -- (0,1) node[midway, above left] {$Q_1$};
    \draw (0,1) -- (1,1) node[midway, below] {$ Q_2$};
    \draw (1,1) -- (3,0) node[midway, above right] {$Q_1$};
    \draw (3,0) -- (-1,0) node[midway,below] {$Q_1^3 Q_2$};
    \draw (0,1) -- (0,2-2*\x) node[midway, left] {$ Q_2$};
    \draw (0,2- 2*\x) arc (270:90:\x);
    \draw (0,2.0) -- (0,3);
    \draw (-1,0) -- (-2,-.5);
    \draw (1,1) -- (\x,2-\x) node[midway, above right] {$Q_2$};
    \draw (\x,2-\x) -- (\x,3);
    \draw (3,0) -- (4.5,-.5);
    \draw (\x,2-\x) -- (-1,2-\x);
\end{tikzpicture}}} \right] =\\ &\mspace{100mu}\sum_{\lambda,\mu} (-Q_2)^{|\mu|} (-Q_1^3 Q_2)^{|\lambda|} q^{\kappa_\mu + 2\kappa_\lambda}  Z \left[ \scalebox{.5}{\raisebox{-.5\height}{\begin{tikzpicture}
        \draw (-1,0) -- (0,1) node[midway, above left] {$Q_1$};
        \draw (0,1) -- (1,1) node[right] {$\mu$};
        \draw  (-1,0)--(1,0) node[right] {$\lambda$};
        \draw (0,1) -- (0,2);
        \draw (-1,0) -- (-2,-.5);\end{tikzpicture}}} \right]Z \left[ \scalebox{.5}{\raisebox{-.5\height}{\begin{tikzpicture}\draw (5,1) -- (7,0) node[midway, above right] {$Q_1$};
        \draw (7,0) -- (4,0) node[left] {$\lambda$};
        \draw (5,1) -- (4,1) node[left] {$\mu$};
        \draw (4,2) -- (4,3);
        \draw (5,1) -- (4,2) node[midway, above right] {$Q_2$};
        \draw (7,0) -- (8.5,-.5);
        \draw (4,2) -- (3,2) node[left] {$\emptyset$};
    \end{tikzpicture}}} \right]
\end{split}
\end{equation} 

The $\mathbb F_1$ geometry is Fano and contains no decoupled contribution. The non-toric web contains a set of 2 parallel branes following the partition $(2)$. Following the discussion in Section \ref{sec:Zdec_GTP}, there is no decoupled sector and :
\begin{equation}
    Z \left[ \scalebox{.5}{\raisebox{-.5\height}{    \begin{tikzpicture}
        \draw (0,0) -- (-1,1) node[midway, left] {$Q_1$};
        \draw (-1,1) -- (1,1) node[midway, above] {$Q_1 Q_2$};
        \draw (1,1) -- (3,0) node[midway, above right] {$Q_1$};
        \draw (3,0) -- (0,0) node[midway,below] {$Q_1^2 Q_2$};
        \draw (0,0) -- (0,-1);
        \draw (-1,1) -- (-3,2);
        \draw (1,1) -- (0,2);
        \draw (3,0) -- (4.5,-.5);
    \end{tikzpicture}}} \right] =Z \left[ \scalebox{.5}{\raisebox{-.5\height}{\begin{tikzpicture}
    \tikzmath{\x = .1;}
     \draw (-1,0) -- (0,1) node[midway, above left] {$Q_1$};
    \draw (0,1) -- (1,1) node[midway, below] {$ Q_2$};
    \draw (1,1) -- (3,0) node[midway, above right] {$Q_1$};
    \draw (3,0) -- (-1,0) node[midway,below] {$Q_1^3 Q_2$};
    \draw (0,1) -- (0,2-2*\x) node[midway, left] {$ Q_2$};
    \draw (0,2- 2*\x) arc (270:90:\x);
    \draw (0,2.0) -- (0,3);
    \draw (-1,0) -- (-2,-.5);
    \draw (1,1) -- (\x,2-\x) node[midway, above right] {$Q_2$};
    \draw (\x,2-\x) -- (\x,3);
    \draw (3,0) -- (4.5,-.5);
    \draw (\x,2-\x) -- (-1,2-\x);
\end{tikzpicture}}} \right]
\label{eq:HW_F1F3}
\end{equation}
As before, this can be checked order by order in the topological vertex expansion. 

Note that \eqref{eq:HW_F0F2} and \eqref{eq:HW_F1F3} are non-trivial combinatorial equalities (writing $s_\mu = s_\mu (q^{- \rho})$ for short) 
\begin{equation}  
    \sum\limits_{\lambda , \mu} \frac{s_\lambda  s_{\lambda^T}   s_\mu   s_{\mu^T} (Q_1 Q_2)^{|\mu|+|\lambda|} q^{-\frac{1}{2} \kappa_\mu + \frac{1}{2} \kappa_\lambda}   }{\mathcal{R}_{\lambda^T \mu} (Q_1)^2 \mathcal{R}_{\emptyset \emptyset} (Q_2)}  =  \sum\limits_{\lambda , \mu} \frac{s_\lambda   s_{\lambda^T}   s_\mu   s_{\mu^T}  Q_2^{|\mu|}  (Q_1^2 Q_2)^{|\lambda|}q^{ \frac{1}{2} \kappa_\mu + \frac{3}{2} \kappa_\lambda}  }{\mathcal{R}_{\lambda^T \mu} (Q_1)^2 }  {}\label{eq:combinatorialIdentity}
\end{equation}  
\begin{equation}  
    \sum\limits_{\lambda , \mu} \frac{s_\lambda  s_{\lambda^T}   s_\mu   s_{\mu^T} (-Q_1 Q_2)^{|\mu|} (-Q_1^2 Q_2)^{|\lambda|} q^{\kappa_\lambda} }{\mathcal{R}_{\lambda^T \mu} (Q_1)^2}  =  \sum\limits_{\lambda , \mu} \frac{s_\lambda   s_{\lambda^T}   s_\mu   s_{\mu^T}  (- Q_2)^{|\mu|}  (-Q_1^3 Q_2)^{|\lambda|}q^{\kappa_\mu + 2\kappa_\lambda}  }{\mathcal{R}_{\lambda^T \mu} (Q_1)^2 \mathcal{R}_{\lambda^T \emptyset} (Q_1 Q_2) \mathcal{R}_{\mu^T \emptyset} (Q_2)} \label{eq:combinatorialIdentity2}
\end{equation} 
The first one follows directly from \cite[Eq. (4.32)]{Bershtein:2018zcz}, which is based on the Chern–Simons-modified Nakajima–Yoshioka blowup relations \cite{Nakajima:2005fg}. 
It would be interesting to find similar proof of \eqref{eq:combinatorialIdentity2} not relying on Hanany-Witten moves and topological vertex.

\subsubsection{An example with parallel branes following a non trivial partition} 

\begin{figure}[t]
    \centering
    \begin{subfigure}[c]{.495\textwidth}
        \centering
        \scalebox{.75}{
        \begin{tikzpicture}[every node/.style={inner sep=2pt}]
            \draw (-1,-.5) -- (0,0) -- node[midway,above left,inner sep=0pt] {$Q_2$} (1,1) --node[midway,left] {$Q_2$} (1,2) -- (-1,2);
            \draw (1,2) --node[midway,above left,inner sep=0pt] {$Q_2$} (2,3) -- (2,4);
            \draw (2,3) --node[midway,above] {$Q_1$} (4,3) -- (4,4);
            \draw (4,3) --node[midway,above right,inner sep=0pt] {$Q_2$} (5,2) -- node[midway,right] {$
            Q_2$} (5,1) --node[midway,above] {$Q_1Q_2^2$} (1,1);
            \draw (5,1) --node[midway,above right,inner sep=0pt] {$Q_2$} (6,0) -- (7,-.5);
            \draw (0,0) --node[midway,above] {$Q_1 Q_2^4$} (6,0);
            \draw (5,2) -- (7,2);
        \end{tikzpicture}}
        \caption{}\label{fig:web_SU3_toric}
    \end{subfigure}
    \begin{subfigure}[c]{.495\textwidth} 
        \centering 
        \scalebox{.75}{
        \begin{tikzpicture}[every node/.style={inner sep=2pt}]
        \tikzmath{\ee = .1;};
        \draw (\ee,0) -- (\ee,1-\ee) --node[midway,below left,inner sep=0pt] {$Q_2^2$} (-2,3) -- (-3,3.5);
        \draw (-2,3) --node[midway,above] {$Q_2$} (-1,3) --node[midway,above right,inner sep=0pt] {$Q_2$} (0,2) -- node[midway,below] {$Q_1$}(2,2) --node[midway,above left,inner sep=0pt] {$Q_2$} (3,3) --node[midway,above] {$Q_2$} (4,3) -- (5,3.5);
        \draw (-1,3) --node[midway,right] {$Q_2$} (-1,4) --node[midway,above] {$Q_1Q_2^2$} (3,4) --node[midway,left] {$Q_2$} (3,3);
        \draw (-1,4) -- (-2,5);
        \draw (3,4) -- (4,5);
        \draw (2-\ee,0) -- (2-\ee,1-\ee) --node[midway,below right,inner sep=0pt] {$Q_2^2$} (4,3);
        \draw  (\ee,1-\ee)--node[midway,below] {$Q_1$} (2-\ee,1-\ee);
        \draw (0,2) -- (0,1+\ee) arc(90:270:\ee) -- (0,0);
        \draw (2,2) -- (2,1+\ee) arc(90:-90:\ee) -- (2,0);
        \end{tikzpicture}}
        \caption{}\label{fig:web_SU3_wd}
    \end{subfigure}
    \caption{Two brane webs related by a Hanany-Witten transition. The first is toric with a set of two parallel branes following $\lambda = (1^2)$. The second is a GTP with a set of four parallel branes following $\lambda=  (2^2)$.}
    \label{fig:HW_wd_par}
\end{figure}
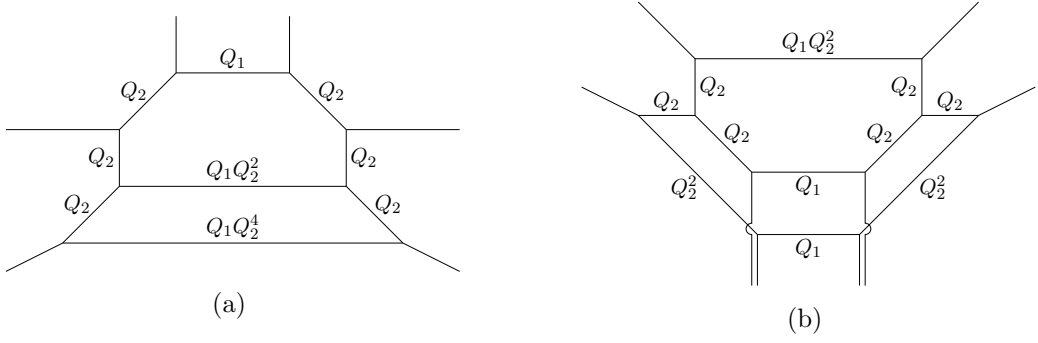

Let us consider an example which tests the prescription for $Z_\parallel$ from Section \ref{sec:Zdec_GTP}. We consider the toric brane web $P_{(a)}$ shown in Figure \ref{fig:web_SU3_toric}, which realizes the 5d $\operatorname{SU}(3)$ $N_f = 2$ gauge theory. The assignment of Kähler parameters is chosen to make the contribution of the parallel branes easily identifiable. The web is toric, with one set of 2 parallel branes with the trivial $(1^2)$ partition, therefore :
\begin{equation}
     Z_\parallel[P_{(a)}] =\operatorname{PExp}\left[\frac{qQ_1}{(1-q)^2}\right] 
\end{equation}

After performing Hanany-Witten transition on these two parallel branes, we get the brane web in figure \ref{fig:web_SU3_wd}. This is now a GTP, with a set of 4 parallel branes following the partition $(2^2)$. As explained in Section \ref{sec:Zdec_GTP}, we have : 
\begin{equation}\label{eq:SU3_wd_par}
    Z_\parallel[P_{(b)}] =\operatorname{PExp}\left[\frac{2qQ_1}{(1-q)^2}\right] 
\end{equation}

We can then check numerically that :
\begin{equation}
    \frac{Z_{\rm{top}}[X_{(a)}]}{Z_\parallel[P_{(a)}]} = \frac{Z_{\rm{top}}[X_{(b)}]}{Z_\parallel[P_{(b)}]}
\end{equation}
Alternatively, we can isolate the parallel brane contribution by setting $Q_2$ to $0$. In the toric case, we see that 
\begin{equation}
    Z_{\rm{top}}[X_{(a)}] (Q_2=0) = \operatorname{PExp}\left[\frac{qQ_1}{(1-q)^2}\right] = Z_\parallel[P_{(a)}] \,.
\end{equation}
On the other hand, 
\begin{equation}
    Z_{\rm{top}}[X_{(b)}] (Q_2=0) = \operatorname{PExp}\left[\frac{2qQ_1}{(1-q)^2}\right]
\end{equation}
in agreement with \eqref{eq:SU3_wd_par}.

\subsection{Flops}
\label{sec:flops}

Flops are transitions in which a $(-1,-1)$-curve is blown down to a conifold singularity and then blown up in a different way. In string theory, flops connect the K\"ahler cones of birationally equivalent Calabi-Yau threefolds, across which physics varies smoothly \cite{Witten:1993yc,Aspinwall:1993nu}. In the context of the topological vertex, they were discussed in \cite{Iqbal:2004ne}, and the flop invariance of the topological vertex was established in \cite{Konishi:2006ev}. Unlike a Hanany-Witten move, a flop changes the K\"ahler parameters, and therefore the low energy effective theory.

However, the topological string partition function can be analytically continued through the transition. Concretely, one can equate the partition functions of the two geometries after factoring out the contribution from the curve being flopped and identifying carefully the Kähler parameters. 

For example, the local $\mathbb F_1$ geometry contains a $(-1,-1)$-curve (one of the sections of $\mathbb F_1 \to \mathbb P^1$). When flopped, it gives rise to a Calabi-Yau threefold whose only compact divisor is a $\mathbb P^2$. The partition functions for these two geometries therefore satisfy :
\begin{equation}
    \frac{1}{\operatorname{PExp}\left(-\frac{q Q_2}{(1-q)^2}\right)} Z \left[ \scalebox{.5}{\raisebox{-.42\height}{
    \begin{tikzpicture}
        \draw (-1,-1) -- (0,0) -- (0,1) node[midway,left] {$Q_1$} -- (-1,2);
        \draw (0,1) -- (1,1) node[midway,above] {$Q_2$} -- (1,2);
        \draw (1,1) -- (2,0) node[midway,right] {$Q_1$} -- (4,-1);
        \draw (2,0) -- (0,0) node[midway,below] {$Q_1Q_2$};
    \end{tikzpicture}}} \right]
    = \frac{1 }{\operatorname{PExp}\left(-\frac{q Q_2^{-1}}{(1-q)^2}\right)}
    Z \left[ \scalebox{.5}{\raisebox{-.42\height}{
    \begin{tikzpicture}
        \draw (-1,-1) -- (0,0) -- (0,1) node[midway,left] {$Q_1$} -- (1,0) node[midway,right] {$Q_1$} -- (0,0) node[midway,below] {$Q_1$};
        \draw (1,0) -- (3,-1);
        \draw (0,1) -- (-.5,2) node[midway,right] {$Q_2^{-1}$} -- (-.5,3);
        \draw (-.5,2) -- (-1.5,2);
    \end{tikzpicture}}} \right]
\end{equation}

Starting from local $dP_2$, there are two curves which can be flopped, leading to geometries containing a compact divisor isomorphic to $\mathbb{P}^1 \times \mathbb P^1$ and $dP_1$ respectively. At the level of partition functions, we get :
\begin{subequations}
\begin{equation}
\frac{1 }{\operatorname{PExp}\left(-\frac{q Q_2}{(1-q)^2}\right)}
Z \left[ \scalebox{.5}{\raisebox{-.42\height}{
    \begin{tikzpicture}
        \draw (-1,-1) -- (0,0) -- (0,1) node[midway,left] {$Q_3$} -- (-1,1);
        \draw (0,1) -- (1,2) node[midway,left] {$Q_2$} -- (1,3);
        \draw (1,2) -- (2,2) node[midway,above] {$Q_1$}-- (3,3);
        \draw (2,2) -- (2,0) node[midway,right] {$Q_2 Q_3$} -- (3,-1);
        \draw (2,0) -- (0,0) node[midway,below] {$Q_1 Q_2$};
    \end{tikzpicture}}} \right]=
  \frac{1 }{\operatorname{PExp}\left(-\frac{q Q_2^{-1}}{(1-q)^2}\right)}
  Z \left[ \scalebox{.5}{\raisebox{-.42\height}{
    \begin{tikzpicture}
        \draw (-1,-1) -- (0,0) -- (0,1) node[midway, left] {$Q_3$} -- (-1,2) node[midway,above right] {$Q_2^{-1}$} -- (-1,3);
        \draw (-1,2) -- (-2,2);
        \draw (0,1) -- (1,1) node[midway,above] {$Q_1$}-- (2,2);
        \draw (1,1) -- (1,0) node[midway,right] {$Q_3$} -- (2,-1);
        \draw (1,0) -- (0,0) node[midway,below] {$Q_1$};
    \end{tikzpicture}}} \right]
\end{equation}
\begin{equation}
\frac{1 }{\operatorname{PExp}\left(-\frac{q Q_3}{(1-q)^2}\right)}
Z \left[ \scalebox{.5}{\raisebox{-.42\height}{
    \begin{tikzpicture}
        \draw (-1,-1) -- (0,0) -- (0,1) node[midway,left] {$Q_3$} -- (-1,1);
        \draw (0,1) -- (1,2) node[midway,left] {$Q_2$} -- (1,3);
        \draw (1,2) -- (2,2) node[midway,above] {$Q_1$}-- (3,3);
        \draw (2,2) -- (2,0) node[midway,right] {$Q_2 Q_3$} -- (3,-1);
        \draw (2,0) -- (0,0) node[midway,below] {$Q_1 Q_2$};
    \end{tikzpicture}}} \right]
=
\frac{1 }{\operatorname{PExp}\left(-\frac{q Q_3^{-1}}{(1-q)^2}\right)}
Z \left[ \scalebox{.5}{\raisebox{-.42\height}{
    \begin{tikzpicture}
        \draw (-1,-1.5) -- (-1,-.5) -- (-2,-.5);
        \draw (-1,-.5) -- (0,0) node[midway,below] {$Q_3^{-1}$} -- (1,1) node[midway,left] {$Q_2$} -- (1,2);
        \draw (1,1) -- (2,1) node[midway,above] {$Q_1$} -- (3,2);
        \draw (2,1) -- (2,0) node[midway,right] {$Q_2$} -- (3,-1);
        \draw (2,0) -- (0,0) node[midway,below] {$Q_1 Q_2$};
    \end{tikzpicture}}} \right]
\end{equation}
\label{eq:flopdP1}
\end{subequations}

\section{Hanany--Witten Transitions and Mori cones}
\label{sec:HWMori}

A direct consequence of the discussion of Hanany-Witten moves in Section \ref{sec:HWtopologicalVertex} is the fact that Gopakumar-Vafa invariants for distinct geometries can be related, if these geometries are associated to the same 5d SCFT. We illustrate this point on the simplest examples of the $E_1$ and $\tilde{E}_1$ theories in this section, translating into geometry the insights from the SCFT. The starting point of the discussion is the fact that the GTP geometry associated to the Hirzebruch surfaces $\mathbb{F}_n$ is $E_1$ for $n$ even and $\tilde{E}_1$ for $n$ odd. In particular, $\mathbb{F}_0$ and $\mathbb{F}_2$ realize the same SCFT. However, the local $\mathbb{F}_n$ Calabi-Yau threefolds are all pairwise distinct, and indeed 
\begin{equation}
        Z_{\text{top}}(K_{\mathbb F_0})    \neq    Z_{\text{top}}(K_{\mathbb F_2}) , \quad  Z_{\text{top}}(K_{\mathbb F_1})    \neq    Z_{\text{top}}(K_{\mathbb F_3}) .
\end{equation}
The difference in the first case lies in the external parallel branes, and the correct equality is \eqref{eq:HW_F0F2} which we write symbolically as 
\begin{equation} \label{eq:relationF0F2}
   Z_{\text{top}} \left[  \raisebox{-.5\height}{ \scalebox{.8}{  
\begin{tikzpicture}[x=.5cm,y=.5cm] 
\draw[gridline] (-1,-0.5)--(-1,2.5); 
\draw[gridline] (0,-0.5)--(0,2.5); 
\draw[gridline] (1,-0.5)--(1,2.5); 
\draw[gridline] (-1.5,0)--(1.5,0); 
\draw[gridline] (-1.5,1)--(1.5,1); 
\draw[gridline] (-1.5,2)--(1.5,2); 
\draw[ligne] (-1,2)--(0,0); 
\draw[ligne] (0,0)--(1,0); 
\draw[ligne] (1,0)--(0,2); 
\draw[ligne] (0,2)--(-1,2); 
\draw[ligne] (0,2)--(0,0); 
\draw[ligne] (-1,2)--(1,0); 
\node[bd] at (0,0) {}; 
\node[bd] at (0,1) {}; 
\node[bd] at (1,0) {}; 
\node[bd] at (0,2) {}; 
\node[bd] at (-1,2) {}; 
\end{tikzpicture}}} \right] = \frac{ Z_{\text{top}} \left[ \raisebox{-.5\height}{ \scalebox{.8}{  
\begin{tikzpicture}[x=.5cm,y=.5cm] 
\draw[gridline] (-1,-0.5)--(-1,2.5); 
\draw[gridline] (0,-0.5)--(0,2.5); 
\draw[gridline] (1,-0.5)--(1,2.5); 
\draw[gridline] (-1.5,0)--(1.5,0); 
\draw[gridline] (-1.5,1)--(1.5,1); 
\draw[gridline] (-1.5,2)--(1.5,2); 
\draw[ligne] (-1,2)--(0,0); 
\draw[ligne] (0,0)--(1,2); 
\draw[ligne] (1,2)--(-1,2); 
\draw[ligne] (0,0)--(0,2); 
\draw[ligne] (-1,2)--(0,1)--(1,2); 
\node[bd] at (0,0) {}; 
\node[bd] at (1,2) {}; 
\node[bd] at (-1,2) {}; 
\node[bd] at (0,2) {}; 
\node[bd] at (0,1) {}; 
\end{tikzpicture}}} \right] }{Z_{\text{top}} \left[ \raisebox{-.5\height}{ \scalebox{.8}{  
\begin{tikzpicture}[x=.5cm,y=.5cm] 
\draw[gridline] (-1,1.5)--(-1,2.5); 
\draw[gridline] (0,1.5)--(0,2.5); 
\draw[gridline] (1,1.5)--(1,2.5); 
\draw[gridline] (-1.5,2)--(1.5,2); 
\draw[ligne] (-1,2)--(1,2); 
\draw[ligne] (1,2)--(-1,2); 
\node[bd] at (1,2) {}; 
\node[bd] at (-1,2) {}; 
\node[bd] at (0,2) {}; 
\end{tikzpicture}}} \right] }
\end{equation}
In the second case, there are no parallel branes, so we expect for the pair $\mathbb{F}_1$-$\mathbb{F}_3$ that  
\begin{equation}
\label{eq:equalityF1F3GTP}
     Z_{\text{top}} \left[  \raisebox{-.5\height}{ \scalebox{.8}{  
\begin{tikzpicture}[x=.5cm,y=.5cm] 
\draw[gridline] (0,-0.5)--(0,3.5); 
\draw[gridline] (1,-0.5)--(1,3.5); 
\draw[gridline] (2,-0.5)--(2,3.5); 
\draw[gridline] (-0.5,0)--(2.5,0); 
\draw[gridline] (-0.5,1)--(2.5,1); 
\draw[gridline] (-0.5,2)--(2.5,2); 
\draw[gridline] (-0.5,3)--(2.5,3); 
\draw[ligne] (1,2)--(0,0); 
\draw[ligne] (0,0)--(1,0); 
\draw[ligne] (1,0)--(2,3); 
\draw[ligne] (2,3)--(1,2); 
\draw[ligne] (1,0)--(1,2); 
\draw[ligne] (0,0)--(1,1)--(2,3); 
\node[bd] at (0,0) {}; 
\node[bd] at (1,0) {}; 
\node[bd] at (1,1) {}; 
\node[bd] at (2,3) {}; 
\node[bd] at (1,2) {}; 
\end{tikzpicture}}}  \right] =  Z_{\text{top}} \left[   \raisebox{-.5\height}{ \scalebox{.8}{  
\begin{tikzpicture}[x=.5cm,y=.5cm] 
\draw[gridline] (0,-0.5)--(0,3.5); 
\draw[gridline] (1,-0.5)--(1,3.5); 
\draw[gridline] (2,-0.5)--(2,3.5); 
\draw[gridline] (-0.5,0)--(2.5,0); 
\draw[gridline] (-0.5,1)--(2.5,1); 
\draw[gridline] (-0.5,2)--(2.5,2); 
\draw[gridline] (-0.5,3)--(2.5,3); 
\draw[ligne] (0,2)--(1,0); 
\draw[ligne] (1,0)--(2,3); 
\draw[ligne] (2,3)--(0,3); 
\draw[ligne] (0,3)--(0,2); 
\draw[ligne] (1,0)--(1,2)--(2,3); 
\draw[ligne] (0,2)--(1,1)--(2,3); 
\draw[ligne] (0,2)--(1,2); 
\node[bd] at (1,0) {}; 
\node[bd] at (2,3) {}; 
\node[bd] at (0,3) {}; 
\node[bd] at (0,2) {}; 
\node[bd] at (1,1) {}; 
\node[bd] at (1,2) {}; 
\node[wd] at (1,3) {}; 
\end{tikzpicture}}} \right] 
\end{equation}
This is indeed the content of \eqref{eq:HW_F1F3}. 
However, the right hand side is \emph{not} equal to $Z_{\text{top}}(K_{\mathbb F_3})$. 
In the rest of this section, we aim to make the above results more precise and extract their geometric content. We first setup some notations regarding Hirzebruch surfaces. 

\paragraph{Conventions for Hirzebruch surfaces}
 
 For the Hirzebruch surface \(\mathbb F_n\), let \(f_n\) denote the fiber class and let \(b_n\) denote the distinguished section, which we also call the base
class. Their intersections are 
\begin{equation}
\label{eq:FnIntersection}
        f_n^2=0,
    \qquad
    b_n^2=-n,
    \qquad
    f_n\cdot b_n=1.
\end{equation}
For a curve $C$ of genus $g(C)$ in $\mathbb{F}_n$, the adjunction formula is 
\begin{equation}
\label{eq:adjunction}
    2 g (C) -2 = C \cdot (C + K_{\mathbb{F}_n}) \, . 
\end{equation}
Using this for $C = b_n$ and $C = f_n$ which are both of genus 0 gives the canonical class 
\begin{equation}
    K_{\mathbb{F}_n} = - 2 b_n - (n+2) f_n \, . 
\end{equation}
The associated parameters are denoted $Q_{b_n}$ and $Q_{f_n}$.
The Mori cone is 
\begin{equation}
    \NEbar(\mathbb F_n) = \bR_{\geq 0} f_n\oplus \bR_{\geq 0} b_n.
\end{equation} 
and we write curve classes as 
\begin{equation}
\label{eq:curveClassFn}
    \beta  = d_{f_n} \, f_n + d_{b_n} \, b_n \, ,  \qquad Q^{\beta} = Q_{f_n}^{d_{f_n}} Q_{b_n}^{d_{b_n}} \, . 
\end{equation}
Using \eqref{eq:adjunction}, the genus of a curve in the class \eqref{eq:curveClassFn} is 
\begin{equation}
\label{eq:genusFormulaFn}
    g (d_{f_n} , d_{b_n}) =  \frac{1}{2} (d_{b_n}  -1) (2 d_{f_n}  - n d_{b_n} -2)  \, . 
\end{equation}
This formula provides a bound on the possible non-vanishing GV invariants. Namely, 
\begin{equation}
\label{eq:implicationGenus}
    r > g(\beta) \qquad \Longrightarrow  \qquad \GV_{r,\beta} = 0 \, . 
\end{equation}
We will see below (see Section \ref{subsec: Mori cone F1 F3}) that the converse is not true.

\subsection{The \texorpdfstring{$\mathbb{F}_0$}{F0} - \texorpdfstring{$\mathbb{F}_2$}{F2} pair} 
\label{subsec: Mori cone F0 F2}

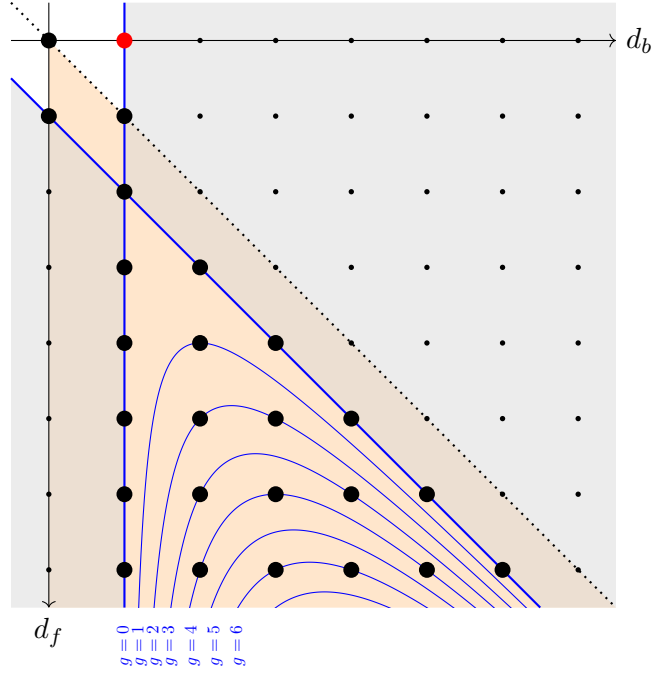
\begin{figure}
    \centering
\begin{tikzpicture}[scale=1] 
\fill[orange!20] (0,0) -- (7.5,-7.5) -- (0,-7.5) -- cycle; 
\fill[gray!30, fill opacity=0.5] (-0.5,-7.5) -- (-.5,-.5) -- (1,-2) -- (1,-7.5) -- cycle; 
\fill[gray!30, fill opacity=0.5] (1,.5)--(7.5,.5)--(7.5,-7.5)--(6.5,-7.5)--(1,-2) --cycle; 
  \draw[->] (-0.5,0) -- (7.5,0) node[right] {$d_b$};
  \draw[->] (0,0.5) -- (0,-7.5) node[below] {$d_f$};
  \draw[thick, blue, domain=-0.5:7.5, samples=100, smooth]
    plot (1, -\x);
  \draw[thick,dotted, domain=-0.5:7.5, samples=100, smooth]
    plot (\x, -\x);
  \draw[thick, blue, domain=-0.5:6.5, samples=100, smooth]
    plot (\x, -\x-1);
\begin{scope}
  \clip (-.5,.5) rectangle (7.5,-7.5);
  \draw[blue, domain=1.1:7, samples=200, smooth]
    plot (\x, {-1/(\x - 1) - \x - 1});
  \draw[blue, domain=1.2:7, samples=200, smooth]
    plot (\x, {-2/(\x - 1) - \x - 1});
  \draw[blue, domain=1.2:7, samples=200, smooth]
    plot (\x, {-3/(\x - 1) - \x - 1});
  \draw[blue, domain=1.2:7, samples=200, smooth]
    plot (\x, {-4/(\x - 1) - \x - 1});
  \draw[blue, domain=1.2:7, samples=200, smooth]
    plot (\x, {-5/(\x - 1) - \x - 1});
  \draw[blue, domain=1.2:7, samples=200, smooth]
    plot (\x, {-6/(\x - 1) - \x - 1});
  \draw[blue, domain=1.2:7, samples=200, smooth]
    plot (\x, {-7/(\x - 1) - \x - 1});
\end{scope}
\node[rotate=90] at (1,-8) {\color{blue} \scalebox{.6}{$g=0$}};
\node[rotate=90] at (1.2,-8) {\color{blue} \scalebox{.6}{$g=1$}};
\node[rotate=90] at (1.4,-8) {\color{blue} \scalebox{.6}{$g=2$}};
\node[rotate=90] at (1.6,-8) {\color{blue} \scalebox{.6}{$g=3$}};
\node[rotate=90] at (1.9,-8) {\color{blue} \scalebox{.6}{$g=4$}};
\node[rotate=90] at (2.2,-8) {\color{blue} \scalebox{.6}{$g=5$}};
\node[rotate=90] at (2.5,-8) {\color{blue} \scalebox{.6}{$g=6$}};

  \foreach \i in {-0,...,7} {
    \foreach \j in {-7,...,0} {
      \fill[black] (\i,\j) circle (1pt);
    }
  }  
  \fill[black] (0,0) circle (3pt);
  \fill[red] (1,0) circle (3pt);
  \fill[black] (0,-1) circle (3pt);
  \fill[black] (1,-1) circle (3pt);
  \fill[black] (1,-2) circle (3pt);
  \fill[black] (1,-3) circle (3pt);
  \fill[black] (1,-4) circle (3pt);
  \fill[black] (1,-5) circle (3pt);
  \fill[black] (1,-6) circle (3pt);
  \fill[black] (1,-7) circle (3pt);
  \fill[black] (2,-3) circle (3pt);
  \fill[black] (2,-4) circle (3pt);
  \fill[black] (2,-5) circle (3pt);
  \fill[black] (2,-6) circle (3pt);
  \fill[black] (2,-7) circle (3pt);
  \fill[black] (3,-4) circle (3pt);
  \fill[black] (3,-5) circle (3pt);
  \fill[black] (3,-6) circle (3pt);
  \fill[black] (3,-7) circle (3pt);
  \fill[black] (4,-5) circle (3pt);
  \fill[black] (4,-6) circle (3pt);
  \fill[black] (4,-7) circle (3pt);
  \fill[black] (5,-6) circle (3pt);
  \fill[black] (5,-7) circle (3pt);
  \fill[black] (6,-7) circle (3pt);
\end{tikzpicture}
    \caption{The dots represent curve classes in the $\mathbb{F}_2$ Mori cone. The orientation of the plot is chosen so that it agrees with the orientation of Table \ref{tab:GV_F2}.  The gray regions are those excluded by the adjunction formula ($g(C) < 0$), the white regions are allowed, $g(C) \geq 0$. The blue curves correspond to constant genus of given value. The orange region, below the dotted line, shows the image of the $\mathbb{F}_0$ Mori cone under the map \eqref{eq:mapF0F2}. The only point curve class outside it is in red. }
    \label{fig:F2Mori}
\end{figure}

Use the notations introduced in Figure \ref{fig:HW_F0F2F1F3}. Then 
\begin{equation}
\label{eq:identificationF0F2}
    \begin{array}{lll}
        Q_{f_0} = Q_1 & & Q_{b_0} =Q_1  Q_2 \\Q_{f_2} = Q_1 & \qquad  & Q_{b_2} =  Q_2  
    \end{array}
\end{equation}
With this identification, the relation \eqref{eq:HW_F0F2} becomes  
\begin{equation}
    \label{F2 over F0 parallel factor}
    \frac{
    Z_{\text{top}}(K_{\mathbb F_2};Q_{f_2}, Q_{b_2})
    }{
    Z_{\text{top}}(K_{\mathbb F_0};Q_{f_0}, Q_{b_0})
    }
    =   \frac{
    Z_{\text{top}}(K_{\mathbb F_2};Q_1,Q_2)
    }{
    Z_{\text{top}}(K_{\mathbb F_0};Q_1,Q_1Q_2)
    }
    =
    \operatorname{PExp}\!\left(\frac{qQ_2}{(1-q)^2}\right).
\end{equation}
We now explain this identity in terms of Mori cones and curve classes. The identification \eqref{eq:identificationF0F2} gives 
\begin{equation}
\label{eq:mapF0F2}
   d_{f_2} = d_{f_0} + d_{b_0} \, , \qquad d_{b_2} = d_{b_0} \, . 
\end{equation}
This is compatible with the intersection form \eqref{eq:FnIntersection}. 
The image of the Mori cone of \(\mathbb F_0\) is the subcone 
\begin{equation}
     \{(d_{f_2},d_{b_2})\in \bZ_{\geq 0}^2\mid d_{f_2} \geq d_{b_2} \}.
\end{equation} 
The remaining classes in the Mori cone of \(\mathbb F_2\) are precisely
those with $d_{f_2} < d_{b_2}$.  These are the classes which are present in \(K_{\mathbb F_2}\) but are not
obtained from \(K_{\mathbb F_0}\) under the Hanany--Witten identification.
Hence \eqref{F2 over F0 parallel factor}  predicts
\begin{equation}
    \label{GV F0 F2 relation}
    \GV_{r,(d_{f_0},d_{b_0})}(K_{\mathbb F_0})
    =
    \GV_{r,(d_{f_0}+d_{b_0},d_{b_0})}(K_{\mathbb F_2})
\end{equation}
for all classes in the image cone.  This is exactly
the pattern visible in Tables~\ref{tab:GV_F0} and~\ref{tab:GV_F2}, see also Figure \ref{fig:F2Mori}. For instance, the class \((3,2)\) in \(K_{\mathbb F_0}\) maps to the class
\((5,2)\) in \(K_{\mathbb F_2}\).

It remains to understand the classes of \(K_{\mathbb F_2}\) lying outside
the image cone. The claim is that the only primitive GV contribution in this region is the
class $b_2 = (0,1)$
\begin{equation}
    \label{GV parallel F2}
    \GV_{r,(0,1)}(K_{\mathbb F_2})
    =
    -\delta_{r,0}.
\end{equation}
This is precisely the inverse conifold contribution. There is a good geometric reason why only this class can have a non-zero GV invariant. 

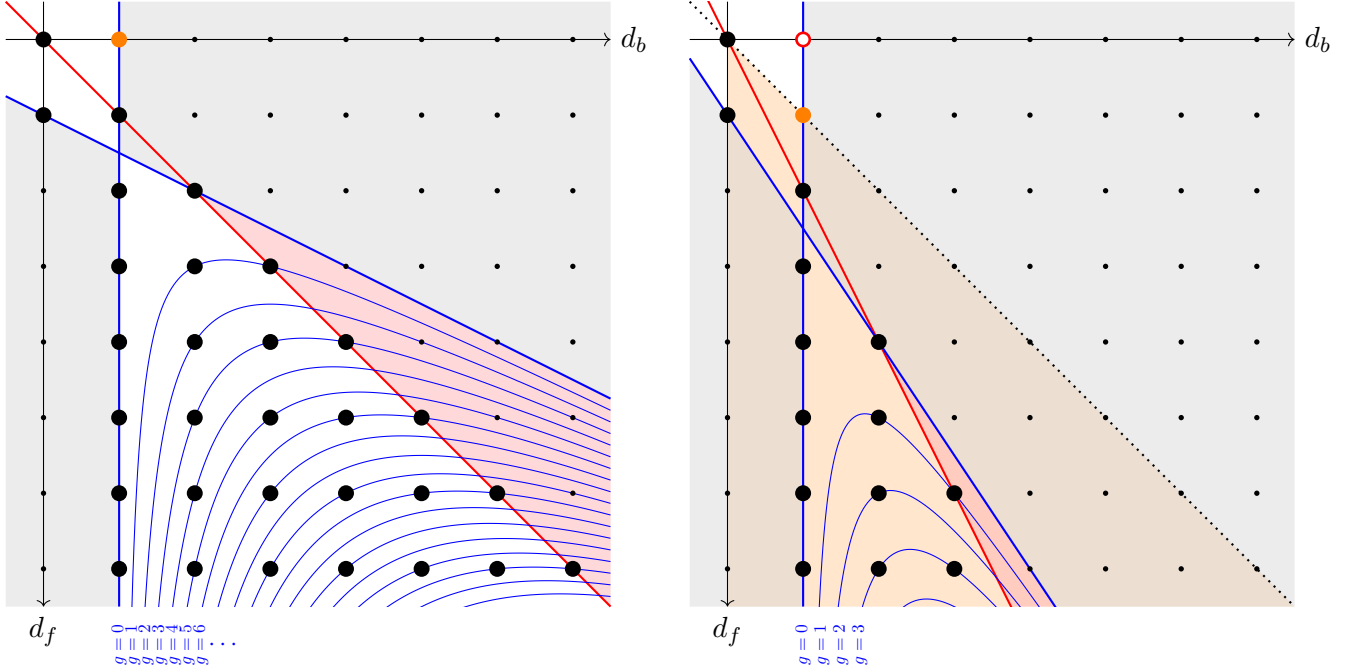
\begin{figure}
    \centering
    \hspace*{-1cm}\begin{tabular}{cc} 
        \begin{tikzpicture}[scale=1] 
\fill[gray!30, fill opacity=0.5] (-0.5,-7.5) -- (-.5,-.75) -- (1,-1.5) -- (1,-7.5) -- cycle; 
\fill[gray!30, fill opacity=0.5] (1,.5)--(7.5,.5)--(7.5,-4.75)--(1,-1.5)--cycle; 
\fill[red!30, fill opacity=0.5] (2,-2)--(7.5,-4.75)--(7.5,-7.5)--cycle; 
  \draw[->] (-0.5,0) -- (7.5,0) node[right] {$d_b$};
  \draw[->] (0,0.5) -- (0,-7.5) node[below] {$d_f$};
  \draw[thick, red, domain=-0.5:7.5, samples=100, smooth]
    plot (\x, -\x);
  \draw[thick, blue, domain=-0.5:7.5, samples=100, smooth]
    plot (1, -\x);
\begin{scope}
  \clip (-.5,.5) rectangle (7.5,-7.5);
  \draw[thick, blue, domain=-0.5:7.5, samples=200, smooth]
    plot (\x, { - 1/2* \x - 1});
  \draw[blue, domain=1.1:7.5, samples=200, smooth]
    plot (\x, {-1/(\x - 1) - 1/2* \x - 1});
  \draw[blue, domain=1.2:7.5, samples=200, smooth]
    plot (\x, {-2/(\x - 1) - 1/2* \x - 1});
  \draw[blue, domain=1.2:7.5, samples=200, smooth]
    plot (\x, {-3/(\x - 1) - 1/2* \x - 1});
  \draw[blue, domain=1.2:7.5, samples=200, smooth]
    plot (\x, {-4/(\x - 1) -  1/2* \x - 1});
  \draw[blue, domain=1.2:7.5, samples=200, smooth]
    plot (\x, {-5/(\x - 1) -  1/2*  \x - 1});
  \draw[blue, domain=1.2:7.5, samples=200, smooth]
    plot (\x, {-6/(\x - 1) -  1/2* \x - 1});
  \draw[blue, domain=1.2:7.5, samples=200, smooth]
    plot (\x, {-7/(\x - 1) -  1/2* \x - 1});
  \draw[blue, domain=1.2:7.5, samples=200, smooth]
    plot (\x, {-8/(\x - 1) -  1/2* \x - 1});
  \draw[blue, domain=1.2:7.5, samples=200, smooth]
    plot (\x, {-9/(\x - 1) -  1/2* \x - 1});
  \draw[blue, domain=1.2:7.5, samples=200, smooth]
    plot (\x, {-10/(\x - 1) -  1/2* \x - 1});
  \draw[blue, domain=1.2:7.5, samples=200, smooth]
    plot (\x, {-11/(\x - 1) -  1/2* \x - 1});
  \draw[blue, domain=1.2:7.5, samples=200, smooth]
    plot (\x, {-12/(\x - 1) -  1/2* \x - 1});
  \draw[blue, domain=1.2:7.5, samples=200, smooth]
    plot (\x, {-13/(\x - 1) -  1/2* \x - 1});
  \draw[blue, domain=1.2:7.5, samples=200, smooth]
    plot (\x, {-14/(\x - 1) -  1/2* \x - 1});
  \draw[blue, domain=1.2:7.5, samples=200, smooth]
    plot (\x, {-15/(\x - 1) -  1/2* \x - 1});
  \draw[blue, domain=1.2:7.5, samples=200, smooth]
    plot (\x, {-16/(\x - 1) -  1/2* \x - 1});
  \draw[blue, domain=1.2:7.5, samples=200, smooth]
    plot (\x, {-17/(\x - 1) -  1/2* \x - 1});
\end{scope}
\node[rotate=90] at (1,-8) {\color{blue} \scalebox{.6}{$g=0$}};
\node[rotate=90] at (1.18,-8) {\color{blue} \scalebox{.6}{$g=1$}};
\node[rotate=90] at (1.36,-8) {\color{blue} \scalebox{.6}{$g=2$}};
\node[rotate=90] at (1.54,-8) {\color{blue} \scalebox{.6}{$g=3$}};
\node[rotate=90] at (1.72,-8) {\color{blue} \scalebox{.6}{$g=4$}};
\node[rotate=90] at (1.9,-8) {\color{blue} \scalebox{.6}{$g=5$}};
\node[rotate=90] at (2.08,-8) {\color{blue} \scalebox{.6}{$g=6$}};
\node at (2.4,-8) {\color{blue} \footnotesize $\cdots$};
  \foreach \i in {-0,...,7} {
    \foreach \j in {-7,...,0} {
      \fill[black] (\i,\j) circle (1pt);
    }
  }  
  \fill[orange] (1,0) circle (3pt);
  \fill[black] (0,0) circle (3pt);
  \fill[black] (0,-1) circle (3pt);
  \fill[black] (1,-1) circle (3pt);
  \fill[black] (1,-2) circle (3pt);
  \fill[black] (1,-3) circle (3pt);
  \fill[black] (1,-4) circle (3pt);
  \fill[black] (1,-5) circle (3pt);
  \fill[black] (1,-6) circle (3pt);
  \fill[black] (1,-7) circle (3pt);
  \fill[black] (2,-2) circle (3pt);
  \fill[black] (2,-3) circle (3pt);
  \fill[black] (2,-4) circle (3pt);
  \fill[black] (2,-5) circle (3pt);
  \fill[black] (2,-6) circle (3pt);
  \fill[black] (2,-7) circle (3pt);
  \fill[black] (3,-3) circle (3pt);
  \fill[black] (3,-4) circle (3pt);
  \fill[black] (3,-5) circle (3pt);
  \fill[black] (3,-6) circle (3pt);
  \fill[black] (3,-7) circle (3pt);
  \fill[black] (4,-4) circle (3pt);
  \fill[black] (4,-5) circle (3pt);
  \fill[black] (4,-6) circle (3pt);
  \fill[black] (4,-7) circle (3pt);
  \fill[black] (5,-5) circle (3pt);
  \fill[black] (5,-6) circle (3pt);
  \fill[black] (5,-7) circle (3pt);
  \fill[black] (6,-6) circle (3pt);
  \fill[black] (6,-7) circle (3pt);
  \fill[black] (7,-7) circle (3pt);
\end{tikzpicture} & \begin{tikzpicture}[scale=1] 
\fill[orange!20] (0,0) -- (7.5,-7.5) -- (0,-7.5) -- cycle; 
\fill[gray!30, fill opacity=0.5] (-0.5,-7.5) -- (-.5,-.25) -- (1,-2.5) -- (1,-7.5) -- cycle; 
\fill[gray!30, fill opacity=0.5] (1,.5)--(7.5,.5)--(7.5,-7.5)--(4.33,-7.5)--(1,-2.5) --cycle; 
\fill[red!30, fill opacity=0.5] (2,-4)--(4.33,-7.5)--(3.75,-7.5) --cycle; 
  \draw[->] (-0.5,0) -- (7.5,0) node[right] {$d_b$};
  \draw[->] (0,0.5) -- (0,-7.5) node[below] {$d_f$};
  \draw[thick, blue, domain=-0.5:7.5, samples=100, smooth]
    plot (1, -\x);
  \draw[thick,dotted, domain=-0.5:7.5, samples=100, smooth]
    plot (\x, -\x);
\begin{scope}
  \clip (-.5,.5) rectangle (7.5,-7.5);
  \draw[thick, red, domain=-0.5:7, samples=200, smooth]
    plot (\x, { - 2* \x});
  \draw[thick, blue, domain=-0.5:7, samples=200, smooth]
    plot (\x, { - 3/2* \x - 1});
  \draw[blue, domain=1.1:7, samples=200, smooth]
    plot (\x, {-1/(\x - 1) - 3/2* \x - 1});
  \draw[blue, domain=1.2:7, samples=200, smooth]
    plot (\x, {-2/(\x - 1) - 3/2* \x - 1});
  \draw[blue, domain=1.2:7, samples=200, smooth]
    plot (\x, {-3/(\x - 1) - 3/2* \x - 1});
  \draw[blue, domain=1.2:7, samples=200, smooth]
    plot (\x, {-4/(\x - 1) -  3/2* \x - 1});
  \draw[blue, domain=1.2:7, samples=200, smooth]
    plot (\x, {-5/(\x - 1) -  3/2*  \x - 1});
  \draw[blue, domain=1.2:7, samples=200, smooth]
    plot (\x, {-6/(\x - 1) -  3/2* \x - 1});
\end{scope}
\node[rotate=90] at (1,-8) {\color{blue} \scalebox{.6}{$g=0$}};
\node[rotate=90] at (1.25,-8) {\color{blue} \scalebox{.6}{$g=1$}};
\node[rotate=90] at (1.5,-8) {\color{blue} \scalebox{.6}{$g=2$}};
\node[rotate=90] at (1.75,-8) {\color{blue} \scalebox{.6}{$g=3$}};
  \foreach \i in {-0,...,7} {
    \foreach \j in {-7,...,0} {
      \fill[black] (\i,\j) circle (1pt);
    }
  }  
  \fill[black] (0,0) circle (3pt);
  \fill[black] (0,-1) circle (3pt);
  \fill[red] (1,0) circle (3pt);
  \fill[white] (1,0) circle (2pt);
  \fill[orange] (1,-1) circle (3pt);
  \fill[black] (1,-2) circle (3pt);
  \fill[black] (1,-3) circle (3pt);
  \fill[black] (1,-4) circle (3pt);
  \fill[black] (1,-5) circle (3pt);
  \fill[black] (1,-6) circle (3pt);
  \fill[black] (1,-7) circle (3pt);
  \fill[black] (2,-4) circle (3pt);
  \fill[black] (2,-5) circle (3pt);
  \fill[black] (2,-6) circle (3pt);
  \fill[black] (2,-7) circle (3pt);
  \fill[black] (3,-6) circle (3pt);
  \fill[black] (3,-7) circle (3pt);
\end{tikzpicture}
    \end{tabular}
    \caption{\textit{Left:}  $\mathbb{F}_1$ Mori cone. The grey zone corresponds to the curves classes excluded by $g \geq 0$. The red zone is eliminated by the flop of the curve represented by the orange dot. The orange dot, is the only dot above the red line which can have a non-vanishing GV invariant. \\ \textit{Right:} $\mathbb{F}_3$ Mori cone. In orange is the image of the $\mathbb{F}_1$ cone, and the blue and red lines are also the images of those in Figure \ref{fig:F1Mori} by Hanany-Witten. The red dot, similarly to the red dot in Figure \ref{fig:F2Mori}, is a curve class one would expect to host a non-zero GV invariant. However, we argue that this is not the case.   }
    \label{fig:F1Mori}
\end{figure}

\paragraph{Genus bound.}

Let \(C\subset \mathbb F_2\) be an irreducible curve in the class \eqref{eq:curveClassFn}. Its genus is $(d_{b_2} -1)(d_{f_2} - d_{b_2} - 1) \geq 0$ by \eqref{eq:adjunction}, so the only solution in the Mori cone of $\mathbb F_2$ with $d_{f_2} < d_{b_2}$ is the class $b_2$.  
Thus, among the classes outside the image of the \(\mathbb F_0\) Mori
cone, the only class which can be represented by an irreducible curve is \(b_2\) itself.

This explains geometrically why the extra factor in
\eqref{F2 over F0 parallel factor} is supported only on the class
\((0,1)\). All other classes have no irreducible representative. Their possible reducible or multiple-cover contributions are already accounted for by the exponential structure of the GV expansion.
The primitive GV invariant outside the Hanany-Witten image is therefore only $\GV_{0,(0,1)}(K_{\mathbb F_2})=-1$, with all higher-genus invariants in this primitive class vanishing.

On Figure \ref{fig:F2Mori}, we have also indicated the hyperbolas of constant genus \eqref{eq:genusFormulaFn}. One can check in Table \ref{tab:GV_F2} that all the GV invariants which are allowed by the bound \eqref{eq:genusFormulaFn} to be non-vanishing are indeed non-vanishing. We will see in the next section that this is not always the case in general. 

\paragraph{The Fano property. } Local mirror symmetry suggests \cite{chiang1999localmirrorsymmetrycalculations} that the 5d SCFT doesn't capture the curve classes that have zero intersection with the anticanonical divisor. The curves corresponding to parallel branes are of this type. The example of $K_{\mathbb{F}_0}$ and $K_{\mathbb{F}_2}$ illustrates this point:
\begin{itemize}
    \item $F_0$ is Fano, and it was shown that instanton partition function from field theory side captures all enumerative information of $K_{\mathbb{F}_0}$ \cite{nekrasov2003seiberg, Eguchi:2003sj}, i.e. $Z_{\text{top}}(K_{\mathbb{F}_0}) = Z_{\text{5d SCFT}}$.
    \item $F_2$ is not Fano and the curve corresponding to parallel brane has zero intersection with the anti canonical divisor. The enumerative invariants from these curves are not captured by the $5d$ SCFT, see \eqref{eq:relationF0F2}.
\end{itemize}
Hence there seems to be a close relation between the surface $CY3$ $K_B$ with base $B$ Fano or not and the existence of decoupled sector of spacetime physics by compactifying M theory on $K_B$. This echoes the remarks made in the introduction.

\subsection{The \texorpdfstring{$\mathbb{F}_1$}{F1} - \texorpdfstring{$\mathbb{F}_3$}{F3} pair}  
\label{subsec: Mori cone F1 F3}

We now consider the Hanany--Witten move relating the ordinary local
\(\mathbb F_1\) geometry to the generalized toric geometry, which we call $X_3^{\mathrm{wd}}$, associated to \(\mathbb F_3\) as shown in Figure~\ref{fig:HW_F0F2F1F3}. In contrast with the
\(K_{\mathbb F_0}\)--\(K_{\mathbb F_2}\) pair, the final diagram contains a
white dot. The white dot modifies the local vertex prescription and removes
the contribution which, in an ordinary toric diagram, would be associated
with a pair of parallel external legs.

The discussion below assumes that the generalized toric diagram admits the
local topological vertex description with white dots reviewed above. Under
this assumption, the topological string partition function of $X_3^{\mathrm{wd}}$ agrees with that of \(K_{\mathbb F_1}\) after the Hanany--Witten identification of K\"ahler parameters:
\begin{equation}
    \label{F1 F3 wd identity}
    Z_{\text{top}}
    (X_3^{\mathrm{wd}};Q_f,Q_b)
    =
    Z_{\text{top}}
    (K_{\mathbb F_1};Q_f,Q_fQ_b).
\end{equation} 
Equivalently, there is no additional parallel-leg factor analogous to the one appearing in the \(K_{\mathbb F_0}\)--\(K_{\mathbb F_2}\) comparison. This is illustrated in Figure \ref{fig:F1Mori}.

For $K_{\mathbb{F}_1}$, we naturally use the notations \eqref{eq:curveClassFn}. For $X_3^{\mathrm{wd}}$, we also use the numerical basis \((f_3,b_3)\) suggested by the local \(\mathbb F_3\) charts. Since \(X_3^{\mathrm{wd}}\) is not an ordinary toric local surface, the following cone should be understood as the numerical cone of compact curve classes determined by the generalized toric description:
\begin{equation}
    \label{Mori cone F3 wd}
    \NEbar_{\mathrm{num}}(X_3^{\mathrm{wd}})
    =
    \bZ_{\geq 0} f_3
    \oplus
    \bZ_{\geq 0} b_3.
\end{equation}
In particular, the two internal curves labelled by the same K\"ahler
monomial $Q_2$ in Figure~\ref{fig:HW_F0F2F1F3} are treated as having the same numerical degree \(b_3\).

The Hanany--Witten move then provides an identification exactly similar to \eqref{eq:mapF0F2}, namely 
\begin{equation}
    d_{f_3} = d_{f_1} + d_{b_1} \, , \qquad  d_{b_3} =   d_{b_1} \label{mori cone map from HW move with white dot} 
\end{equation}
Thus the image of the Mori cone of \(\mathbb F_1\) inside the numerical cone
of the generalized \(K_{\mathbb F_3}\)   is
\begin{equation}
     \{(d_{f_3},d_{b_3})\in \bZ_{\geq 0}^2\mid d_{f_3} \geq d_{b_3} \}.
\end{equation} 
The extra numerical classes in $X_3^{\mathrm{wd}}$ satisfy $d_{f_3}  < d_{b_3}$. Imposing that the genus given by \eqref{eq:genusFormulaFn} is non negative implies again $(d_{f_3},d_{b_3}) = (0,1)$.  
Among the numerical classes outside the Hanany--Witten image, this is the only class which can be represented by an irreducible curve.

One could therefore expect a possible extra contribution supported on this class. In the generalized diagram with a white dot, however, this contribution cancels.
Indeed, setting \(Q_f\to 0\) isolates the part of the topological string
partition function supported only in \(Q_b\)-degree. The $Q_1\to 0$ limit of \eqref{eq:combinatorialIdentity2} gives the identity
\begin{equation}
    \sum_{\mu}\ \frac{s_\mu(q^{-\rho})\,s_{\mu^T}(q^{-\rho})\,(-Q_b)^{|\mu|}\,q^{\kappa_\mu}}
{\mathcal R_{\mu^T\emptyset}(Q_b)} = 1 \, . 
\end{equation}
This means the naive $(1,-3)$-curve sum $\sum_\mu s_\mu s_{\mu^T}(-Q_b)^{|\mu|}q^{\kappa_\mu}$ is cancelled against the extra $\mathcal R_{\mu^T\emptyset}(Q_b)^{-1}$ that the white-dot rule supplies. 
This example also clarifies the role of the white dot in the generalized toric vertex. The numerical Mori-cone analysis shows that there is one possible extra primitive class, namely \((0,1)\), but the vertex calculation then shows that the two curves of this numerical degree contribute with opposite framing and cancel.

\paragraph{Pattern of non-vanishing GV invariants. }
One thing remains to be explained in the results shown in Tables \ref{tab:GV_F1}--\ref{tab:GV_GTPF3}.  Take for instance in $\mathbb{F}_1$ the classes  $5 b_1 + 4 f_1$ and $6 b_1 + 4 f_1$. The genus formula gives that their genera are $2$ and $0$ respectively, so one could expect there to be non-zero GV invariants in these classes up to these genera. But this is not what is shown in Table \ref{tab:GV_F1}. 

These anomalous vanishings have a geometric explanation in the fact that the geometries can be flopped. Using the results of Section \ref{sec:flops}, one sees that the topological string partition function of $K_{\mathbb F_1}$ agrees with that of its flop (the geometry whose only compact divisor is a $\mathbb P^2$) once the conifold factor of the flopped curve is removed and the K\"ahler parameters are identified as indicated there. GV invariants are therefore conserved through the flop, away from the class of the flopped curve itself. The flopped curve is the section $b_1$, and the flop acts on curve classes by $b_1 \mapsto - b_1$, while $f_1 + b_1$ is identified with the hyperplane class of the $\mathbb P^2$; decomposing $\beta = d_{f_1} f_1 + d_{b_1} b_1 = d_{f_1}(f_1+b_1) + (d_{b_1} - d_{f_1}) \, b_1$ shows the effect of the flop on the degrees.
This implies that a class with $d_{b_1} > d_{f_1}$ lies outside of the Mori cone in the flopped geometry, and the conservation of GV invariants implies that the classes must have vanishing GV invariants at all genus. This additional bound is depicted in red in Figure \ref{fig:F1Mori}. A similar argument can be made for $\mathbb{F}_3$, giving the exclusion red zone in Figure \ref{fig:F1Mori} (right).
 
These observations lead us to conjecture the converse to the implication \eqref{eq:implicationGenus}. Namely, if a class $\beta$ lies in the Mori cone of the geometry under consideration and in the Mori cone of all its possible flops, then generically $r \leq g(\beta)$ implies $\GV_{r , \beta} \neq 0$.  More precisely, calling $M$ the intersection of the Mori cones of all geometries that can be reached using successions of flops, an interesting question is whether\footnote{Note this is a kind of the converse to the $BPS$ conjecture, see \cite{Pandharipande:2011jz}.  } 
\begin{equation}
\label{eq:conj}
   \# \{ (r , \beta) \in  \mathbb{N} \times M | \,  r \leq g(\beta) \quad \text{and} \quad \GV_{r , \beta} = 0 \} < \infty \, . 
\end{equation}

\subsection{Distinguishing \texorpdfstring{$K_{\mathbb{F}_3}$}{KF3} and \texorpdfstring{$X_3^{\text{wd}}$}{X3wd}}
\label{sec:F3vsGTP}

In this last subsection, we comment on the distinction between the geometries $K_{\mathbb{F}_3}$ and the generalized toric geometry associated to the GTP on the right hand side of \eqref{eq:equalityF1F3GTP}. This is elaborated in \cite{alexeev2025non,BCL}, to which we refer for the details. 

We need to introduce some notations to make the discussion precise. Let $\mathfrak{D}$ be a generalized toric diagram. We say it is toric if it has no white dot. Given a toric diagram, we can construct the associated three-dimensional fan $\Sigma_\mathfrak{D}$, and given a fan $\Sigma$, we can construct the associated toric variety $X[\Sigma]$. When $\mathfrak{D}$ is a toric diagram, we call for short $X[\mathfrak{D}] := X[\Sigma_\mathfrak{D}]$.  When $\mathfrak{D}$ is a non-toric generalized toric diagram with one white dot, what we mean by $X[\mathfrak{D}]$ is the generic fiber of a 1-parameter family of threefolds (we call $\alpha \in \mathbb{P}^1$ the parameter), as explained in \cite{alexeev2025non,BCL}. The fiber above $\alpha = 0$ is $X[\overline{\mathfrak{D}}]$, where $\overline{\mathfrak{D}}$ is the generalized toric diagram $\mathfrak{D}$ with white dot removed.   

For the discussion of the present section, this can be illustrated in the following diagram: 
\begin{equation}
    \begin{tikzpicture}
    \node at (-2.9,1.4) {$K_{\mathbb{F}_3}$};
    \node at (.1,1.4) {$X_3^{\text{wd}}$};
        \node at (-3,0) {$X \left[   \raisebox{-.5\height}{ \scalebox{.8}{  
\begin{tikzpicture}[x=.5cm,y=.5cm] 
\draw[gridline] (0,-0.5)--(0,3.5); 
\draw[gridline] (1,-0.5)--(1,3.5); 
\draw[gridline] (2,-0.5)--(2,3.5); 
\draw[gridline] (-0.5,0)--(2.5,0); 
\draw[gridline] (-0.5,1)--(2.5,1); 
\draw[gridline] (-0.5,2)--(2.5,2); 
\draw[gridline] (-0.5,3)--(2.5,3); 
\draw[ligne] (0,2)--(1,0); 
\draw[ligne] (1,0)--(2,3); 
\draw[ligne] (0,2)--(1,2)--(2,3); 
\draw[ligne] (0,2)--(1,1)--(2,3); 
\draw[ligne] (1,2)--(1,0); 
\node[bd] at (1,0) {}; 
\node[bd] at (1,1) {}; 
\node[bd] at (1,2) {}; 
\node[bd] at (2,3) {}; 
\node[bd] at (0,2) {}; 
\end{tikzpicture}}} \right]$};
\node at (-1.5,0) {$\neq$};
        \node (1) at (0,0) {$X \left[   \raisebox{-.5\height}{ \scalebox{.8}{  
\begin{tikzpicture}[x=.5cm,y=.5cm] 
\draw[gridline] (0,-0.5)--(0,3.5); 
\draw[gridline] (1,-0.5)--(1,3.5); 
\draw[gridline] (2,-0.5)--(2,3.5); 
\draw[gridline] (-0.5,0)--(2.5,0); 
\draw[gridline] (-0.5,1)--(2.5,1); 
\draw[gridline] (-0.5,2)--(2.5,2); 
\draw[gridline] (-0.5,3)--(2.5,3); 
\draw[ligne] (0,2)--(1,0); 
\draw[ligne] (1,0)--(2,3); 
\draw[ligne] (2,3)--(0,3); 
\draw[ligne] (0,3)--(0,2); 
\draw[ligne] (0,2)--(1,2)--(2,3); 
\draw[ligne] (0,2)--(1,1)--(2,3); 
\draw[ligne] (1,2)--(1,0); 
\node[bd] at (1,0) {}; 
\node[bd] at (1,1) {}; 
\node[bd] at (1,2) {}; 
\node[bd] at (2,3) {}; 
\node[bd] at (0,3) {}; 
\node[bd] at (0,2) {}; 
\node[wd] at (1,3) {}; 
\end{tikzpicture}}} \right]$};
        \node (2) at (5,0) {$X \left[   \raisebox{-.5\height}{ \scalebox{.8}{  
\begin{tikzpicture}[x=.5cm,y=.5cm] 
\draw[gridline] (0,-0.5)--(0,3.5); 
\draw[gridline] (1,-0.5)--(1,3.5); 
\draw[gridline] (2,-0.5)--(2,3.5); 
\draw[gridline] (-0.5,0)--(2.5,0); 
\draw[gridline] (-0.5,1)--(2.5,1); 
\draw[gridline] (-0.5,2)--(2.5,2); 
\draw[gridline] (-0.5,3)--(2.5,3); 
\draw[ligne] (0,2)--(1,0); 
\draw[ligne] (1,0)--(2,3); 
\draw[ligne] (2,3)--(0,3); 
\draw[ligne] (0,3)--(0,2); 
\draw[ligne] (0,2)--(1,2)--(2,3); 
\draw[ligne] (0,2)--(1,1)--(2,3); 
\draw[ligne] (1,2)--(1,0); 
\node[bd] at (1,0) {}; 
\node[bd] at (1,1) {}; 
\node[bd] at (1,2) {}; 
\node[bd] at (2,3) {}; 
\node[bd] at (0,3) {}; 
\node[bd] at (0,2) {}; 
\end{tikzpicture}}} \right]$};
        \node (3) at (0,-3) {$X \left[   
  \raisebox{-.5\height}{ \scalebox{.8}{  
\begin{tikzpicture}[x=.5cm,y=.5cm] 
\draw[gridline] (0,-0.5)--(0,3.5); 
\draw[gridline] (1,-0.5)--(1,3.5); 
\draw[gridline] (2,-0.5)--(2,3.5); 
\draw[gridline] (-0.5,0)--(2.5,0); 
\draw[gridline] (-0.5,1)--(2.5,1); 
\draw[gridline] (-0.5,2)--(2.5,2); 
\draw[gridline] (-0.5,3)--(2.5,3); 
\draw[ligne] (0,2)--(1,0); 
\draw[ligne] (1,0)--(2,3); 
\draw[ligne] (2,3)--(0,3); 
\draw[ligne] (0,3)--(0,2); 
\node[bd] at (1,0) {}; 
\node[bd] at (2,3) {}; 
\node[bd] at (0,3) {}; 
\node[bd] at (0,2) {}; 
\node[wd] at (1,3) {}; 
\end{tikzpicture}}} \right]$};
        \node (4) at (5,-3) {$X \left[   
  \raisebox{-.5\height}{ \scalebox{.8}{  
\begin{tikzpicture}[x=.5cm,y=.5cm] 
\draw[gridline] (0,-0.5)--(0,3.5); 
\draw[gridline] (1,-0.5)--(1,3.5); 
\draw[gridline] (2,-0.5)--(2,3.5); 
\draw[gridline] (-0.5,0)--(2.5,0); 
\draw[gridline] (-0.5,1)--(2.5,1); 
\draw[gridline] (-0.5,2)--(2.5,2); 
\draw[gridline] (-0.5,3)--(2.5,3); 
\draw[ligne] (0,2)--(1,0); 
\draw[ligne] (1,0)--(2,3); 
\draw[ligne] (2,3)--(0,3); 
\draw[ligne] (0,3)--(0,2); 
\node[bd] at (1,0) {}; 
\node[bd] at (2,3) {}; 
\node[bd] at (0,3) {}; 
\node[bd] at (0,2) {}; 
\end{tikzpicture}}} \right]$};
    \draw[->] (1)--(2);
    \draw[->] (1)--(3);
    \draw[->] (2)--(4);
    \draw[->] (3)--(4);
    \end{tikzpicture}
\end{equation}
The horizontal arrows correspond to the $\alpha \rightarrow 0$ limit and the vertical arrows correspond to a resolution of the singularities. On the top left corner, we have the two distinct geometries $K_{\mathbb{F}_3}$ and $X_3^{\text{wd}}$. The one which corresponds to the 5d SCFT data is $X_3^{\text{wd}}$, which is physically motivated as the compact divisors can be shrunk. The local $\mathbb{F}_3$ geometry hosts many more curves, as can be seen by computing its GV invariants, shown in Table \ref{tab:GV_GTPF3} (top). Our interpretation is that these curves arise from the non-ample canonical divisor in the $\mathbb{F}_3$ surface; the non-ampleness implies that it can not take part in the 5d SCFT. Equivalently, this is the reason for the large number of non-zero entries in Table \ref{tab:GV_GTPF3} (top). One can understand the white dot geometry $X_3^{\text{wd}}$ as what remains of $K_{\mathbb{F}_3}$ when the curves that obstruct the SCFT limit are removed. 

\begin{table}[t]
    \centering          
        \makebox[\textwidth][c]{\begin{tabular}{c||c|c|c|c|c|c}
        $d_f \backslash d_b $ & 0 & 1 & 2 & 3 & 4 & 5 \\\hline\hline
        0 &  & 1 & -1 & 2, -1 & -7, 11, -6, 1 & 31, -106, 161, -130, 56, -12, 1 \\\hline
        1 & -2 & 2 & -2 & 6, -2 & -26, 32, -14, 2 & 136, -392, 506, -352, 134, -26, 2 \\\hline
        2 &  & 3 & -3 & 10, -3 & -52, 56, -22, 3 & 322, -817, 945, -602, 215, -40, 3 \\\hline
        3 &  & 5 & -4 & 14, -4 & -82, 80, -30, 4 & 572, -1316, 1414, -856, 296, -54, 4 \\\hline
        4 &  & 7 & -12 & 25, -5 & -133, 111, -38, 5 & 952, -1955, 1937, -1117, 377, -68, 5 \\\hline
        5 &  & 9 & -40, 9 & 63, -15 & -270, 196, -55, 6 & 1771, -3188, 2798, -1468, 467, -82, 6
        \end{tabular}}

\vspace{1cm}
        
\makebox[\textwidth][c]{\begin{tabular}{c||c|c|c|c|c}
 $d_f \backslash d_b $ & 0 & 1 & 2 & 3 & 4  \\\hline\hline
0 &  &  &  &  &    \\\hline
1 & -2 & 1 &  &  &    \\\hline
2 &  & 3 &  &  &    \\\hline
3 &  & 5 &  &  &   \\\hline
4 &  & 7 & -6 &  &   \\\hline
5 &  & 9 & -32, 9 &  &    \\\hline
6 &  & 11 & -110, 68, -12 & 27, -10 &    \\\hline
7 & & 13 & -288, 300, -116, 15 & 286, -288, 108, -14 & \\\hline
8 & & 15 & -644, 988, -628, 176, -18 & 1651, -2938, 2353, -992, 212, -18  & -192, 231, -102, 15 
\end{tabular}}
    \caption{\emph{Top:} GV invariants for the local $\mathbb F_3$ geometry. \emph{Bottom:} GV invariants for the GTP geometry associated to $\mathbb F_3$.}
    \label{tab:GV_GTPF3}
    \end{table}

\section{Rank-1 SCFTs, Del Pezzo and Generalized del Pezzo surfaces}
\label{sec:rank1}

The rank-1 $E_n$ 5d SCFTs provide the simplest yet non trivial example where the strategy outlined in the introduction can be applied. In this section, we consider the range $2 \leq n \leq 8$ (since Section \ref{sec:HWMori} has dealt with $E_1$ and $\tilde{E}_1$). They can be engineered from M-theory on local del Pezzo surfaces, $K_{dP_{n}}$ \cite{Morrison:1996xf}. Recall that $dP_{n}$ is obtained by blowing up $n$ points at generic position on $\mathbb{P}^2$. Since one can pick at most three points on $\mathbb{P}^2$ in such a way that there is a toric action fixing them, $dP_{0 \leq n \leq 3}$ is toric but $dP_{4 \leq n \leq 8}$ is not toric. However, the same SCFT can be engineered on local \textit{generalized} del Pezzo surfaces, local GdPs for short (there are several distinct surfaces of type $GdP_n$ for any fixed $n$). We refer to Appendix \ref{app:delPezzo} for details about the geometry of $dP_n$ and $GdP_n$ surfaces.  The \emph{toric} surfaces of type $GdP_{4 \leq n \leq 6}$ are iterated blow-ups of $\mathbb{P}^2$ at toric points. There are no toric surfaces of type $GdP_{7 \leq n \leq 8}$. However, for any $n$, there are $GdP_n$ surfaces for which the local model is a GTP threefold.\footnote{Strictly speaking, this last statement is conjectural. } In the following, we denote by $GdP_{n}$ such a surface.   
As discussed above, we can compute the GV invariants of $K_{dP_{n}}$ from these using \eqref{eq:main2}: 
\begin{equation}
    \label{Gdp and dP via SCFT}
 \forall \, 2 \leq n \leq 8 \, , \qquad  \frac{Z_{\mathrm{top}}(K_{GdP_n})}
     {Z_{\parallel}(K_{GdP_n})}
=
Z_{E_n\,\mathrm{SCFT}}
=
Z_{\mathrm{top}}(K_{dP_n}).
\end{equation}
provided the K\"ahler parameters on both sides are identified via the same $5d$ SCFT they give rise to. The first equality follows from the discussion in Appendix \ref{app:GVcomplexStructureVariation}.  The second equality uses the physical identification with the same
$E_n$ SCFT together with the absence, for a Fano surface $dP_n$, of the anticanonical-degree-zero sector discussed in Appendix~\ref{app:GVcomplexStructureVariation}. The left hand side of \eqref{Gdp and dP via SCFT} is computed using the topological vertex as reviewed in Section \ref{sec:topologicalVertex}. The results are presented in Section \ref{sec:topologicalVertexEn}.

Before turning to this, we describe complementary mirror-symmetry calculations that will be used to check progressively finer specializations of our results. In Section~\ref{sec:localMirrorSymmetry}, a one-parameter local mirror-symmetry calculation is performed at the massless point. Consequently, it does not distinguish curve classes with the same anticanonical degree. It determines the sums
\begin{equation}
\label{eq:GVsums}
      n_d^r
    =
    \sum_{\substack{\beta\in\NEbar(S)\\-K_S\cdot\beta=d}}
    \GV_{r,\beta}(K_S),  
\end{equation}
for $r=0$ and arbitrary anticanonical degree $d$.
We then restore the K\"ahler parameters in the directions
orthogonal to the canonical class. This refines the preceding calculation by organizing the invariants of $dP_8$ into $E_8$ Weyl orbits.
Specializing these parameters then gives the anticanonical-degree sums for the lower del Pezzo surfaces. Higher-genus information can in principle be obtained recursively from the holomorphic-anomaly equations.

Note that the invariants for $K_{dP_4}$ can be efficiently calculated by combining \eqref{Gdp and dP via SCFT} with the results of Section \ref{sec:flops}. It is worth mentioning that $dP_4$ is the only del Pezzo surface to which current local mirror symmetry techniques seem not to apply directly, as its defining equations in weighted projective spaces are complicated, see \cite{bauer2020mathfraks5equivariantsyzygiesdel}.

\subsection{Local mirror symmetry and its mass-dependent refinement}
\label{sec:localMirrorSymmetry}

\paragraph{The massless one-parameter specialization.}
Local mirror symmetry at the massless point reduces the K\"ahler moduli of $K_{dP_8}$ to a single parameter $t$ in the anticanonical direction. Under this specialization, every effective class $\beta$ contributes through the monomial
\[
    Q^{-K_{dP_8}\cdot\beta},
    \qquad
    Q=e^{2\pi i t}.
\]
The calculation therefore determines the genus-zero invariants only after summing over all curve classes of fixed anticanonical degree, see \eqref{eq:GVsums}. 

Following \cite{chiang1999localmirrorsymmetrycalculations}, these
one-parameter genus-zero invariants can be extracted by embedding the local
geometry into a compact elliptically fibered Calabi--Yau threefold and then
taking the limit in which the elliptic fiber has infinite volume. The
relevant local periods satisfy the Picard--Fuchs equation
\begin{equation}
    \cL_{dP_8} \Pi = 0
\end{equation}
with \footnote{One way to derive the PF operator\cite{Katz:1999xq} is noticing that $K_{dP_8}$ can be realized by $(2,2)$ GLSM with the following field contents:
\begin{itemize}
    \item $U(1)$ gauge fields with complexified FI parameter $t=i\zeta + \frac{\theta}{2\pi}$
    \item $6$ chiral multiplets with $U(1)$ gauge charge $(\ell_i)_{i=1,..,6} = (-6,1,1,2,3,-1)$.
\end{itemize}
With above specified spectrum, we can write down the PF operator:
\begin{equation}\label{GKZ for dP8 GLSM}
    \cL = \prod_{i:\ell_{i}>0}\prod_{j=0}^{\ell_i-1}(\ell_i \theta + j)- z\prod_{i:\ell_i<0}\prod_{j=0}^{-1-\ell_i}(\ell_i \theta - j)  
\end{equation}
where $z=e^{2\pi i t_{\mathrm{UV}}}$ is the algebraic coordinate
near the large-complex-structure point. The flat coordinate $t(z)$
is obtained from the period ratio, and
$Q=e^{2\pi i t(z)}$ is related to $z$ by the mirror map. \eqref{GKZ for dP8 GLSM} specified to above matter contents simplified to \eqref{E8 PF}, after factoring out common components. \\ Another derivation is provided in Appendix \ref{app:delPezzo}. 
}:  
\begin{equation}\label{E8 PF}
   \cL_{dP_8} = \theta^3 + 12z (6\theta +5)(6\theta +1)\theta, \quad \theta = z\partial_z.  
\end{equation}
The fundamental solutions to the equation $\cL_{dP_8} \Pi = 0$ are defined uniquely by specifying the leading logarithmic behaviors as follows
\begin{equation}
\label{eq:periodsLocaldP8}
    \begin{split}
        \Pi_0 (z) &= 1 \\
        \Pi_1 (z) &= \log z  + \Pi_1^{\text{pert}} (z)   \\
        \Pi_2 (z)&= \frac{1}{2} (\log z)^2 + \Pi_1^{\text{pert}} (z) \log z + \Pi_2^{\text{pert}} (z) \\
    \end{split}
\end{equation}
where $\Pi_{1,2}^{\text{pert}} (z)$ are power series in $z$. Explicitly, we find 
\begin{equation}
    \begin{split}
        \Pi_1^{\text{pert}} (z) &= -60 z + 6930 z^2 - 1361360 z^3 + 334639305 z^4 + O(z^5) \\ 
        \Pi_2^{\text{pert}} (z) &= -252 z + 35361 z^2 - 7374592 z^3 + \frac{7465226529}{4}z^4 + O(z^5)   
    \end{split}
\end{equation}
Introduce the standard notations
\begin{equation}
    t(z) = \frac{1}{2 \pi i} \Pi_1 (z) \, , \qquad t_D(z) = \frac{1}{(2 \pi i)^2} \Pi_2 (z) \, . 
\end{equation}
and 
\begin{equation}
    Q(z) = e^{2 \pi i \, t(z)} = e^{\Pi_1 (z)} \, . 
\end{equation}
The mirror map involves computing the inverse series expansion $z(Q)$, which can be substituted into $t_D(z)$ to obtain
$t_D(z(Q))$ as a series in $Q$. Explicitly, one finds 
\begin{equation}
    Q(z)
    =
    z-60z^2+8730z^3-1813160z^4+\mathcal O(z^5) 
\end{equation}
and 
\begin{equation}
    z(Q) = Q + 60 Q^2 - 1530 Q^3 + 274160 Q^4 + O(Q^5) \, . 
\end{equation}
Local mirror symmetry implies the genus zero invariants \eqref{eq:GVsums} can be obtained as 
\begin{equation}
\label{enumerative from local mirror}
    \partial_t^2 t_D(t)
    =
    1-
    \sum_{d=1}^{\infty}
    d^3 n_d^0
    \frac{Q^d}{1-Q^d}.
\end{equation}
Equation~\eqref{enumerative from local mirror} determines $n_d^0$ to arbitrary anticanonical degree, and Table \ref{tab:genus0acGV} displays the first twelve degrees.   

Since $dP_8$ is Fano, it has no nonzero effective class of anticanonical degree zero. Thus no parallel-brane sector of the type identified in Appendix~\ref{app:GVcomplexStructureVariation} needs to be removed from $Z_{\mathrm{top}} (K_{dP_8})$. This is consistent with identifying the genus-zero anticanonical sums above with the massless $E_8$ SCFT data. Notice, however, that Fano-ness alone does not derive the full equality of partition functions; it only rules out the particular decoupled sector supported on anticanonical-degree-zero curves.

\begin{table}[t]
    \centering
    \begin{tabular}{|c|c|}\hline 
    $d$ & $n_d^0$ \\ \hline 
 1 & 252 \\
 2 & -9252 \\
 3 & 848628 \\
 4 & -114265008 \\
 5 & 18958064400 \\
 6 & -3589587111852 \\
 7 & 744530011302420 \\
 8 & -165076694998001856 \\
 9 & 38512679141944848024 \\
 10 & -9353163584375938364400 \\
 11 & 2346467355966572489025540 \\
 12 & -604657435721239536237491472 \\ \hline 
    \end{tabular}
    \caption{Genus 0 anticanonical GV sums for $dP_8$ from \eqref{enumerative from local mirror}.  }
    \label{tab:genus0acGV}
\end{table}

\paragraph{Restoring the $E_8$ mass parameters.}
We now restore the dependence on the K\"ahler parameters orthogonal to the canonical class. At genus zero, the resulting
mass-dependent generating function determines the individual GV invariants of $K_{dP_8}$ up to the action of the $E_8$ Weyl group.

For $dP_8$, the orthogonal complement of the canonical class in
$H_2(dP_8;\bZ)$ is the negative-definite $E_8$ root lattice. For a curve class $\beta$ of anticanonical degree $d=-K_{dP_8}\cdot\beta$,  define its component in the $E_8$ lattice by
\begin{equation}
    v_\beta = \beta-(K_{dP_8}\cdot\beta)K_{dP_8}.
\end{equation}
We introduce a mass parameter $\mathbf m = (m_1 , \dots , m_8) \in\Gamma_{E_8}^\ast\otimes_{\bZ}\bC$ conjugate to this component. We denote by $(- , -)$ the natural pairing $\Gamma_{E_8}^{*} \times \Gamma_{E_8} \rightarrow \mathbb{C}$.    The variable $Q$ continues to record the anticanonical degree, whereas $\mathbf m$ distinguishes different $E_8$ Weyl orbits at fixed degree. Then \eqref{enumerative from local mirror} is the $\mathbf m\rightarrow 0$ limit\footnote{This is due to the fact that the GLSM we used to derive PF system can only probe the $m=0$.  } of the refined series   
\begin{equation}
\label{Weyl orbit of genus 0 dP8 oth}
Y_{dP_8}(Q;\mathbf m)
= 1+
\sum_{\substack{\beta\in\NEbar(dP_8)\\ \beta\neq0}}
(K_{dP_8}\cdot\beta)^3
\GV_{0,\beta}(K_{dP_8}) 
\frac{
Q^{-K_{dP_8}\cdot\beta}
e^{2\pi i(\mathbf m,v_\beta)}
}{
1-
Q^{-K_{dP_8}\cdot\beta}
e^{2\pi i(\mathbf m,v_\beta)}
}.
\end{equation}
The Weyl group of $E_8$ naturally acts on the root lattice $\Gamma_{E_8}$, so the sum above can be organized into $E_8$ orbits of elements in $\Gamma_{E_8}$.\footnote{Recall that elements of $\Gamma_{E_8}$ have norm squared in $2 \mathbb{N}$, and the Weyl group action preserves the norm, but does not necessarily act transitively on the set of elements of a given norm. The small orbits are fully characterized by the norm squared of their elements and their cardinality. } So \eqref{Weyl orbit of genus 0 dP8 oth} can be organized into 
\begin{equation}
\label{Weyl orbit of genus 0 dP8 1st}
Y_{dP_8}(Q;\mathbf m)
=
1-
\sum_{d=1}^{\infty}
d^3
\sum_{\mathcal O\in\Gamma_{E_8}/W(E_8)}
a_{\mathcal O}^d\,
\chi_{\mathcal O}(Q^d,\mathbf m),
\end{equation}
where 
\begin{equation}
    \chi_{\mathcal O}(Q,\mathbf{m})
    =
    \sum_{\mathbf v\in\mathcal O}
    \frac{
        Qe^{2\pi i(\mathbf m,\mathbf v)}
    }{
        1-Qe^{2\pi i(\mathbf m,\mathbf v)}
    }.
\end{equation}
and the $a_{\mathcal{O}}^d$ are integers. The first few are given in Table \ref{tab:EstringIntegers 1st}.
Comparing \eqref{Weyl orbit of genus 0 dP8 oth} and \eqref{Weyl orbit of genus 0 dP8 1st} shows how these integers are determined: by Weyl invariance, $\GV_{0,\beta}(dP_8)$ depends only on the degree $d = -K_{dP_8} \cdot \beta$ and on the orbit $\mathcal{O}$ of $v_\beta$, so that
\begin{equation}
    a_{\mathcal O}^d
    =
    \GV_{0,\beta}(K_{dP_8})
\end{equation}
for any class $\beta$ of degree $d$ whose associated lattice vector lies in $\mathcal{O}$. In Table \ref{tab:EstringIntegers 1st}, each orbit is labelled by the square norm\footnote{Since $K_{dP_8}^{\perp}$ is negative definite, we use the
positive-definite convention $|\mathbf v|^2 := -\,\mathbf v\cdot\mathbf v $. } of its elements together with its cardinality $|\mathcal{O}|$, which characterize the small orbits uniquely.

Thus the mass-dependent genus-zero series contains more information than
the one-parameter Picard--Fuchs equation. The latter is recovered by
setting $\mathbf m=0$ and summing over all Weyl orbits:
\begin{equation}
    n_d^0
    =
    \sum_{\mathcal O}
    |\mathcal O|\,a_{\mathcal O}^d.
\end{equation}
For example, at degree one and two Table~\ref{tab:EstringIntegers 1st} gives
\[
    n_1^0=12+240=252, \qquad n_2^0
    =
    -132-240\times20-2160\times2
    =
    -9252,
\]
in agreement with Table~\ref{tab:genus0acGV}.

\begin{table}[t]
    \centering
\begin{tabular}{cc|ccc}
    $|\mathbf v|^2$ & $|\mathcal O|$
    & $d=1$ & $d=2$ & $d=3$ \\ \hline
    0 & 1     & 12 & $-132$ & 4068 \\
    2 & 240   & 1  & $-20$  & 927 \\
    4 & 2160  &    & $-2$   & 180 \\
    6 & 6720  &    &        & 27 \\
    8 & 17280 &    &        & 3
\end{tabular}
    \caption{Coefficients of the smallest Weyl orbits for $E_8$, taken from \cite[Table 1]{Minahan_1997}. Each row corresponds to an orbit, which is specified by the norm squared $|\mathbf{v}|^2$ of one of its element, and its cardinality $|\mathcal{O}|$.  }
    \label{tab:EstringIntegers 1st}
\end{table}

The mass-dependent $dP_8$ series can also be specialized to the massless
points of the lower del Pezzo theories. We restrict here to
$3\leq n\leq8$, for which the specialization below is directly expressed
in terms of the standard $E_n$ lattice data\footnote{As this method relies on the Weyl orbit structure. It applies to the $dP_n$ surfaces with second homology lattice is $K_{dP_n}\oplus \Lambda_{E_n}$, where $\Lambda_{E_n}$ is the $E_n$ root lattice. This is true for $dP_{n\geq 3}$, that's why we focus on $3\leq n\leq 8$ here. Here we see $dP_3$ is special. It is toric, its Mori cone is not simplicial, and its second homology lattice orthogonal to canonical class is the $E_3$ root lattice.}.

Since $K_{dP_n}^2=9-n$, the classical term of the unnormalized genus-zero Yukawa coupling is $9-n$. It is convenient to divide the full coupling by this number so that its constant term is one, as in the $dP_8$ convention. With this normalization, the specialization reads
\begin{align}\label{GV series based on Weyl orbit}
   Y_{dP_8} ( Q ; \mathbf{m}) \left|_{\begin{array}{c}
       m_1 = \ldots = m_{n-1} = 0  \\
       m_n  =\ldots = m_7 = \frac{1}{9-n}\\ 
       m_8 = \frac{8-n}{9-n}
  \end{array}}\right. &= 1 -   \frac{1}{9-n}  \sum_{d=1}^{+\infty}d^3 n_d^0 (dP_n)\frac{Q^d}{1-Q^d} 
 \end{align} 

The explicit calculations above concern genus zero. Higher-genus
information is not contained in the Picard--Fuchs equation
\eqref{E8 PF} alone. It can be obtained either using holomorphic anomaly equations, or E string techniques.   We do not repeat them here. Instead, we use the available E-string results as independent checks of the invariants obtained from the topological vertex. In particular, for $dP_4$ we first remove the degree-zero parallel-brane sector from the generalized toric partition function and then compare the sums \eqref{eq:GVsums} with the corresponding E-string data at low degree. Thus the topological vertex provides the curve-class-refined invariants, while local mirror symmetry provide progressively stronger checks after specialization.

\subsection{Topological Vertex computation}
\label{sec:topologicalVertexEn}

We now compute explicitly $Z_{5d}$ for the rank 1 SCFTs and carefully identify the contributions from the topological string partition and the decoupled sectors. We relate in some cases these expressions with Nekrasov instanton counting functions. The main formula, which we unpack for each SCFT in the subsections below, is 
\begin{equation}  \label{eq:ZtopZnek}
    Z_{\mathrm{top}} (\tX_n ; Q) = Z_{\parallel}  (\tX_n ; Q) \; \underbrace{Z_{\mathrm{Pert}} (Q) \; Z_{\mathrm{Inst}}^{\mathrm{SU}(2)} (Q) }_{Z_{5d} (Q)} 
\end{equation}
where $X_n$ is any GTP threefold engineering the $E_n$ SCFT. Here, $Z_{\mathrm{Inst}}^{\mathrm{SU}(2)}$ is the Nekrasov instanton partition function, and we have added a superscript to emphasize it should be computed with gauge group SU(2). Note that the notion of $ Z_{\parallel}  (\tX_n)$ is well-defined since $X_n$ has an associated brane web by assumption. The terms which depend on the choice of $X_n$ have been explicitly indicated in the formula above. It has been noted that $Z_{\mathrm{Inst}}^{\mathrm{SU}(2)}$ can in some cases be related to the U(2) instanton partition function by means of a multiplicative so-called U(1) term:   
\begin{equation} \label{eq:U1term}
   Z_{\mathrm{Inst}}^{\mathrm{SU}(2)} (Q) \stackrel{?}{=}  \frac{Z_{\mathrm{Inst}}^{\mathrm{U}(2)} (Q)}{Z^{\mathrm{U}(1)} (Q)}  \, . 
\end{equation}
However there is no fundamental reason why such a U(1) term should exist in general\footnote{In the (cohomological) $4d$ instanton counting, a similar $U(1)$ factor can be derived using $qq$ character of \cite{Nekrasov:2015wsu}.}. 
In the following subsections, we can exhibit such a term only when $\tX_n$ is toric, which suggests no such rewriting exists for non-toric cases in general. This is related to the fact that the $U(2)$ instanton counting \cite{Nekrasov:2003rj} works mathematically for any number of fundamental hypermultiplets 
$N_f$, but the 4d interpretation suggests it may have bad analytical properties when $N_f>4$, as the $4d$ theory is not well defined in the UV. In 5d, the validity of \eqref{eq:U1term} is controlled by the sign of $N_f + 2 |k| - 4$ \cite[Sec. 3.4.4]{Hwang:2014uwa}. We expect trivial U(1) term when this quantity is $<0$ (i.e. the $E_1$, $\tilde{E_1}$, $E_2$ and $E_3$ cases), non-trivial term when it is $=0$ (i.e. the $E_4$ and $E_5$), and equation \eqref{eq:U1term} does not hold when it is $>0$ (i.e. $E_6$, $E_7$ and $E_8$).   

Actually, the Nekrasov instanton partition function $Z_{\mathrm{Inst}}^{\mathrm{SU}(2)} (Q)$ can in principle be computed using the Sp(1) ADHM formalism with gauge group $\mathrm{O}(k)$ \cite{Hayashi:2013qwa,Bergman:2013aca,Hwang:2014uwa}. It should be possible to analytically relate this computation to the topological vertex; in this paper, we don't attempt to do this and leave this question for future work.

\subsubsection{\texorpdfstring{$E_1$}{E1} and \texorpdfstring{$\tilde{E}_1$}{E1tilde}}

The $E_1$ SCFT is engineered on the local Hirzebruch surfaces $\mathbb{F}_0$ and $\mathbb{F}_2$, which are related by a Hanany-Witten move. The topological vertex computation and the removal of the parallel-brane factor for this pair were presented in detail in Sections \ref{sec:HWtopologicalVertex} and \ref{subsec: Mori cone F0 F2}, and the resulting GV invariants are collected in Tables \ref{tab:GV_F0} and \ref{tab:GV_F2}. Similarly, the $\tilde{E}_1$ SCFT arises from the local $\mathbb{F}_1$ surface and from the GTP geometry obtained from it by a Hanany-Witten move. This pair was analyzed in Sections \ref{sec:HWtopologicalVertex} and \ref{subsec: Mori cone F1 F3}, with the GV invariants collected in Tables \ref{tab:GV_F1} and \ref{tab:GV_GTPF3}.

Before moving to the next theories, we pause to rewrite the topological string partition function in the form \eqref{eq:U1term}. The same kind of rephrasing is then applied to all the rank-1 theories below. Start from \eqref{eq:FmZtop}. Changing $\lambda \rightarrow \lambda^T$ and using \eqref{eq:Nlambdalambda}, \eqref{eq:symmetryNfactor} and \eqref{eq:symSfactor} to write 
\begin{equation}
  \left( \frac{s_{\lambda^T} (q^{-\rho}) s_{\nu} (q^{-\rho})}{ \mathcal{N}_{\lambda  \nu} (Q_1 ; q) }\right)^2 = \frac{Q_1^{-(|\lambda| + |\nu|)}}{\mathcal{N}_{\lambda  \nu} (Q_1 ; q) \mathcal{N}_{\nu \lambda  } ( Q_1^{-1} ; q) \mathcal{N}_{\lambda  \lambda } (1 ; q)  \mathcal{N}_{\nu \nu} (1 ; q) } \, , 
\end{equation}
this can be rephrased into the form 
\begin{equation}
\label{eq:ZtopFm}
     Z \left[ \mathbb{F}_m \right] 
    = \operatorname{PExp} \left[  2 Q_1\frac{q}{(1-q)^2} \right] \sum\limits_{\lambda , \nu}   \frac{ Q_1^{m |\nu|} q^{\frac{m}{2} (\kappa_\nu + \kappa_\lambda)} ((-1)^m Q_2 Q_1^{-1})^{|\lambda| + |\nu|}  }{ \mathcal{N}_{\lambda  \nu} (Q_1 ; q) \mathcal{N}_{\nu \lambda  } ( Q_1^{-1} ; q) \mathcal{N}_{\lambda  \lambda } (1 ; q)  \mathcal{N}_{\nu \nu} (1 ; q)  } \, . 
\end{equation} 

We can now trade the K\"ahler parameters for field theory variables: write the SU(2) Cartan as $\mathbf{s} = (s , s^{-1})$ with $s = \sqrt{Q_1}$ and call $\fq = (-\sqrt{Q_1})^m \frac{Q_2}{Q_1}$ the instanton counting variable. The sign in $\fq$ corresponds in gauge theory to the discrete theta angle $\theta \in \{0 , \pi\}$. 
\begin{itemize}
    \item For the $E_1$ theory, we get 
    \begin{equation}
\label{eq:NekPureSU2}
 Z \left[ \mathbb{F}_0 \right] 
    = \operatorname{PExp} \left[  2 Q_1\frac{q}{(1-q)^2} \right]     \sum\limits_{\lambda_1 , \lambda_2}    \frac{ \fq^{|\lambda_1| + |\lambda_2|} }{\mathcal{N}_{\lambda_1 \lambda_2} (s^{-2} , q) \mathcal{N}_{\lambda_2 \lambda_1} (s^2 , q) \mathcal{N}_{\lambda_1 \lambda_1} (1 , q) \mathcal{N}_{\lambda_2 \lambda_2} (1 , q)}
\end{equation}
\item For $\widetilde{E}_1$, we have an additional contribution in the sum from an effective Chern-Simons interaction   
\begin{equation}
\label{eq:NekE1tilde}
\begin{split}
    Z \left[ \mathbb{F}_1 \right] 
    =& \operatorname{PExp} \left[  2 Q_1\frac{q}{(1-q)^2} \right] \sum\limits_{\lambda_1 , \lambda_2} z_{CS}^{(k=1)}(\mathbf{s} , q , \blambda) \times \\ &    \qquad  \qquad  \qquad  \qquad     \frac{  \fq^{|\lambda_1| + |\lambda_2|} }{\mathcal{N}_{\lambda_1 \lambda_2} (s^{-2} , q) \mathcal{N}_{\lambda_2 \lambda_1} (s^2 , q) \mathcal{N}_{\lambda_1 \lambda_1} (1 , q) \mathcal{N}_{\lambda_2 \lambda_2} (1 , q)} 
\end{split}
\end{equation}
where (see \eqref{eq:zCS})
\begin{equation}
    z_{CS}^{(k)}(\mathbf{s} , q , \blambda) = \left( s^{|\lambda_2| - |\lambda_1| } q^{\frac{1}{2} (\kappa_{\lambda_1} + \kappa_{\lambda_2})} \right)^k \, . 
\end{equation}
\end{itemize} 
The PExp factor in \eqref{eq:ZtopFm} is the perturbative contribution of the vector multiplet. So this perfectly agrees with \eqref{eq:ZtopZnek} and \eqref{eq:U1term} with trivial U(1) factor.

\subsubsection{\texorpdfstring{$E_2$}{E2}}

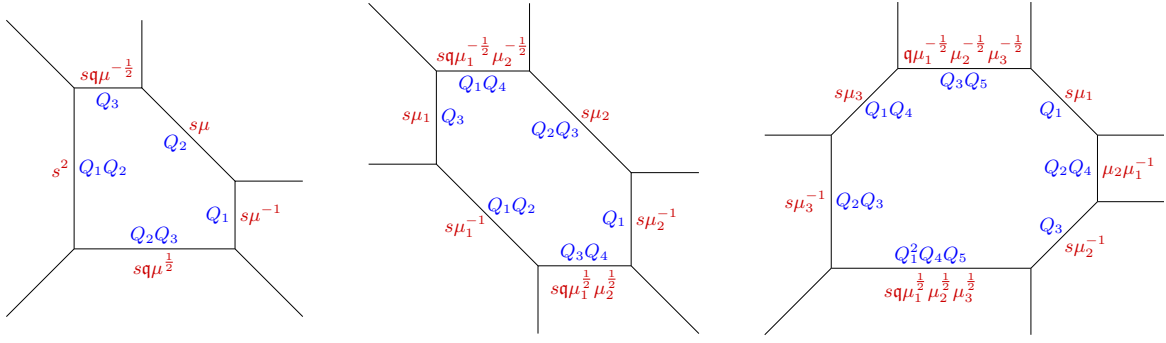
\begin{figure}[t]
\centering
    \hspace*{-1cm}\scalebox{.8}{\raisebox{-.5\height}{\begin{tikzpicture}[styleQ/.style={font=\footnotesize, text=blue},
    styles/.style={font=\footnotesize, text=red!80!black},every node/.style={inner sep=2pt}]
    \tikzmath{\x=-.3*1.4;\y=1.1*1.4;\z=.8*1.4;}
        \draw (-\z,-\y-\z) -- (0,-\y) -- (0,\x+\y) node[midway, right, styleQ] {$Q_1 Q_2$}  node[midway, left, styles] {$s^2$} -- (-\z,\x+\y+\z);
        \draw  (0,\x+\y) --  (\x + \y,\x+\y) node[midway, below, styleQ] {$Q_3$} node[midway, above, styles] {$s \fq \mu^{- \frac{1}{2}}$} -- (\x + \y,\x+\y + \z);
        \draw (\x + \y,\x+\y) -- (\x + 2*\y,\x) node[midway, below left,inner sep=0pt, styleQ] {$Q_2$}  node[midway, above right,inner sep=0pt, styles] {$s \mu$};
        \draw (\z + \x + 2*\y,\x) -- (\x + 2*\y,\x) -- (\x + 2*\y,-\y) node[midway, right, styles] {$s \mu^{-1}$} node[midway, left, styleQ] {$Q_1$};
        \draw (\x + 2*\y+\z, -\y-\z) -- (\x + 2*\y, -\y) -- (0,-\y) node[midway, below, styles] {$s \fq \mu^{\frac{1}{2}}$}  node[midway, above, styleQ] {$Q_2 Q_3$};
    \end{tikzpicture}} }  \hspace{.5cm}
    \scalebox{.8}{\raisebox{-.5\height}{\begin{tikzpicture}[styleQ/.style={font=\footnotesize, text=blue},
    styles/.style={font=\footnotesize, text=red!80!black},every node/.style={inner sep=2pt}]
    \tikzmath{\x=-.1*1.4;\y=1.2*1.4;\z=.8*1.4;}
        \draw (-\z,0) -- (0,0) -- (0,\x+\y) node[midway, right, styleQ] {$Q_3$}  node[midway, left, styles] {$s \mu_1$} -- (-\z,\x+\y+\z);
        \draw  (0,\x+\y) --  (\x + \y,\x+\y) node[midway, below, styleQ] {$Q_1 Q_4$}  node[midway, above, styles] {$s \fq \mu_1^{-\frac{1}{2}} \mu_2^{-\frac{1}{2}}$} -- (\x + \y,\x+\y + \z);
        \draw (\x + \y,\x+\y) -- (\x + 2*\y,\x) node[midway, above right,inner sep=0pt, styles] {$s \mu_2$} node[midway, below left,inner sep=0pt, styleQ] {$Q_2 Q_3$};
        \draw (\z + \x + 2*\y,\x) -- (\x + 2*\y,\x) -- (\x + 2*\y,-\y) node[midway, left, styleQ] {$Q_1$}  node[midway, right, styles] {$s \mu_2^{-1}$};
        \draw (\x + 2*\y+\z, -\y-\z) -- (\x + 2*\y, -\y) -- (\y, -\y) node[midway, above, styleQ] {$Q_3 Q_4$}  node[midway, below, styles] {$s \fq \mu_1^{\frac{1}{2}} \mu_2^{\frac{1}{2}}$}-- (\y, -\y-\z);
        \draw (\y, -\y) -- (0,0)  node[midway, above right,inner sep=0pt, styleQ] {$Q_1 Q_2$} node[midway, below left,inner sep=0pt, styles] {$s \mu_1^{-1}$};
    \end{tikzpicture}  }}    \hspace{.5cm} 
    \scalebox{.8}{\raisebox{-.5\height}{\begin{tikzpicture}[scale=1.1, styleQ/.style={font=\footnotesize, text=blue}, styles/.style={font=\footnotesize, text=red!80!black},every node/.style={inner sep=2pt}]
        \coordinate (V1) at (1,0);
        \coordinate (V6) at (1,1);
        \coordinate (V2) at (0,2);
        \coordinate (V7) at (-2,2);
        \coordinate (V4) at (-3,1);
        \coordinate (V3) at (-3,-1);
        \coordinate (V5) at (0,-1);
        \draw (V1) -- (V6) node[midway, left, styleQ] {$Q_2 Q_4$} node[midway, right, styles] {$\mu_2 \mu_1^{-1}$};
        \draw (V6) -- (V2) node[midway, below left,inner sep=0pt, styleQ] {$Q_1$} node[midway, above right,inner sep=0pt,inner sep=0pt, styles] {$s \mu_1$};
        \draw (V2) -- (V7) node[midway, below, styleQ] {$Q_3 Q_5$} node[midway, above, styles] {$\fq \mu_1^{- \frac{1}{2}} \mu_2^{- \frac{1}{2}} \mu_3^{- \frac{1}{2}}$};
        \draw (V7) -- (V4) node[midway, below right,inner sep=0pt, styleQ] {$Q_1 Q_4$} node[midway, above left,inner sep=0pt, styles] {$s \mu_3$};
        \draw (V4) -- (V3) node[midway, right, styleQ] {$Q_2 Q_3$} node[midway, left, styles] {$s \mu_3^{-1}$};
        \draw (V3) -- (V5) node[midway, above, styleQ] {$Q_1^2 Q_4 Q_5$} node[midway, below, styles] {$s \fq  \mu_1^{ \frac{1}{2}} \mu_2^{ \frac{1}{2}} \mu_3^{ \frac{1}{2}}$};
        \draw (V5) -- (V1) node[midway, above left,inner sep=0pt, styleQ] {$Q_3$} node[midway, below right,inner sep=0pt, styles] {$s \mu_2^{-1}$};
        \draw (V1) -- (2,0);
        \draw (V6) -- (2,1);
        \draw (V2) -- (0,3);
        \draw (V7) -- (-2,3);
        \draw (V4) -- (-4,1);
        \draw (V3) -- (-4,-2);
        \draw (V5) -- (0,-2);
    \end{tikzpicture}}}
    \caption{Brane web realizing the toric local del Pezzo $dP_2$, $dP_3$ and toric Generalized del Pezzo $GdP_4$ surfaces. Kähler parameters are shown in blue and gauge theory parameters in red.}
    \label{fig:branes_GdP4}
\end{figure}

The $E_2$ theory is engineered on the local $dP_2$, which is toric; its brane web is the first one shown in Figure \ref{fig:branes_GdP4}. Applying the topological vertex to this web, one finds
\begin{eqnarray}
    Z &=& \sum\limits_{\lambda , \mu} \left[ \frac{s_{\lambda^T} s_{\mu^T}}{\mathcal{R}_{\lambda^T \mu} (Q_1 Q_2)} \right] \left[ \frac{s_{\lambda} s_{\mu} \mathcal{R}_{\lambda^T \emptyset} (Q_1)  \mathcal{R}_{\mu \emptyset} (Q_2) }{\mathcal{R}_{\lambda^T \mu} (Q_1 Q_2)} \right] (Q_2 Q_3)^{|\lambda|} q^{\frac{1}{2} \kappa_{\lambda}} (- Q_3)^{|\mu|} \label{eq:E2nekN} \\ 
    &=& \operatorname{PExp} \left( \frac{(2 Q_1 Q_2 - Q_1 - Q_2)  q}{(1-q)^2} \right)\sum\limits_{\lambda , \mu}  \frac{\mathcal{N}_{\mu \emptyset} (Q_2^{-1}) \mathcal{N}_{\lambda \emptyset} (Q_1)}{\mathcal{N}_{\lambda \mu} (Q_1 Q_2) \mathcal{N}_{ \mu \lambda} (Q_1^{-1} Q_2^{-1}) \mathcal{N}_{ \lambda \lambda } (1) \mathcal{N}_{ \mu \mu } (1) } \left( \frac{Q_3}{Q_1} \right)^{|\lambda| + |\mu|} \, .  \nonumber
\end{eqnarray}

The corresponding gauge theory is $5d$ $SU(2)$ with $N_f = 1$, and the expression above is the associated Nekrasov partition function. This becomes manifest once the Nekrasov factors are traded for their symmetrized counterparts \eqref{eq:NFactor_sym}. The partition function reads, in the geometric variables,
\begin{align}\label{eq:E2nek_symQ}
    Z = \operatorname{PExp} \left( \frac{(2 Q_1 Q_2 - Q_1 - Q_2) q}{(1-q)^2} \right) \sum\limits_{\lambda , \mu} & \left( Q_1^{-1/2} Q_3 \right)^{|\lambda|} \left( Q_1^{-1} Q_2^{-1/2} Q_3 \right)^{|\mu|}  q^{\frac{\kappa_\lambda + \kappa_\mu}{4}} \\ &  \times \frac{\tilde{\mathcal{N}}_{\lambda \emptyset} (Q_1) \tilde{\mathcal{N}}_{\mu \emptyset} (Q_2^{-1})}{\tilde{\mathcal{N}}_{\lambda \mu} (Q_1 Q_2) \tilde{\mathcal{N}}_{\mu \lambda} (Q_1^{-1} Q_2^{-1}) \tilde{\mathcal{N}}_{\lambda \lambda} (1) \tilde{\mathcal{N}}_{\mu \mu} (1)} \, . \nonumber 
\end{align}

To exhibit the gauge theory, introduce the Coulomb parameter $s$, the hypermultiplet mass fugacity $\mu_h$ and the instanton fugacity $\fq$,
\begin{equation}\label{eq:E2dict}
    s = \sqrt{Q_1 Q_2} \, , \qquad \mu = \sqrt{Q_2/Q_1} \, , \qquad \fq = Q_1^{-3/4} Q_2^{-1/4} Q_3 \, .
\end{equation}
These relations are determined using the labelling in Figure \ref{fig:branes_GdP4}. 
The perturbative prefactor is exactly the $N_f = 1$ one-loop contribution \eqref{eq:ZpertSU2},
\begin{equation} 
   Z_{\rm{pert}} =  \operatorname{PExp} \left(   \frac{(2 Q_1 Q_2 - Q_1 - Q_2)q}{(1-q)^2} \right) =   \operatorname{PExp} \left(   \frac{(2 s^2 - s (\mu + \mu^{-1}))q}{(1-q)^2} \right)   \, .
\end{equation}
Finally the instanton monomial factorizes symmetrically,
\begin{equation}\label{eq:E2weight}
    \left( Q_1^{-1/2} Q_3 \right)^{|\lambda|} \left( Q_1^{-1} Q_2^{-1/2} Q_3 \right)^{|\mu|} = \fq^{|\lambda| + |\mu|} \, s^{\frac{1}{2}(|\lambda| - |\mu|)} \, ,
\end{equation}
so that \eqref{eq:E2nek_symQ} takes the gauge-theory form
\begin{equation}\label{eq:E2nek_symabs}
    Z = Z_{\rm{pert}} \sum\limits_{\lambda_1 , \lambda_2} \fq^{|\lambda_1| + |\lambda_2|}  z_{CS}^{(k= \frac{1}{2})}(\mathbf{s} , q , \blambda)\frac{\prod\limits_{i} \tilde{\mathcal{N}}_{\lambda_i \emptyset}( \frac{1}{\mu s_i} )  }{\prod\limits_{i,j} \tilde{\mathcal{N}}_{\lambda_i \lambda_j}( \frac{s_j}{s_i})} \, . 
\end{equation}
Recall that in the U(2) lift, gauge invariance requires $k + \frac{N_f}{2} \in \mathbb{Z}$ and here we choose $k = \frac{1}{2}$. In particular, the U(1)-factor in \eqref{eq:U1term} is again trivial here.

\subsubsection{\texorpdfstring{$E_3$}{E3}}

Now consider the second web of Fig \ref{fig:branes_GdP4}, realizing the $E_3$ theory. We find 
\begin{eqnarray}
    Z &=& \sum\limits_{\lambda , \mu} \left[ \frac{s_{\lambda} s_{\mu} \mathcal{R}_{\lambda^T \emptyset} (Q_1 Q_2)  \mathcal{R}_{\mu \emptyset} (Q_3) }{\mathcal{R}_{\lambda^T \mu} (Q_1 Q_2 Q_3)} \right] \left[ \frac{s_{\lambda^T} s_{\mu^T} \mathcal{R}_{\lambda^T \emptyset} (Q_1)  \mathcal{R}_{\mu \emptyset} (Q_2 Q_3) }{\mathcal{R}_{\lambda^T \mu} (Q_1 Q_2 Q_3)} \right] (-Q_3 Q_4)^{|\lambda|} (-Q_1 Q_4)^{|\mu|} \nonumber \\ 
    &=& \operatorname{PExp} \left[ (2 Q_1 Q_2 Q_3 - Q_1 - Q_3 - Q_1 Q_2 - Q_2 Q_3)\frac{  q}{(1-q)^2} \right] \times \\ & & \qquad  \qquad   \sum\limits_{\lambda , \mu}  \frac{\mathcal{N}_{\mu \emptyset} (Q_3^{-1}) \mathcal{N}_{\lambda \emptyset} (Q_1 Q_2) \mathcal{N}_{\mu^T \emptyset} (Q_2 Q_3) \mathcal{N}_{\lambda^T \emptyset} (Q_1^{-1})}{\mathcal{N}_{\lambda \mu} (Q_1 Q_2 Q_3) \mathcal{N}_{ \mu \lambda} (Q_1^{-1} Q_2^{-1} Q_3^{-1}) \mathcal{N}_{ \lambda \lambda } (1) \mathcal{N}_{ \mu \mu } (1) } \left( \frac{Q_4}{Q_2} \right)^{|\lambda| + |\mu|} \, .
\end{eqnarray}

The gauge theory phase is $SU(2)$ with $N_f = 2$. The four numerator factors are the hypermultiplet contributions. Two of them carry a transpose: $\mathcal{N}_{\lambda \emptyset}(Q_1 Q_2)$ and $\mathcal{N}_{\mu \emptyset}(Q_3^{-1})$ are of \emph{fundamental} type, whereas $\mathcal{N}_{\lambda^T \emptyset}(Q_1^{-1})$ and $\mathcal{N}_{\mu^T \emptyset}(Q_2 Q_3)$ are of \emph{antifundamental} type.
For $SU(2)$ the fundamental and antifundamental representations are equivalent. This becomes clear using the symmetrized Nekrasov factors \eqref{eq:symmetryNfactor}: 
\begin{equation}\label{eq:E3nek_sym}
    Z = Z_{\rm{pert}} \sum\limits_{\lambda , \mu} \fq^{|\lambda| + |\mu|} \, \frac{ \tilde{\mathcal{N}}_{\lambda \emptyset}(Q_1 Q_2) \, \tilde{\mathcal{N}}_{\lambda \emptyset}(Q_1) \, \tilde{\mathcal{N}}_{\mu \emptyset}(Q_3^{-1}) \, \tilde{\mathcal{N}}_{\mu \emptyset}(Q_2^{-1} Q_3^{-1}) }{ \tilde{\mathcal{N}}_{\lambda \mu}(s^2) \, \tilde{\mathcal{N}}_{\mu \lambda}(s^{-2}) \, \tilde{\mathcal{N}}_{\lambda \lambda}(1) \, \tilde{\mathcal{N}}_{\mu \mu}(1) } \, , \qquad \fq = - \frac{Q_4}{\sqrt{Q_2}} \, .
\end{equation}
Note that the minus sign in $\fq$ here can not be attributed to the theta angle; it can be seen as a physically irrelevant choice of branch-cut when identifying the K\"ahler parameters with the gauge theory quantities, due to the presence of square roots:
\begin{equation}\label{eq:E3dict}
    s = \sqrt{Q_1 Q_2 Q_3} \, , \qquad \mu_1 = \sqrt{\frac{Q_3}{Q_1 Q_2}  } \, , \qquad \mu_2 = \sqrt{\frac{Q_2 Q_3}{Q_1}} \, ,
\end{equation}
\eqref{eq:E3nek_sym} is the $SU(2)$, $N_f = 2$ Nekrasov partition function. The perturbative prefactor is exactly \eqref{eq:ZpertSU2} with $N_f = 2$. We see again that the U(1)-factor in \eqref{eq:U1term} is again trivial here, in agreement with the observations in \cite{Bergman:2013aca}. 
 
\subsubsection{\texorpdfstring{$E_4$}{E4}}

The $E_4$ theory is $5d$ $SU(2)$ with $N_f = 3$ fundamental hypermultiplets, engineered by the local del Pezzo $dP_4$. Since $dP_4$ is not toric, we use a generalized toric presentation with five K\"ahler parameters $Q_1 , \dots , Q_5$, with the assignment of Figure~\ref{fig:branes_GdP4}. Computing the topological vertex and organizing the result as an instanton sum, one finds
\begin{equation}\label{eq:E4nek_symabs}
\begin{split}
        Z = & \operatorname{PExp} \left[ \frac{\mu_2}{\mu_1} \frac{q}{(1-q)^2} \right] \operatorname{PExp} \left[ \Big( 2 s^2 - \sum_{f=1}^3 (\mu_f + \mu_f^{-1}) s \Big) \frac{q}{(1-q)^2} \right]  \\ & \times \sum\limits_{\lambda_1 , \lambda_2} \fq^{|\lambda_1| + |\lambda_2|} z_{CS}^{(k= - \frac{1}{2})}(\mathbf{s} , q , \blambda) \frac{\prod\limits_{i=1,2} \prod\limits_{f=1,2,3} \tilde{\mathcal{N}}_{\lambda_i \emptyset}( \frac{1}{\mu_f s_i} )  }{\prod\limits_{i,j=1,2} \tilde{\mathcal{N}}_{\lambda_i \lambda_j}( \frac{s_j}{s_i})} \, . 
\end{split}
\end{equation}
with the translation read directly from the brane web 
\begin{equation}
\begin{split}
     &   s = \sqrt{Q_1 Q_2 Q_3 Q_4} \, , \quad  \fq = Q_1^{\frac{3}{4}} Q_3^{\frac{1}{4}} Q_5 \, , \\ &   \mu_1 = \sqrt{\frac{Q_1}{Q_2 Q_3 Q_4}} \, , \quad \mu_2 = \sqrt{\frac{Q_1 Q_2 Q_4}{Q_3}} \, , \quad \mu_3 = \sqrt{\frac{Q_1 Q_4}{Q_2 Q_3}} \, . 
\end{split}
\end{equation} 
Note that we have split the PExp factor in front of \eqref{eq:E4nek_symabs} into two parts. One is the universal perturbative contribution \eqref{eq:ZpertSU2}. The other one can be attributed to the pair of \textit{horizontal} external parallel branes in Figure \ref{fig:branes_GdP4}. We expect, however, to also see a contribution from the \textit{vertical} external parallel branes. It is indeed hidden in the sum over partitions in \eqref{eq:E4nek_symabs}, which is really a U(2) Nekrasov instanton function. The actual SU(2) Nekrasov instanton function should be obtained after removing a so-called U(1) term, which in this case has been studied in \cite[Eq. (9)]{Bergman:2013aca} and reads 
$Z^{\mathrm{U}(1)} = \operatorname{PExp}\!\left[  \frac{\fq}{\sqrt{\mu_1 \mu_2 \mu_3}} \;  \frac{q }{(1-q)^2} \right]$
so that 
\begin{equation}\label{eq:E4parallel}
 \frac{Z_{\parallel}}{Z^{\mathrm{U}(1)}} = \frac{\operatorname{PExp}\!\left[ \left( \frac{\mu_2}{\mu_1} + \frac{\fq}{\sqrt{\mu_1 \mu_2 \mu_3}} \right) \frac{q }{(1-q)^2} \right]}{\operatorname{PExp}\!\left[  \frac{\fq}{\sqrt{\mu_1 \mu_2 \mu_3}} \;  \frac{q }{(1-q)^2} \right]  }   = \operatorname{PExp}\!\left[  \frac{\mu_2}{\mu_1}   \frac{q }{(1-q)^2} \right]
\end{equation} 
is exactly the prefactor in \eqref{eq:E4nek_symabs} so that \eqref{eq:U1term} holds.

\paragraph{Computation of $dP_4$ GV invariants. }
The expression \eqref{eq:E4nek_symabs} gives the GV invariants for $GdP_4$ for any degree. From it, we can extract invariants for $dP_4$ using the argument above, but the matching of curve classes is not obvious. Hence in the following we focus on anticanonical degree. We proceed in two steps. First, we specialize the parameters to a 2-dimensional family as shown at the top of Table \ref{tab:GVdP4}. Then, in order to speed up the computation, we use a flopped geometry,\footnote{This allows to speed up the computation as the GV invariants appear at lower degrees in $Q_1$. } for which the partition function reads 
\begin{equation}
    Z = \operatorname{PExp} \left[ \left(  Q_1 Q_2^2 - 3 Q_2 (1+Q_1) + 2Q_1 \right) \frac{q}{(1-q)^2} \right] \sum\limits_{\lambda , \mu} \frac{\mathcal{N}_{\mu,0}^2 (Q_2) \mathcal{N}_{\lambda,0}^2 (Q_1 Q_2) \mathcal{N}_{\mu^T,0} (Q_1 Q_2) \mathcal{N}_{\lambda^T,0} (Q_2)}{\mathcal{N}_{\lambda,\mu} (Q_1)\mathcal{N}_{\mu,\lambda} (Q_1^{-1})\mathcal{N}_{\lambda,\lambda} (1) \mathcal{N}_{\mu,\mu} (1)}
\end{equation}
From there, using an obvious generalization of equation \eqref{eq:flopdP1}, we obtain the results of Table \ref{tab:GVdP4}. 
To our knowledge these invariants have not been tabulated before.

\begin{table}[t]
    \centering 
 \scalebox{.8}{\raisebox{-.5\height}{\begin{tikzpicture}[every node/.style={inner sep=2pt}]
    \tikzmath{\x=.7;\y=.6;\z=.8;}
        \draw (-\z,0) -- (0,0) -- (0,\x+\y) node[midway, left] {$Q_1Q_2$} -- (-\z,\x+\y+\z);
        \draw  (0,\x+\y) --  (\x + \y,\x+\y) node[midway, above] {$Q_1Q_2$} -- (\x + \y,\x+\y + \z);
        \draw (\x + \y,\x+\y) -- (\x + 2*\y,\x) node[midway, above right,inner sep=0pt] {$Q_1$};
        \draw (\z + \x + 2*\y,\x) -- (\x + 2*\y,\x) -- (\x + 2*\y,0) node[midway, right] {$Q_2$} -- (\z + \x + 2*\y,0);
        \draw (\x + 2*\y,0) -- (\x + \y, -\y) node[midway, below right,inner sep=0pt] {$Q_1$};
        \draw (\x + \y, -\y-\z) -- (\x + \y, -\y) -- (\y, -\y) node[midway, below] {$Q_2$}-- (\y, -\y-\z);
        \draw (\y, -\y) -- (0,0) node[midway, below left] {$Q_1$};
    \end{tikzpicture}}}
    
    \vspace{.5cm}
    
 \begin{tabular}{c|ccccccccc}
$d_1 \backslash d_2 $&  0 & 1  & 2 & 3 & 4 & 5 & 6 & 7 & 8 \\ \hline 
0 &   &  {\color{red}$\mathbf{-2}$} &   &   &  &   &  &  & \\
1 &   3  &  6  &  1  &  &  &  &  &  & \\
2 &   &  $-4$  &  $-6$  &  &  &  &  &  & \\
3 &   &  &  $9$  & $ 6 $ &  &  &  &  & \\
4 &  &  &  $-4$  &  $-24$  &$  -12$  &  &  &  & \\
5 &   &  &  &  $30$  &  $75, -6$  & $ 30$  &  &  & \\
6 &  &  &  & $-12$ & $-162,21$ & $-252,42$ & $-84,7$ &  & \\
7 &  &  &  &  & $147,-24$ & $798, -244,18$ & $903,-294,27$ & $252,-48$ & 
    \end{tabular} 

    \vspace{.5cm}
    
    \hspace*{-2cm}\begin{tabular}{c|cccccccccc}
     $d \backslash g$  &  $0$ & $1$ & $2$ & $3$ & $4$ & $5$ & $6$ & $7$ & $8$ & $9$\\ \hline
     0 & & & & & & & & & & \\
     1 & $10$ & & & & & & & & & \\
     2 & $-10$ & & & & & & & & & \\
     3 & $15$ & & & & & & & & & \\
     4 & $-40$ & & & & & & & & & \\
     5 & $135$ & $-6$ & & & & & & & & \\
     6 & $-510$ & $70$ & & & & & & & & \\
     7 & $2100$ & $-610$ & $45$ & & & & & & & \\
     8 & $-9280$ & $4695$ & $-870$ & $55$ & & & & & & \\
     9 & $43245$ & $-34160$ & $11615$ & $-1920$ & $130$ & & & & & \\
     10 & $-209870$ & $241114$ & $-129920$ & $39250$ & $-6670$ & $565$ & $-16$ & & & \\
     11 & $1052700$ & $-1671400$ & $1311405$ & $-622200$ & $187035$ & $-35070$ & $3775$ & $-180$ & & \\
     12 & $-5427360$ & $11455975$ & $-12371920$ & $8464045$ & $-3910910$ & $1240720$ & $-266600$ & $37005$ & $-2980$ & $105$
\end{tabular}
    \caption{\textit{Top:} Brane web used to compute the invariants. \textit{Middle:} GV invariants for the brane web above. The red $-2$ coefficient corresponds to the two pairs of parallel branes with K\"ahler parameter $Q_2$. We only show the degrees up to 8, but we can compute many more. \textit{Bottom:} Summing the black numbers over the lines, one gets the GV invariants for local $dP_4$ in anticanonical degree and arbitrary genus.   }
    \label{tab:GVdP4}
\end{table}

\subsubsection{\texorpdfstring{$E_5$}{E5}}

The computation for the $E_5$ theory is very similar to the one performed for $E_4$ in the previous subsection, so we keep things brief. 
The toric variety that we use to realize the $E_5$ theory contains $4$ $(-1,-1)$ curves of degree $1$ and $4$ $(0,-2)$ curves of degree $0$. We need to introduce a second Kähler parameter $Q_2$ for the $(0,-2)$ curves, which discriminate between curves of the same degree and will be set to $1$ to produce the GV invariants. This brane web with its assignment of Kähler parameters is shown at the top of Table \ref{tab:GVdP5}.
After decoupling the contributions from parallel semi-infinite 5-branes, we recover the invariants computed in \cite{Katz:1999xq} in the limit $Q_2\to 1$ : 
\begin{equation}
 \frac{Z_{GdP_5}(Q_1,Q_2)}{\operatorname{PExp}\left(4 \frac{q Q_2}{(1-q)^2}\right)}  \overset{Q_2\to 1}{\longrightarrow} Z_{dP_5}(Q_1)
\end{equation}
This gives the relation between the middle and bottom parts of Table \ref{tab:GVdP5}.

\begin{table}[t]
\centering  
\begin{tabular}{c|ccccccc}
 $d_1 \backslash d_2$ & 0& 1 & 2 & 3 & 4 & 5 & 6 \\
\hline
0 &  & {\color{red}$\mathbf{-4}$} &  &  &  &  &\\
1 & $4$ & $8$ & $4$ &  &  & & \\
2 &  & $-4$ & $-12$ & $-4$ &  & & \\
3 &  &  & $12$ & $24$ & $12$ & &\\
4 &  &  & $-4$ & $-48$ & $-88,\,5$ & $-48$ & $-4$ \\
5 &  &  &  & $40$ & $240,-24$ & $400,\,-48$& $240,-24$ \\
6 &  &  &  & $-12$ & $-312, 42$ & $-1344, \, 312, \, -16 $& $-2100, \, 572, \, -48$ \\
\end{tabular}\vspace{.5cm}
  \scalebox{.8}{\raisebox{-.5\height}{   \begin{tikzpicture}[every node/.style={inner sep=2pt}]
    \tikzmath{\x=1;\y=1.2;\z=1;}
        \draw (-\z,0) -- (0,0) -- (0,\x) node[midway,left] {$Q_2$} -- (-\z,\x);
        \draw (0,\x) -- (\y,\x+\y) node[midway, above left ,inner sep=0pt] {$Q_1$};
        \draw (\y,\x+\y+\z) --  (\y,\x+\y) --  (\x + \y,\x+\y) node[midway,above] {$Q_2$} -- (\x + \y,\x+\y + \z);
        \draw (\x + \y,\x+\y) -- (\x + 2*\y,\x) node[midway, above right,inner sep=0pt] {$Q_1$};
        \draw (\z + \x + 2*\y,\x) -- (\x + 2*\y,\x) -- (\x + 2*\y,0) node[midway, right] {$Q_2$} -- (\z + \x + 2*\y,0);
        \draw (\x + 2*\y,0) -- (\x + \y, -\y) node[midway, below right,inner sep=0pt] {$Q_1$};
        \draw (\x + \y, -\y-\z) -- (\x + \y, -\y) -- (\y, -\y) node[midway, below] {$Q_2$}-- (\y, -\y-\z);
        \draw (\y, -\y) -- (0,0) node[midway, below left,inner sep=0pt] {$Q_1$};
    \end{tikzpicture}}}  \hspace{2cm} \begin{tabular}{c|ccc}
       $d$  &  $g= 0$ & 1 & 2 \\ \hline
       0 & & & \\
       1  & $16$ & & \\
       2 & $-20$ & & \\
       3 &  $48$ & & \\
       4 & $-192$ & $5$ & \\
       5 & $960$ & $-96$ & \\
       6 & $-5436$ & $1280$ & $-80$ 
    \end{tabular}
\caption{\textit{Top:} Brane web used to compute the invariants. \textit{Middle:} GV invariants for the brane web above. The red $-4$ coefficient corresponds to the four pairs of parallel branes with K\"ahler parameter $Q_2$. We only show the degrees up to 6, but we can compute many more. \textit{Bottom:} Summing over the lines, one gets the GV invariants for local $dP_5$ in anticanonical degree and arbitrary genus. This is partly contained in \cite[p. 50]{chiang1999localmirrorsymmetrycalculations}. }
\label{tab:GVdP5}
\end{table}

\subsubsection{\texorpdfstring{$E_6$}{E6}}

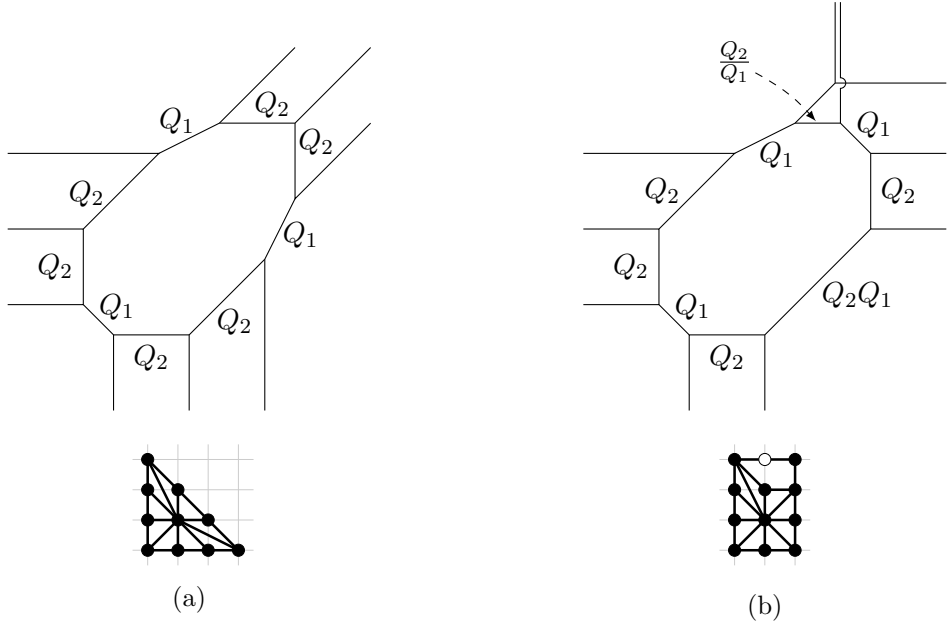
\begin{figure}[t]
    \centering
    \begin{subfigure}[t]{0.495\textwidth}
        \centering
        \begin{tikzpicture}
        \tikzmath{\x=1;\y=.4;\z=1;}
        \draw (-\z,0) -- (0,0) -- (0,\x) node[midway, left] {$Q_2$}-- (-\z ,\x);
        \draw (0,\x) -- (\x, 2*\x) node[midway,xshift=-13pt,yshift=-1pt] {$Q_2$} -- (-\z,2*\x);
        \draw (\x,2*\x) -- (\x +2*\y,2*\x +\y)node[pos=.3, above] {$Q_1$} -- (\x +2*\y+\z,2*\x +\y+\z);
        \draw (\x +2*\y,2*\x +\y) -- (2*\x +2*\y,2*\x +\y)node[pos=.7,yshift=7pt] {$Q_2$} -- (2*\x +2*\y+\z,2*\x +\y+\z);
        \draw (2*\x +2*\y,2*\x +\y) -- (2*\x +2*\y,\x+\y)node[pos=.25,xshift=8pt] {$Q_2$}-- (2*\x +2*\y+\z,\x +\y+\z);
        \draw (2*\x +2*\y,\x+\y) -- (2*\x +\y,\x-\y)node[pos=.6, xshift=9pt] {$Q_1$} --  (2*\x +\y,-\y- \z);
        \draw (2*\x +\y,\x-\y) -- (\x +\y,-\y) node[pos=.4, yshift=-12pt] {$Q_2$} -- (\x +\y,-\y-\z);
        \draw  (\x +\y,-\y) --  (\y,-\y)node[midway, below] {$Q_2$} -- (\y,-\y-\z);
        \draw (\y,-\y) -- (0,0)node[midway,xshift=7pt,yshift=5pt] {$Q_1$};
        \end{tikzpicture}
        
\vspace{.4cm}

        \scalebox{.8}{  
\begin{tikzpicture}[x=.5cm,y=.5cm] 
\draw[gridline] (0,-0.5)--(0,3.5); 
\draw[gridline] (1,-0.5)--(1,3.5); 
\draw[gridline] (2,-0.5)--(2,3.5); 
\draw[gridline] (3,-0.5)--(3,3.5); 
\draw[gridline] (-0.5,0)--(3.5,0); 
\draw[gridline] (-0.5,1)--(3.5,1); 
\draw[gridline] (-0.5,2)--(3.5,2); 
\draw[gridline] (-0.5,3)--(3.5,3); 
\draw[ligne] (0,1)--(0,0); 
\draw[ligne] (0,0)--(1,0); 
\draw[ligne] (1,0)--(2,0); 
\draw[ligne] (2,0)--(3,0); 
\draw[ligne] (3,0)--(2,1); 
\draw[ligne] (2,1)--(1,2); 
\draw[ligne] (1,2)--(0,3); 
\draw[ligne] (0,3)--(0,2); 
\draw[ligne] (0,2)--(0,1); 
\node[bd] at (0,0) {}; 
\node[bd] at (1,0) {}; 
\node[bd] at (2,0) {}; 
\node[bd] at (3,0) {}; 
\node[bd] at (2,1) {}; 
\node[bd] at (1,2) {}; 
\node[bd] at (0,3) {}; 
\node[bd] at (0,2) {}; 
\node[bd] at (0,1) {}; 
\node[bd] at (1,1) {};
\foreach \n in {1,...,3} {\draw[ligne] (1,1) -- (\n,0);\draw[ligne] (1,1) -- (0,3-\n);\draw[ligne] (1,1) -- (3-\n,\n);};
\end{tikzpicture}}
        \caption{}\label{fig:branesGdP6Toric}
    \end{subfigure}
    \begin{subfigure}[t]{0.495\textwidth}
        \centering
        \begin{tikzpicture}
        \tikzmath{\x=1;\y=.4;\z=1;\ee=.07;}
        \draw (-\z,0) -- (0,0) -- (0,\x) node[midway, left] {$Q_2$}-- (-\z ,\x);
        \draw (0,\x) -- (\x, 2*\x) node[pos=.5, xshift=-13pt] {$Q_2$} -- (-\z,2*\x);
        \draw (\x,2*\x) -- (\x +2*\y,2*\x +\y) node[pos=.7, below] {$Q_1$} -- (2*\x +\y-\ee,3*\x -\ee) -- (2*\x+\y-\ee,3*\x +\z);
        \draw (\x +2*\y,2*\x +\y)-- node[pos=.5,inner sep=0pt](p1) {} (2*\x +\y,2*\x +\y)--(2*\x +\y,3*\x - 2*\ee) arc (-90:90:\ee) -- (2*\x +\y,3*\x +\z) ;
        \draw(2*\x +\y-\ee,3*\x -\ee) -- (2*\x +2*\y+\z,3*\x -\ee);
        \draw  (2*\x +\y,2*\x +\y) -- (2*\x +2*\y,2*\x ) node[midway, xshift=7pt,yshift=5pt] {$Q_1$} --(2*\x +2*\y+\z,2*\x);
        \draw (2*\x +2*\y,2*\x ) -- (2*\x +2*\y,\x ) node[midway, right] {$Q_2$} -- (2*\x +2*\y+\z,\x );
        \draw (2*\x +2*\y,\x) -- (\x +\y,-\y) node[pos=.5,xshift=15pt,yshift=-5pt] {$Q_2Q_1$} -- (\x +\y,-\y-\z);
        \draw  (\x +\y,-\y) --  (\y,-\y)node[midway, below] {$Q_2$} -- (\y,-\y-\z);
        \draw (\y,-\y) -- (0,0)node[midway,xshift=7pt,yshift=5pt] {$Q_1$};
        \node[inner sep=1pt]  at (\x ,2*\x +3*\y) (l1) {$\frac{Q_2}{Q_1}$};
        \draw[-latex,dashed,bend left =10] (l1) to (p1);
        \end{tikzpicture}

\vspace{.4cm}

\rotatebox{90}{\scalebox{.8}{  
\begin{tikzpicture}[x=.5cm,y=.5cm] 
\draw[gridline] (0,-0.5)--(0,2.5); 
\draw[gridline] (1,-0.5)--(1,2.5); 
\draw[gridline] (2,-0.5)--(2,2.5); 
\draw[gridline] (3,-0.5)--(3,2.5); 
\draw[gridline] (-0.5,0)--(3.5,0); 
\draw[gridline] (-0.5,1)--(3.5,1); 
\draw[gridline] (-0.5,2)--(3.5,2); 
\draw[ligne] (0,1)--(0,0); 
\draw[ligne] (0,0)--(1,0); 
\draw[ligne] (1,0)--(2,0); 
\draw[ligne] (2,0)--(3,0); 
\draw[ligne] (3,0)--(3,2); 
\draw[ligne] (3,2)--(2,2); 
\draw[ligne] (2,2)--(1,2); 
\draw[ligne] (1,2)--(0,2); 
\draw[ligne] (0,2)--(0,1); 
\node[bd] at (0,0) {}; 
\node[bd] at (1,0) {}; 
\node[bd] at (2,0) {}; 
\node[bd] at (3,0) {}; 
\node[bd] at (3,2) {}; 
\node[bd] at (2,2) {}; 
\node[bd] at (1,2) {}; 
\node[bd] at (0,2) {}; 
\node[bd] at (0,1) {}; 
\node[bd] at (1,1) {}; 
\node[bd] at (2,1) {};
\node[wd] at (3,1) {};
\foreach \y in {0,...,3} \draw[ligne] (1,1) -- (\y,2);
\foreach \y in {0,...,2} \draw[ligne] (1,1) -- (\y,0);
\foreach \y in {0,2} \draw[ligne] (1,1) -- (\y,1);
\draw[ligne] (2,0) -- (2,1) -- (3,2);
\end{tikzpicture}}}

        \caption{}\label{fig:branes_GdP6_nonToric}
    \end{subfigure}
    \caption{Brane webs realizing the toric (a) and generalized-toric (b) Generalized del Pezzo $GdP_6$.}
    \label{fig:branes_GdP6}
\end{figure}

This is the last case where there is a \textit{toric} Generalized Del Pezzo surface (see Figure \ref{fig:branesGdP6Toric}). However, this toric $GdP_6$ is not like the one studied above. In particular, it cannot be split into two strips. By a Hanany-Witten transition, we also obtain \textit{non-toric} $GdP_6$ surfaces (one example is shown in Figure \ref{fig:branes_GdP6_nonToric}), which can be split as two strips. As before, a second parameter $Q_2$ should be introduced and set to $1$ at the end. This yields the choice of Kähler parameters shown in Figure \ref{fig:branes_GdP6}, and we recover the Gopakumar-Vafa invariants of the local $dP_6$ by factoring out the contribution from parallel branes and taking the limit $Q_2 \to 1$ (see Table \ref{tab:GV_E6}).

The assignment of Kähler parameters in the non-toric case can be obtained by keeping track of the values in the toric case through the Hanany-Witten transition or directly by computing the degree with respect to the anti-canonical divisor, and adding suitable powers of $Q_2$.
From the non-toric brane web, we can use the third row of Table \ref{tab:TV} to resolve the non-toric vertex, then split the web in two strips, along the two inner horizontal edges. Using the formula for strips, we get:\footnote{The expression in the PExp is the sum of the contributions of the horizontal branes from the strips. The left strip contributes $3+3+9=15$ terms, namely $(2Q_2+Q_2^2)+(Q_1^2 Q_2^2 + \frac{Q_2}{Q_1} + Q_1 Q_2^3) - (Q_2 + 2 Q_1) (1+Q_2+Q_2^2)$, and the right one $1+1+4=6$ terms, namely $Q_1^2 Q_2^2 + Q_2 - Q_1 (1+Q_2)^2$.   }
\begin{equation}\label{eq:E6_NekFactors}
\begin{split}
Z_{\textrm{top}} =&  \operatorname{PExp}\left[\frac{q\left(2 Q_1^2 Q_2^2-Q_2^3-3 Q_1 Q_2^2-4 Q_1 Q_2-3 Q_1+Q_1 Q_2^3+2 Q_2+\frac{Q_2}{Q_1}\right) }{(1-q)^2}\right] \times\\ & 
\sum_{\lambda,\mu}(-Q_2)^{|\mu|}\left(\frac{Q_2}{Q_1}\right)^{|\lambda|} q^{-\kappa_\lambda + \frac{1}{2} \kappa_{\mu}}\frac{\mathcal N_{\lambda,\emptyset} (Q_1)^2 \mathcal N_{\lambda,\emptyset} (Q_1 Q_2)^2\mathcal N_{\lambda,\emptyset} (Q_1 Q_2^2) \mathcal N_{\emptyset,\mu}(Q_1)\mathcal N_{\emptyset,\mu}(Q_1 Q_2)^2\mathcal N_{\emptyset,\mu} (Q_1 Q_2^2)^2
}{\mathcal N_{\lambda,\lambda}(1) \mathcal N_{\lambda,\mu}(Q_1^2Q_2^2)^2\mathcal N_{\mu,\mu}(1)\mathcal N_{\emptyset,\lambda} (\frac{Q_2}{Q_1})\mathcal N_{\emptyset,\mu}(Q_1 Q_2^3)}
\end{split}
\end{equation}
 
\paragraph{Generic Masses Version. } In order to understand better the origin of the Nekrasov factors in the above equation \eqref{eq:E6_NekFactors}, we can perform the fully refined computation using gauge theory parameters as follows: 
\begin{equation}
    \raisebox{-.5\height}{\begin{tikzpicture}
        \tikzmath{\x=1;\y=.4;\z=1;\ee=.07;}
        \draw (-\z,0) -- (0,0) -- (0,\x) node[midway, left] {$\frac{\mu_5}{\mu_4}$}-- (-\z ,\x);
        \draw (0,\x) -- (\x, 2*\x) node[pos=.5, xshift=-13pt] {$\frac{\mu_4}{\mu_3}$} -- (-\z,2*\x);
        \draw (\x,2*\x) -- (\x +2*\y,2*\x +\y) node[pos=.7, below] {$s \mu_3$} -- (2*\x +\y-\ee,3*\x -\ee) -- (2*\x+\y-\ee,3*\x +\z);
        \draw (\x +2*\y,2*\x +\y)-- node[pos=.5,inner sep=0pt](p1) {} (2*\x +\y,2*\x +\y)--(2*\x +\y,3*\x - 2*\ee) arc (-90:90:\ee) -- (2*\x +\y,3*\x +\z) ;
        \draw(2*\x +\y-\ee,3*\x -\ee) -- (2*\x +2*\y+\z,3*\x -\ee);
        \draw  (2*\x +\y,2*\x +\y) -- (2*\x +2*\y,2*\x ) node[midway, xshift=7pt,yshift=5pt] {$s \mu_1$} --(2*\x +2*\y+\z,2*\x);
        \draw (2*\x +2*\y,2*\x ) -- (2*\x +2*\y,\x ) node[midway, right] {$\frac{\mu_2}{\mu_1}$} -- (2*\x +2*\y+\z,\x );
        \draw (2*\x +2*\y,\x) -- (\x +\y,-\y) node[pos=.5,xshift=15pt,yshift=-5pt] {$\frac{s}{\mu_2}$} -- (\x +\y,-\y-\z);
        \draw  (\x +\y,-\y) --  (\y,-\y)node[midway, below] {$\fq \sqrt{\Pi}$} -- (\y,-\y-\z);
        \draw (\y,-\y) -- (0,0)node[midway,xshift=7pt,yshift=5pt] {$\frac{s}{\mu_5}$};
        \node[inner sep=1pt]  at (\x ,2*\x +3*\y) (l1) {$\frac{\fq}{s \sqrt{\Pi}}$};
        \draw[-latex,dashed,bend left =10] (l1) to (p1);
        \end{tikzpicture}}
\end{equation}
Here, $\mu_1 \dots , \mu_5$ correspond to the masses of the 5 hypermultiplets, $s$ and $\fq$ are as before, and we have introduced $\Pi = \mu_1 \cdots \mu_5$ the product of the five masses. Using the strip formula, one finds 
\begin{equation}
\begin{split}
       Z_{\mathrm{top}} =& \sum\limits_{\lambda_1 , \lambda_2} \left[ \frac{ \mathcal{R}_{0 0}( \fq \frac{\mu_3}{\sqrt{\Pi}} ) \mathcal{R}_{0 0}( \fq \frac{\mu_4}{\sqrt{\Pi}} )\mathcal{R}_{0 0}( \fq \frac{\mu_5}{\sqrt{\Pi}} )  \prod_f \mathcal{R}_{\lambda_1 0}( s \mu_f )  \mathcal{R}_{\lambda_2^T 0}( \frac{s}{\mu_f})}{\mathcal{R}_{\lambda_2^T \lambda_1}^2(s^2) \mathcal{R}_{00}( \frac{\mu_2}{\mu_1} ) \mathcal{R}_{00}( \frac{\mu_4}{\mu_3} )\mathcal{R}_{00}( \frac{\mu_5}{\mu_4} ) \mathcal{R}_{00}( \frac{\mu_5}{\mu_3} )\mathcal{R}_{\lambda_1^T 0}( \frac{ \fq}{s \sqrt{\Pi}} )  \mathcal{R}_{\lambda_2^T 0}( \frac{s \fq}{\sqrt{\Pi}} ) } \right]  \\ & s_{\lambda_1} s_{\lambda_1^T} s_{\lambda_2} s_{\lambda_2^T} q^{\kappa_{\lambda_1} - \frac{1}{2} \kappa_{\lambda_2}} \left(- \frac{ \fq}{s \sqrt{\Pi}}  \right)^{|\lambda_1|} \left(  \fq  \sqrt{\Pi}   \right)^{|\lambda_2|} \\ 
=&
\mathrm{PExp}\left[
\frac{
\left(
2s^2
+\frac{\mu_2}{\mu_1}
+\frac{\mu_4}{\mu_3}
+\frac{\mu_5}{\mu_4}
+\frac{\mu_5}{\mu_3}
+\frac{\fq}{\sqrt{\Pi}}\left(s+s^{-1}\right)
-\fq\frac{\mu_3}{\sqrt{\Pi}}
-\fq\frac{\mu_4}{\sqrt{\Pi}}
-\fq\frac{\mu_5}{\sqrt{\Pi}}
-\sum_{f=1}^{5}s\left(\mu_f+\mu_f^{-1}\right)
\right)q
}{(1-q)^2}
\right]
\\[2mm]
&\times
\sum_{\lambda_1,\lambda_2}
\left(- \fq ^{\frac{1}{2}} \Pi^{\frac{1}{4}} \right)^{|\lambda_1|+|\lambda_2|}
\frac{
\displaystyle\prod_{f=1}^{5}
\widetilde{\mathcal N}_{\lambda_1 0}\left(\frac{1}{s \mu_f}\right)\,
\widetilde{\mathcal N}_{\lambda_2 0}
\left(\frac{s}{\mu_f}\right)
}{
\displaystyle
\left[ \widetilde{\mathcal N}_{\lambda_2\lambda_1}^{\,2}(s^2) \widetilde{\mathcal N}_{\lambda_1\lambda_1}(1)\widetilde{\mathcal N}_{\lambda_2\lambda_2}(1) \right]\,
\widetilde{\mathcal N}_{\lambda_1 0}
\left(\frac{\fq}{s\sqrt{\Pi}}\right)\,
\widetilde{\mathcal N}_{\lambda_2 0}
\left(\frac{s\fq}{\sqrt{\Pi}}\right)
} 
\end{split}
\label{eq:ZtopE6refined}
\end{equation}
In the second line, we have rephrased the equation to have a form akin to Nekrasov instanton sum. Somewhat surprisingly, the generalized toric vertex computes effectively the partition function with five fundamental hypermultiplets, together with a sixth fundamental hypermultiplet of multiplicity $-1$ whose mass fugacity $\mu_6 = \sqrt{\Pi}/\fq$ depends on the instanton counting parameter. This is only a formal rewriting of the topological vertex sum applied to the brane web above -- and this also explains why the partition sum carries a factor $\left(- \fq ^{\frac{1}{2}} \Pi^{\frac{1}{4}} \right)^{|\lambda_1|+|\lambda_2|}$ and not the expected $\fq^{|\lambda_1|+|\lambda_2|}$. 
For the parameters of Table \ref{tab:GV_E6}, 
\begin{equation}\label{eq:E6dictCorrect}
    s = Q_1 Q_2\,,\qquad
    (\mu_1,\dots,\mu_5) = \bigl(Q_2^{-1},\,1,\,Q_2^{-1},\,1,\,Q_2\bigr)\,,\qquad
    \Pi = Q_2^{-1}\,,\qquad \fq = Q_2^{3/2}\,,
\end{equation}
this specializes to \eqref{eq:E6_NekFactors}. One can compare with the Nekrasov partition function using the U(2) formalism:  
\begin{equation}
\begin{split}
Z_{\mathrm{Inst}}^{\mathrm{U}(2),\,N_f=5}|_{s_1 s_2 = 1}
={}&
\sum_{\lambda_1,\lambda_2}
\fq^{|\lambda_1|+|\lambda_2|}
s^{\frac{|\lambda_2|-|\lambda_1|}{2}}
q^{\frac{\kappa_{\lambda_1}+\kappa_{\lambda_2}}{4}}
\frac{
\displaystyle
\prod_{f=1}^{5}
\widetilde{\mathcal N}_{\lambda_1 0}
\left(\frac{1}{s\mu_f}\right)
\widetilde{\mathcal N}_{\lambda_2 0}
\left(\frac{s}{\mu_f}\right)
}{
\displaystyle
\widetilde{\mathcal N}_{\lambda_1\lambda_1}(1)\,
\widetilde{\mathcal N}_{\lambda_2\lambda_2}(1)\,
\widetilde{\mathcal N}_{\lambda_1\lambda_2}(s^{-2})\,
\widetilde{\mathcal N}_{\lambda_2\lambda_1}(s^2)
}. \\ 
 = &
\sum_{\lambda_1,\lambda_2}
\left(\frac{Q_2}{Q_1}\right)^{|\lambda_1|}
(-Q_2)^{|\lambda_2|}
q^{\kappa_{\lambda_1}-\frac12\kappa_{\lambda_2}}
 \\ & \qquad \frac{
\mathcal N_{0\lambda_1}^{\,2}(Q_1)\,
\mathcal N_{0\lambda_1}^{\,2}(Q_1Q_2)\,
\mathcal N_{0\lambda_1}(Q_1Q_2^2)
\mathcal N_{\lambda_2 0}(Q_1)\,
\mathcal N_{\lambda_2 0}^{\,2}(Q_1Q_2)\,
\mathcal N_{\lambda_2 0}^{\,2}(Q_1Q_2^2).}{
\mathcal N_{\lambda_1\lambda_1}(1)\,
\mathcal N_{\lambda_2\lambda_2}(1)\,
\mathcal N_{\lambda_2\lambda_1}^{\,2}
\left(Q_1^2Q_2^2\right)
} 
\end{split}
\label{eq:ZnekU(2)Nf=5}
\end{equation}
We see that there are differences which are not removable using a U(1)-factor as in \eqref{eq:U1term}. It would be interesting to understand how to relate \eqref{eq:ZnekU(2)Nf=5} with \eqref{eq:ZtopE6refined} using combinatorial identities. In any case, we expect the proper match \eqref{eq:ZtopZnek} requires performing the instanton computation in the orthosymplectic ADHM formalism (\cite{Bao:2013pwa} contains an expression involving a sum over three partitions).  

\begin{table}[t]
    \centering
\makebox[\textwidth][c]{\begin{tabular}{c||cccccc}
    $d_2\backslash d_1$ & 0 & 1 & 2 & 3 & 4 & 5 \\\hline
    0 &  & 3 &  &  &  &  \\
    1 & {\color{red}-5} & 6 &  &  &  &  \\
    2 & {\color{red}-2} &{\color{blue} 9} & -6 &  &  &  \\
    3 &  & 6 & -12 & 6 &  &  \\
    4 &  & 3 & {\color{blue} -18} & 27 & -12 &  \\
    5 &  &  & -12 & 54 & -72 & 30 \\
    6 &  &  & -6 & {\color{blue} 69,-4} & -204, 15 & 225, -18 \\
    7 &  &  &  & 54 & -360, 30 & 810, -108 \\
    8 &  &  &  & 27 & {\color{blue} -432, 45}  & 1860, -366, 21 \\
    9 &  &  &  & 6 & -360, 30 & 2970, -660, 42 \\
    10 &  &  &  &  & -204, 15 & {\color{blue}3465, -828, 63} 
    \end{tabular}}

    \vspace{.5cm}
    \makebox[\textwidth][c]{\begin{tabular}{c|cccc}
        $d\backslash g$ & 0 & 1 & 2  & 3\\ \hline
        1 & 27 & &\\
        2 & -54& & \\
        3 & 243 & -4 &\\
        4 & -1728 & 135 & \\
        5 & 15255 & -3132 & 189 \\
    \end{tabular}}
    \caption{\emph{Top:} GV invariants for the brane web in Figure \ref{fig:branes_GdP6_nonToric}. The red coefficients correspond to the parallel branes with K\"ahler parameter $Q_2$ and ${Q_2}^2$. The entries in each column are palindromic, with the central entry shown in blue. \emph{Bottom:} Summing over the lines, one gets the GV invariants for local $dP_6$ in anticanonical degree and arbitrary genus.}
    \label{tab:GV_E6}
\end{table}

\subsubsection{\texorpdfstring{$E_7$}{E7}}

\begin{figure}[t]
    \centering
    \scalebox{.8}{
    \begin{tikzpicture}
 \tikzmath{\x=1.5;\y=.8;\z=2;\ee=.1;}
        \draw (-\z,0) -- (0,0) -- (0,\x) node[midway, left] {$Q_2$}-- (-\z ,\x);
        \draw (0,\x) -- (\x, 2*\x) node[midway, xshift=-14pt] {$Q_2$} -- (-\z,2*\x);
        \draw (\x,2*\x) -- (\x +2*\y,2*\x +\y) node[pos=.45,yshift=10pt] {$Q_1$} -- (2*\x +\y-\ee,3*\x -\ee) -- (2*\x+\y-\ee,3*\x+.5*\z);
        \draw (\x +2*\y,2*\x +\y)-- node[pos=.5,inner sep=0pt] (p1) {} (2*\x +\y,2*\x +\y)--(2*\x +\y,3*\x - 2*\ee) arc (-90:90:\ee) -- (2*\x +\y,3*\x +.5*\z) ;
        \draw(2*\x +\y-\ee,3*\x -\ee) -- (2*\x +2*\y+\z,3*\x -\ee);
        \draw  (2*\x +\y,2*\x +\y) -- (2*\x +2*\y,2*\x ) node[midway, yshift=5pt,xshift=8pt] {$Q_1$} --(2*\x +2*\y+\z,2*\x);
        \draw (2*\x +2*\y,2*\x ) -- (2*\x +2*\y,\x ) node[midway, right] {$Q_2$} -- (2*\x +2*\y+\z,\x );
        \draw (2*\x +2*\y,\x) -- (\x +2*\y,0) node[midway, xshift=13pt] {$Q_2$}-- (2*\x +2*\y+\z,0);
        \draw (\x +2*\y,0) -- node[pos=.6,below right,yshift=3pt] {$Q_1$} (\x ,-\y)  -- node[pos=.5,inner sep=0pt](p2){} (\y,-\y) -- (\y,-\x+2*\ee) arc(90:270:\ee) -- (\y,-\x-.5*\z);
        \draw   (\x ,-\y) -- (\y+\ee,-\x+\ee) -- (-\z,-\x+\ee);
        \draw (\y+\ee,-\x+\ee) -- (\y+\ee,-\x-.5*\z);
        \draw (\y,-\y) -- (0,0)node[midway,xshift=-8pt,yshift=-5pt] {$Q_1$}; 
        \node[inner sep=1pt]  at (\x+\y,\x) (l1) {$\frac{Q_2}{Q_1}$};
        \draw[-latex,dashed,bend right =10] (l1) to (p1);
        \draw[-latex,dashed,bend right=10] (l1) to (p2);
    \end{tikzpicture}}\hspace{1cm}
    \raisebox{3cm}{\scalebox{1}{  
    \begin{tikzpicture}[x=.5cm,y=.5cm] 
    \foreach \x in {-1,...,1} {\draw[gridline] (\x,-2.5) -- (\x,2.5);};
    \foreach \y in {-2,...,2} {\draw[gridline](-1.5,\y) -- (1.5,\y); };
    \foreach \x in {-1,...,1} {\node[bd] at (0,\x) {};};
    \foreach \y in {-2,...,2} {\node[bd] at (-1,\y) {};\node[bd] at (1,\y) {};};
    \foreach \y in {-2,...,1} {\draw[ligne] (0,0) -- (1,\y);\draw[ligne] (0,0) -- (-1,-\y);\draw[ligne] (1,\y) -- (1,1+\y);\draw[ligne] (-1,-\y) -- (-1,-1-\y);};
    \foreach \ee in {-1,1} {\draw[ligne] (\ee,2*\ee) -- (-\ee,2*\ee) --(0,\ee) -- (\ee,\ee);\draw[ligne] (0,0) -- (0,\ee);};
    \node[wd] at (0,2) {};
    \node[wd] at (0,-2) {};
    \end{tikzpicture}}}

      \scalebox{.8}{ \begin{tikzpicture}[every node/.style={inner sep=2pt}]
    \tikzmath{\x=1.5;\y=.8;\z=2;\ee=.1;}
        \draw (-\z,0) -- (0,0) -- (0,\x) node[midway, left] {$Q_2$}-- (-\z ,\x);
        \draw (0,\x) -- (\x, 2*\x) node[midway, xshift=-14pt] {$Q_2$} -- (-\z,2*\x);
        \draw (\x,2*\x) --  node[pos=.45,yshift=10pt] {$Q_1$} (\x +2*\y,2*\x +\y) -- (\x +2*\y+\z,2*\x +\y+\z);
        \draw (\x +2*\y,2*\x +\y) --node[pos=.65,above] {$Q_2$} (2*\x +2*\y,2*\x +\y) -- (2*\x +2*\y+\z,2*\x +\y+\z);
        \draw (2*\x +2*\y,2*\x +\y) --node[pos=.35,right] {$Q_2$} (2*\x +2*\y,\x +\y) -- (2*\x +2*\y+\z,\x +\y+\z);
        \draw (2*\x +2*\y,\x +\y) -- node[pos=.45,xshift=-10pt] {$Q_1$}(2*\x +\y,\x -\y) -- (2*\x +\y,\ee)arc(90:270:\ee)--(2*\x +\y,-\x-.5*\z);
        \draw (2*\x +\y,\x -\y) -- node[pos=.5,inner sep=0pt] (p1) {} (\x +2*\y,0) -- (2*\x +\y+\ee,0)-- (2*\x +\y+\z+\ee,\z) ;
        \draw (2*\x +\y+\ee,0) -- (2*\x +\y+\ee,-\x-.5*\z);
        \draw (\x +2*\y,0) -- node[pos=.6,below right,yshift=3pt] {$Q_1$} (\x ,-\y)  -- node[pos=.5,inner sep=0pt](p2) {} (\y,-\y) -- (\y,-\x+2*\ee) arc(90:270:\ee) -- (\y,-\x-.5*\z);
        \draw   (\x ,-\y) -- (\y+\ee,-\x+\ee) -- (-\z,-\x+\ee);
        \draw (\y+\ee,-\x+\ee) -- (\y+\ee,-\x-.5*\z);
        \draw (\y,-\y) -- (0,0)node[midway,xshift=-8pt,yshift=-5pt] {$Q_1$};
        \node[inner sep=1pt]  at (\x+\y,\x) (l1) {$\frac{Q_2}{Q_1}$};
        \draw[-latex,dashed,bend right =10] (l1) to (p1);
        \draw[-latex,dashed,bend right=10] (l1) to (p2);
    \end{tikzpicture}}
    \raisebox{3cm}{\scalebox{1}{  
    \begin{tikzpicture}[x=.5cm,y=.5cm] 
    \foreach \x in {-1,...,3} {\draw[gridline] (\x,-2.5) -- (\x,2.5);\node[bd] at (\x,1-\x) {};};
    \foreach \y in {-2,...,2} {\draw[gridline](-1.5,\y) -- (3.5,\y); \node[bd] at (-1,\y) {};};
    \foreach \x in {0,...,3} {\node[bd] at (\x,1-\x) {};};
    \draw[ligne] (-1,2) -- (3,-2) -- (-1,-2) -- (-1,2);
    \draw[ligne] (-1,-1) -- (0,-1) -- (1,-2) -- (1,-1) -- (2,-1);
    \draw[ligne] (1,-2) -- (-1,2);
    \draw[ligne] (0,1) -- (0,-1);
    \draw[ligne] (1,0) -- (-1,0);
    \draw[ligne] (-1,-1) --(0,0) -- (2,-1);
    \draw[ligne] (1,-1) -- (-1,1);
    \node[bd] at (0,0) {};
    \node[bd] at (0,-1) {};
    \node[bd] at (1,-1) {};  
    \node[bd] at (1,-2) {};
    \node[wd] at (2,-2) {};
    \node[wd] at (0,-2) {};
    \end{tikzpicture}}}
    \caption{Two brane webs realizing generalized-toric $GdP_7$ using only simple crossings. They are related by a Hanany-Witten transition.} 
    \label{fig:branes_GdP7}
\end{figure}
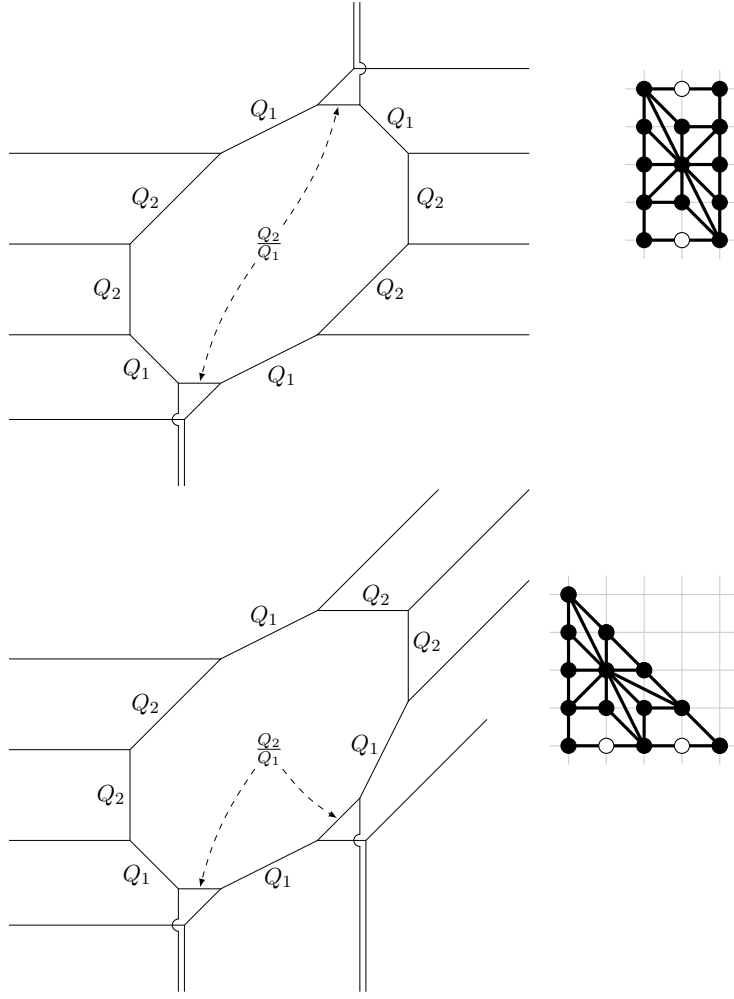

For $n=7$, there are no toric $GdP_7$. There are generalized-toric realizations, including some where the brane web includes only simple crossing. In this situation, the computation is similar to the one performed in the previous section for $E_6$.  The GV invariants, shown in Table \ref{tab:GV_E7}, were computed using the brane web and the assignment of Kähler parameters are shown in the top part of Figure \ref{fig:branes_GdP7}. A second web, shown in the bottom of the Figure, it obtained by Hanany-Witten transition. It allows for a check of Hanany-Witten invariance and of the count of parallel branes discussed in Section \ref{sec:Zdec_GTP} : the first web has two sets of parallel branes following the partition $(1^4)$ with the branes separated by curves of weights $Q_2$ and two following $(2)$; the second has two  sets of parallel branes following the partition $(1^4)$, separated by curves of weight $Q_2$, and one following $(2^2)$, separated by a curve of weight $Q_2^2$. Therefore, we have :
\begin{align}
    Z_\parallel^{(a)} = \operatorname{PExp}\left[\frac{q}{(1-q)^2} \left(6 Q_2 +4Q_2^2 + 2Q_2^3\right)\right] \\
    Z_\parallel^{(b)} = \operatorname{PExp}\left[\frac{q}{(1-q)^2} \left(6 Q_2 +6Q_2^2 + 2Q_2^3\right)\right]
\end{align}

\begin{table}[t]
    \centering
\makebox[\textwidth][c]{\begin{tabular}{c||ccccc}
    $d_2\backslash d_1$ & 0 & 1 & 2 & 3 & 4  \\ \hline\hline
    0 &  & 4 &  &  &    \\
    1 & {\color{red} -6} & 6 &  &  &    \\
    2 & {\color{red} -4} & 12 & -6 &  &    \\
    3 &  {\color{red} -2} & {\color{blue} 12} & -16 & 6 &    \\
    4 &  & 12 & -36 & 36 & -12   \\
    5 &  & 6 & -48 & 108 & -96   \\
    6 &  & 4 & {\color{blue} -60, 3} & 240, -16 & -400, 30   \\
    7 &  &  & -48 & 396, -24 & -1152, 120   \\
    8 &  &  & -36 & 540, -48 & -2568, 411, -18   \\
    9 &  &  & -16 & {\color{blue} 588, -48} &  -4608, 856, -48  \\
    10 &  &  & -6 & 540, -48 & -6912, 1506, -108  \\
    11 &  &  &  & 396, -24 &  -8736, 1968, -144 \\
    12 &  &  &  & 240, -16 & {\color{blue}-9464, 2263, -208, 7}
    \end{tabular}}

    \vspace{.4cm}
    \makebox[\textwidth][c]{\begin{tabular}{c|cccc}
        $d\backslash g$ & 0 & 1 & 2 & 3  \\ \hline
        1 & 56 &  &  \\
        2 & -272 & 3 &  \\
        3 & 3240 & -224 &  \\   
        4 & -58432 & 12045 & -844 & 7  
    \end{tabular}}
    \caption{\emph{Top:} GV invariants for the brane web in Figure \ref{fig:branes_GdP7}. The red coefficients correspond to the parallel branes with K\"ahler parameter $Q_2,{Q_2}^2$ and ${Q_2}^3$. The entries in each column are palindromic, with the central entry shown in blue. \emph{Bottom:} Summing over the lines, one gets the GV invariants for local $dP_7$ in anticanonical degree and arbitrary genus. \\ Note that the entry at $d=4$, $g=1$ disagrees with the corresponding entry in \cite{Katz:1999xq}, which forgets the elliptic multi-cover formula and instead uses the old counting of \cite{Lerche:1996ni}, called $n_d^{*1}$ in \cite{Katz:1999xq}, see equation (2.6) there. The correct counting is $12045 =0+3+ 12042$.}
    \label{tab:GV_E7}
\end{table}

\subsubsection{\texorpdfstring{$E_8$}{E8}}
Generalized-toric realizations of $GdP_8$ exist
(Fig. 8 of \cite{Hayashi:2014wfa}, and Table \ref{tab:GTPs_for_GdPs} below), and in principle the rules of Table \ref{tab:TV} apply. In practice the diagrams are large enough that we have not been able to push the computation to a degree where a comparison with \cite{Katz:1999xq} is possible. We leave this to future work.

\section{Conclusions and outlook}
\label{sec:conclusions}

In this paper, we investigated the topological string partitions of non compact $CY3$ and their relation with $5d$ $\cN=1$ gauge theory via geometric engineering in M-theory. There are many interesting questions to be explored, we list a few of them here: 
\begin{itemize}
    \item We observe that for all del Pezzo surfaces ($dP_n$ with $0 \leq n \leq 8$), the first non trivial genus one GV invariant appears at degree $9-n$ and satisfies $GV_{g=1,9-n}(dP_{n}) = (-1)^n(n-10)$, with the grading provided by the anti-canonical divisor. It would be interesting to prove this geometrically. 
    \item We observe in Tables \ref{tab:GV_F0} and \ref{tab:GV_F1} that 
    \begin{equation}
       \forall r ,d_f \geq 0 \, , \qquad  \GV_{r , d_{f} f_0 + 2 b_0} (\mathbb{F}_0) = \GV_{r , (d_{f} +1) f_1 + 2 b_1} (\mathbb{F}_1) \, . 
    \end{equation}
    We do not know of a physical or mathematical explanation for this. The generalization of the results of \cite{gottsche1996quantum} may be helpful. 
    \item In this paper, we are able to identify $Z_{5d \ \text{SCFT}}(\mathcal{T}_X)$ sector of $Z_{top}[X]$ using local mirror symmetry for local surface $CY3$ (i.e. rank-0 or rank-1 theories). It would be interesting to develop the methods for such an identification. One direction is to generalize local mirror symmetry of \cite{chiang1999localmirrorsymmetrycalculations} to more general cases of non compact $CY3$. 
\end{itemize}
These questions are under our current investigation.

\section*{Acknowledgements}
We thank Cyril Closset, Hans Jockers, Taro Kimura and Andrea Sangiovanni for helpful discussions. The work of PC is supported by the Cluster of Excellence PRISMA++ (EXC 2118/2, Project ID 390831469). The work of QL is supported by École Normale Supérieure - PSL through a CDSN doctoral grant. QL also thanks hospitality of JGU Mainz and support from the short term scientific mission grant of COST action CA22113 THEORY-CHALLENGES.

\appendix

\section{Notations and Conventions}
\label{app:notations}

We gather in this appendix the notations we use throughout the paper. 

\subsection{Geometry and topology}

\paragraph{Basic Geometric Data}
\begin{itemize}
    \item $X$ -- Singular Local Calabi-Yau threefold. 
    \item $\tX \rightarrow X$ -- Non-singular quasi-projective Calabi-Yau resolution of $X$. 
    \item $g \in \mathbb{N}$ -- genus of maps. 
    \item $\beta \in H_2^{\text{cpt}} (\tX,\mathbb{Z})$ -- degree of maps.  
    \item $\NEbar(\tX)$ -- the Mori cone of effective curve classes of $\tX$ . 
    \item $\omega = B + i J$ -- complexified K\"ahler class on $\tX$    
    \item $t_\beta := -i\int_\beta \omega = \int_\beta J - i\int_\beta B$ -- complexified K\"ahler parameter associated to $\beta$. Its real part $\operatorname{Re}(t_{\beta})$ is the volume of $\beta$.
    \item $Q^\beta := e^{-t_\beta} = \exp\!\left(i\int_\beta \omega\right)$ -- exponentiated Novikov/K\"ahler monomial. 
    \end{itemize}

\paragraph{Enumerative Invariants}    
    \begin{itemize}
    \item $\overline{\mathcal{M}}_g (\tX , \beta)$ -- moduli space of stable\footnote{Roughly speaking, automorphism group should be finite. For instance, genus zero requires at least 3 punctures. Genus 1 requires at least 1 puncture. } maps $f : \Sigma_g \rightarrow \tX$ from connected curves of genus $g$ into $\tX$ representing the class $\beta$, i.e. with $[f(\Sigma_g)] = \beta$. 
    \item $N_{g,\beta} (\tX) = \int_{[\overline{\mathcal{M}}_g (\tX , \beta)]^{\text{vir}}} 1$ -- Gromov-Witten invariant.\footnote{We use virtual to ensure that there is no accidental appearance of a curve when hitting a specific complex structure. For instance in $T^* \mathbb{P}^1$, in the HK complex structure, there is a genus 0 curve, but not in other complex structures. So the invariant should be 0. } 
    \item $\mathcal{I}_{n,\beta} (\tX)$ -- moduli space of stable ideal sheaves. Each ideal sheaf determines a subscheme $Y \subset \tX$ with maximal dimensional\footnote{$Y$ can be for instance a curve and isolated points. Imposing $\beta \neq 0$ amounts to requiring the dimension to be non-zero. More precisely, the subscheme $Y$ determined by an ideal sheaf can contain one-dimensional components (curves) together with zero-dimensional ones (isolated or embedded points). $Y_{\text{max}}$ is the union of the one-dimensional components, and its class defines $\beta$; the zero-dimensional components carry no curve class but contribute to $\chi(\cO_Y)$, playing the role of D0-branes bound to the D2-branes wrapping $Y_{\text{max}}$. } component $Y_{\text{max}}$. Then $\mathcal{I}_{n,\beta} (\tX)$ is the moduli space of stable ideal sheaves with $\chi (\mathcal{O}_{Y}) = n$ and $[Y_{\text{max}}] = \beta$. 
    \item $\mathrm{DT}_{n,\beta} (\tX) = \int_{[\mathcal{I}_{n,\beta} (\tX)]^{\text{vir}}} 1$ -- rank-1 Donaldson-Thomas invariant.    
    \item $F_{\textrm{GW}} (\tX ; g_s,Q)$ and $F'_{\textrm{GW}} (\tX ; g_s,Q)$ -- generating function for GW invariants, and its reduced version,
    \begin{equation}
    \label{eq:GenFunctionofGW}
          F_{\GW}(\tX ; g_s,Q) = \sum_{g\geq 0} \sum_{\beta \in H_2(\tX , \mathbb{Z})} N_{g,\beta} (\tX) \,  Q^{\beta} g_s^{2g-2} = F_{\text{const}}(\tX ; g_s) + F_{\text{top}}(\tX ; g_s,Q)
    \end{equation}
    with
    \begin{equation}
          F_{\text{top}}(\tX ; g_s,Q) = \sum_{g\geq 0} \sum_{0 \neq \beta \in H_2(\tX , \mathbb{Z})} N_{g,\beta} (\tX) \,  Q^{\beta} g_s^{2g-2}
    \end{equation}
The reduced generating function is the restriction to nonzero curve classes, $F'_{\GW} (\tX ; g_s,Q) = F_{\text{top}}(\tX ; g_s,Q)$, i.e.\ the constant-map contribution $F_{\text{const}}$ is discarded.
    and the regularized  expression \footnote{Setting $Q \rightarrow 0$ in \eqref{eq:GenFunctionofGW} gives an asymptotic series for  \eqref{eq:Fconst}. }
    \begin{equation}
    \label{eq:Fconst}
          F_{\text{const}}(\tX ; g_s)    = - \frac{\chi (\tX)}{2}   \sum\limits_{k=1}^{\infty}    \frac{ 1}{k} \left( 2 \sin \frac{k g_s}{2} \right)^{-2}
    \end{equation} 
    \item $Z_{\text{top}}(\tX ; g_s,Q) = e^{F_{\text{top}}(\tX ; g_s,Q)}$ -- the GW partition function, which is also the topological string partition function. 
    \item $Z_{\DT} (\tX ; q ,Q) $ and $Z'_{\DT} (\tX ; q ,Q) $ -- the DT partition function
    \begin{equation}
        Z_{\DT} (\tX ; q ,Q) = \sum_{n \in \mathbb{Z}} \sum_{\beta \in H_2(\tX , \mathbb{Z})}  \mathrm{DT}_{n,\beta} (\tX) Q^{\beta} (-q)^n
    \end{equation}
and its reduced version
    \begin{equation}
        Z'_{\DT} (\tX ; q ,Q) =  \frac{ Z_{\DT} (\tX ; q ,Q)}{ Z_{\DT} (\tX ; q ,0)} \, .
    \end{equation}
    \item The GW/DT correspondence:\footnote{It has been proven that (the degree-$0$ statement below was conjectured in \cite{maulik2004gromovwittentheorydonaldsonthomastheory} and proven in \cite{li2006zero,behrend2008symmetric})
\begin{equation}
    \label{D6-D0 BPS states}
  Z_{\DT} (\tX ; q ,0) = M(q)^{\chi(\tX)} =  \prod_{n=1}^{\infty}(1-q^n)^{-n \, \chi(\tX)}  \,. 
\end{equation}
This corresponds to D6-D0 BPS states. }
    \begin{equation}
       Z_{\text{top}}(\tX ; g_s,Q) =   \frac{ Z_{\DT} (\tX ; q ,Q)}{ Z_{\DT} (\tX ; q ,0)}  \, , \qquad  q =  e^{i g_s} \, . 
    \end{equation}
    \item $\GV_{r , \beta} (\tX)$ -- Gopakumar-Vafa invariants, defined by the equality 
    \begin{equation}\label{GW=GV}
        F_{\GW}(\tX ; g_s,Q) = \sum\limits_{r=0}^{\infty} \sum\limits_{k=1}^{\infty} \sum_{\beta \in H_2(\tX , \mathbb{Z})} \GV_{r, \beta} (\tX) \; \frac{Q^{k \beta}}{k} \left( 2 \sin \frac{k g_s}{2} \right)^{2r-2}
    \end{equation}    Here we formally extend the GV notation to degree zero by setting
    \begin{equation}
        \GV_{0,0}(\widetilde X)=-\frac{\chi(\widetilde X)}{2},
\qquad
\GV_{r>0,0}(\widetilde X)=0  \, . 
    \end{equation} 
With this convention, the $\beta=0$ contribution reproduces
$F_{\mathrm{const}}$. Ordinary GV invariants will otherwise refer to
nonzero effective curve classes.
    This can be rewritten as 
    \begin{equation}
\label{eq:GVPlethystic}
Z_{\mathrm{top}}(\tX;q,Q)
=
\operatorname{PExp}\!\left[
\sum_{\substack{\beta\neq 0\\ r\geq 0}}
\GV_{r,\beta}(\tX)\,
\left(\frac{-q}{(1-q)^2}\right)^{1-r}
Q^\beta
\right],
\qquad q=e^{ig_s}.
\end{equation}
    \end{itemize}

\paragraph{Physical Interpretation of the invariants}
    \begin{itemize}
    \item $Z_{\text{top}} (\tX ; g_s , Q) = e^{F_{\text{top}} (\tX ; g_s , Q)}$ -- topological string partition function and free energy. 
    \item Note that the full partition function also includes the constant maps and perturbative contributions
\begin{equation}
\label{eq:GVBPS}
F_{\mathrm{full}} (g_s,Q) = \underbrace{\frac{1}{6 g_s^2} \kappa_{ijk} t_i t_j t_k + \frac{1}{24} c_{2,i} t_i + F_{\text{const}}(g_s)}_{F_{\mathrm{pert}}(g_s,Q)}  + \underbrace{F'_{\GW} (g_s,Q)}_{F_{\mathrm{top}} (g_s,Q)}
\end{equation}  
The last two terms add up to the generating function of GW invariants \eqref{eq:GenFunctionofGW}.  $F_{\mathrm{pert}}(g_s,Q)$ depends on a choice of normalization.  
\item Massive particles in 5d are characterized by their $\operatorname{Spin}(4)$ spins $(j_L , j_R)$. M2 branes wrapping two-cycles in class $\beta$ give rise to such particles, and we denote by $N_{j_L , j_R}^{\beta} (\tX)$ their multiplicities. Then the GV invariants are obtained by taking an index over the $SU(2)_R$ quantum numbers 
\begin{equation}
    \sum\limits_{j_L , j_R} (-1)^{2 j_R} (2 j_R +1) \;  [j_L] \;  N_{j_L , j_R}^{\beta} (\tX)   = \sum\limits_{r \geq 0}  \left( \left[\frac{1}{2} \right] + 2 [0] \right)^{\otimes r} \; \GV_{r,\beta} (\tX) \, ,  
\end{equation}
where the square brackets indicate the corresponding $\mathfrak{su}(2)$ characters. 
\end{itemize}

\paragraph{SCFT and partition functions}

For background on the $5d$ superconformal index and its relation to BPS state counting, see \cite{Kim:2012gu,Iqbal:2012xm}.

    \begin{itemize}
    \item $\mathcal{T}_X = \mathcal{T} [X]$ -- 5d $\mathcal{N}=1$ superconformal field theory obtained from M-theory on $X$. 
\item $R$ -- radius of the $S^1$ used for compactifying from 5d to 4d. 
\item $\epsilon_1,\epsilon_2$ -- equivariant parameters for the $\Omega$-background. They have dimension of inverse length. We also use the exponential versions $q_1 = e^{R \epsilon_1}$,  $q_2 = e^{R \epsilon_2}$
\item $R \epsilon_1 = - R \epsilon_2 = i g_s$ -- parameters for the self-dual $\Omega$-background. 
\item $Z_{\text{Nek}} = Z_{\mathcal{T}}[\mathbb{C}^2_{\epsilon_1 , \epsilon_2} \times S^1_R]$ -- The partition function on $\mathbb{C}^2_{\epsilon_1 , \epsilon_2} \times S^1_R$, which is also called the Nekrasov partition function. 
\end{itemize}

\subsection{Combinatorics}

\paragraph{Partitions. }
Greek letters denote partitions / Young diagrams, e.g. $\lambda = (\lambda_i)_{i = 1 , 2 , \dots}$ with $\lambda_1 \geq \lambda_2 \geq \dots$. We associate $\lambda$ to the Young diagram where the $i$th row has length $\lambda_i$. We call 
\begin{itemize}
\item $\ell (\lambda)$ the length of $\lambda$, so that $\lambda_{\ell(\lambda)} > \lambda_{\ell(\lambda)+1} = 0$. 
    \item $|\lambda| \doteq \sum_{i=1}^{\ell (\lambda)} \lambda_i$. 
    \item $\lambda^T$ the transpose of $\lambda$. 
    \item $\kappa_\lambda \doteq |\lambda| +  \sum_{i=1}^{\ell (\lambda)} \lambda_i (\lambda_i - 2i) = 2 \sum_{(i,j) \in \lambda} (j-i)$. 
    \item The zero partition is denoted $0$. We have $\ell (0) = \kappa_0 = 0$. 
\end{itemize} 

\paragraph{Symmetric functions. }
\begin{itemize}
    \item $e_i (\mathbf{x})$ and $h_i (\mathbf{x})$ -- elementary and complete symmetric functions in infinitely many variables for $i \in \mathbb{N}$, defined by \begin{equation}
    \prod\limits_{i=1}^{\infty} (1 + x_i t) = \sum\limits_{r=0}^{\infty} e_r  (\mathbf{x}) t^r \, , \qquad    \prod\limits_{i=1}^{\infty} \frac{1}{1 - x_i t}  = \sum\limits_{r=0}^{\infty} h_r  (\mathbf{x}) t^r \, . 
\end{equation} 
\item $ s_{\lambda} (\mathbf{x})  = \det (h_{\lambda_i -i+j} (\mathbf{x}))_{1 \leq i,j < \infty} = \det (e_{\lambda_i^T -i+j} (\mathbf{x}))_{1 \leq i,j < \infty} $ -- Schur function.\footnote{We use $e_r = 0$ and $h_r = 0$ for $r<0$. } 
\item $ s_{\lambda / \mu} (\mathbf{x}) = \det (h_{\lambda_i - \mu_j -i+j} (\mathbf{x}))_{1 \leq i,j < \infty}  = \det (e_{\lambda_i^T - \mu_j^T -i+j} (\mathbf{x}))_{1 \leq i,j < \infty}$ -- skew-Schur function. In particular $s_{\lambda / 0} = s_{\lambda}$. Note that $s_{\lambda / \mu} = 0$ if and only if $ \exists i$ such that $\lambda_i < \mu_i$. The skew Schur functions satisfy the Cauchy identities
\begin{equation}
\label{eq:CauchyIdentities}
    \begin{split} 
        \sum\limits_{\lambda} s_{\lambda / \mu} (\mathbf{x}) s_{\lambda / \nu} (\mathbf{y}) &= \prod\limits_{i,j=1}^{\infty} \frac{1}{1-x_i y_j} \sum\limits_{\eta}  s_{\nu / \eta} (\mathbf{x}) s_{\mu / \eta} (\mathbf{y}) \\
    \sum\limits_{\lambda} s_{\lambda / \mu^T} (\mathbf{x}) s_{\lambda^T / \nu} (\mathbf{y}) &= \prod\limits_{i,j=1}^{\infty} (1+x_i y_j) \sum\limits_{\eta}  s_{\nu^T / \eta} (\mathbf{x}) s_{\mu / \eta^T} (\mathbf{y})   
\end{split}
\end{equation}
They convert the infinite sums on the left into finite sums on the right. 
\item $\rho = \left( - \frac{1}{2} , - \frac{3}{2} , - \frac{5}{2} , \dots \right)$ -- Weyl vector. For a given partition $\nu$, when we evaluate $s_{\lambda / \mu}$ at $\mathbf{x} = q^{\nu - \rho} = (q^{\nu_1 + \frac{1}{2}} , q^{\nu_2 + \frac{3}{2}} , \dots )$, it gives a rational function in $q^{\frac{1}{2}}$. We use often the identity 
\begin{equation}
    s_{\lambda^T} (q^{-\rho}) = q^{ \frac{1}{2} \kappa_{\lambda}} s_{\lambda} (q^{-\rho}) \, . \label{eq:symSfactor}
\end{equation}
\end{itemize}

\paragraph{Nekrasov factors. }
The Nekrasov factor is 
\begin{equation}\label{eq:NFactor_finite_product}  
    \mathcal{N}_{\lambda \nu} (Q,q) \doteq \prod\limits_{(i,j) \in \lambda} (1-Q q^{\lambda_i + \nu^T_j -i-j+1}) \prod\limits_{(i,j) \in \nu} (1-Q q^{-\lambda_j^T - \nu_i +i+j-1})   \, . 
\end{equation}
Some useful identities are 
\begin{eqnarray}
    \mathcal{N}_{00} (Q,q) &=&  1 \\ 
    \mathcal{N}_{\lambda \nu} (Q,q) &=&  \mathcal{N}_{\nu^T \lambda^T } (Q,q) \label{eq:transposeNfactor} \\
     \mathcal{N}_{\lambda \nu} (Q,q) &=&  (-Q)^{|\lambda| + |\nu|} q^{\frac{1}{2} (\kappa_{\lambda} - \kappa_{\nu})}  \mathcal{N}_{\nu \lambda } (Q^{-1},q) \, . \label{eq:symmetryNfactor}
\end{eqnarray}
We also define a symmetrized Nekrasov factor $\tilde{\mathcal{N}}$, obtained by replacing every factor $1-Q q^x$ in \eqref{eq:NFactor_finite_product} by $(Q q^x)^{-1/2}-(Q q^x)^{1/2}$,
\begin{equation}\label{eq:NFactor_sym}
    \tilde{\mathcal{N}}_{\lambda \nu} (Q,q) \doteq \prod\limits_{(i,j) \in \lambda} \Big( (Qq^{\lambda_i + \nu^T_j -i-j+1})^{-\frac12}-(Qq^{\lambda_i + \nu^T_j -i-j+1})^{\frac12} \Big) \prod\limits_{(i,j) \in \nu} \Big( (Qq^{-\lambda_j^T - \nu_i +i+j-1})^{-\frac12}-(Qq^{-\lambda_j^T - \nu_i +i+j-1})^{\frac12} \Big) \, .
\end{equation}
Using $(Qq^x)^{-1/2}-(Qq^x)^{1/2} = (Qq^x)^{-1/2}(1-Qq^x)$ per box together with $\sum_{(i,j)} x = \tfrac{1}{2}(\kappa_\lambda - \kappa_\nu)$ (the cross terms $\sum_j \lambda^T_j \nu^T_j$ of the two products cancelling), one finds that $\tilde{\mathcal{N}}$ is proportional to the Nekrasov factor,
\begin{equation}\label{eq:NFactor_sym_prop}
    \tilde{\mathcal{N}}_{\lambda \nu} (Q,q) = Q^{-\frac{1}{2}(|\lambda| + |\nu|)} \, q^{-\frac{1}{4}(\kappa_\lambda - \kappa_\nu)} \, \mathcal{N}_{\lambda \nu} (Q,q) \, .
\end{equation}
Each factor is antisymmetric under $Qq^x \to (Qq^x)^{-1}$, so the reflection symmetry \eqref{eq:symmetryNfactor} takes the simpler form
\begin{equation}\label{eq:symmetryNfactorSym}
    \tilde{\mathcal{N}}_{\lambda \nu} (Q,q) = (-1)^{|\lambda| + |\nu|} \, \tilde{\mathcal{N}}_{\nu \lambda } (Q^{-1},q) \, ,
\end{equation}
the residual sign counting the total number of boxes.
We also introduce the $\mathcal{R}$-factor
\begin{equation} 
    \mathcal{R}_{\lambda \nu} (Q,q) \doteq \frac{ \mathcal{N}_{\lambda^T \nu } (Q,q)}{\operatorname{PExp} \left( \frac{q Q}{(1-q)^2} \right)} = \prod\limits_{i,j=1}^{\infty} (1-Q q^{-\lambda_i - \nu_j +i+j-1})   \, . 
    \label{eq:RtoN}
\end{equation}
In particular, for trivial partitions, we have 
\begin{equation}
     \mathcal{R}_{00} (Q,q) = \operatorname{PExp} \left( - \frac{Qq}{(1-q)^2}\right) \, . \label{eq:R00}
\end{equation}
Note that the expression here is an infinite product, while the Nekrasov factor is a finite product. The "gauge theoretic" form of the Nekrasov factors, valid for $Q \neq 1$, is \footnote{We have used the formula, valid for any function $f$ not vanishing over the integers (see \cite[(57)]{Konishi:2003qq}): 
\begin{equation}
    \prod\limits_{i,j=1}^{\infty} \frac{f(\lambda_i - \nu_j +j-i)}{f(j-i)} = \prod\limits_{(i,j) \in \lambda} \frac{1}{f(\lambda_i + \nu_j^T -i-j+1)} \prod\limits_{(i,j) \in \nu} \frac{1}{f(-\lambda^T_j - \nu_i +i+j-1)} \, . 
\end{equation} }
\begin{equation}  
    \frac{1}{\mathcal{N}_{\lambda \nu} (Q,q)} = \prod\limits_{i,j=1}^{\infty} \frac{1-Q q^{\lambda_i - \nu_j +j-i}}{1-Q q^{j-i}}  = \prod\limits_{i,j=1}^{\infty} \frac{\sinh \left( \frac{\log Q}{2} + \frac{\log q}{2} (\lambda_i - \nu_j +j-i) \right)}{\sinh \left( \frac{\log Q}{2} + \frac{\log q}{2} (j-i) \right)}  \, . \label{eq:NekrasovFactorGaugeForm}
\end{equation}
When $Q \rightarrow 1$ we have  
\begin{equation}
    \frac{1}{\mathcal{N}_{\lambda \lambda} (1,q)}   = \prod\limits_{1 \leq i \neq j < \infty}\frac{\sinh \left(  \frac{\log q}{2} (\lambda_i - \lambda_j +j-i) \right)}{\sinh \left( \frac{\log q}{2} (j-i) \right)}  = (-1)^{|\lambda|} s_{\lambda} (q^{-\rho}) s_{\lambda^T} (q^{-\rho}) \, .  \label{eq:Nlambdalambda}
    \end{equation}

The Cauchy formulas \eqref{eq:CauchyIdentities} can be written in the form 
\begin{equation}
\begin{split}
        \sum\limits_{\lambda} Q^{|\lambda|} s_{\lambda / \mu_1} (q^{-\nu_1 - \rho}) s_{\lambda / \mu_2} (q^{-\nu_2 - \rho}) &=\mathcal{R}^{-1}_{\nu_1  \nu_2} (Q,q)  \sum\limits_{\eta} Q^{|\mu_1| + |\mu_2| - |\eta|} s_{\mu_2 / \eta} (q^{-\nu_1 - \rho})  s_{\mu_1 / \eta} (q^{-\nu_2 - \rho}) \\ 
        \sum\limits_{\lambda} Q^{|\lambda|} s_{\lambda / \mu_1^T} (q^{-\nu_1 - \rho}) s_{\lambda^T / \mu_2} (q^{-\nu_2 - \rho}) &= \mathcal{R}_{\nu_1  \nu_2} (-Q,q)  \sum\limits_{\eta} Q^{|\mu_1| + |\mu_2| - |\eta|} s_{\mu_2^T / \eta} (q^{-\nu_1- \rho})  s_{\mu_1 / \eta^T} (q^{-\nu_2 - \rho}) 
\end{split}
\label{eq:CauchyIdentitiesR}
\end{equation}
Here again, the left hand side is an infinite sum while the right hand side is a finite sum.

\section{Instanton counting}
\label{app:instantoncounting}

\subsection{Review of the localization computation}
\label{subsec: instanton partition function}
We review the $5d$ Nekrasov partition function \cite{nekrasov2003seiberg} of a $U(N)$ gauge theory with $N_f$ fundamental hypermultiplets on $\bC^2_{\e_1,\e_2}\times S^1$. Set $q_a = e^{R\e_a}$ ($a=1,2$), Coulomb fugacities $s_j = e^{R a_j}$ ($j=1,\dots,N$) and flavour fugacities $\mu_f = e^{R m_f}$ ($f=1,\dots,N_f$), and let $\fq$ be the instanton counting parameter. The partition function factorizes into a perturbative (one-loop) and an instanton part,
\begin{equation}\label{eq:ZNekUN}
    Z_{\rm{Nek}} = Z_{\rm{pert}}\, Z_{\rm{inst}} \, .
\end{equation}
The perturbative part is
\begin{equation}\label{eq:ZpertUN}
    Z_{\rm{pert}} = \prod_{\alpha \in \Delta^+}\, \prod_{a,b \geq 0}\frac{1}{(1-s_\alpha q_1^{a}q_2^{b})(1-s_\alpha q_1^{a+1}q_2^{b+1})}\, \prod_{w \in \mathbf{R}}\, \prod_{a,b \geq 0}\big(1-w(s)\, q_1^{a}q_2^{b}\big) \, ,
\end{equation}
with $s_\alpha = s_i/s_j$ the roots and $w(s) = \mu_f s_j$ the weights of the matter representation $\mathbf{R}$.

By equivariant localization, $Z_{\rm{inst}}$ is a sum over the $T = U(1)^N \times U(1)^{N_f}\times U(1)^2_\e$ fixed points of the instanton moduli space, labelled by $N$-tuples of Young diagrams $\blambda = (\lambda_1,\dots,\lambda_N)$,
\begin{equation}\label{eq:ZinstUN}
      \sum_{\blambda} \fq^{\,|\blambda|}\, \frac{\wedge\cE^\vee}{\wedge\cN^\vee}\bigg|_{\blambda}\, z_{CS} \, , \qquad |\blambda| = \sum_{j=1}^N |\lambda_j| \, ,
\end{equation}
each factor being the K-theoretic Euler class $\wedge(\oplus_i \bC_{a_i})^\vee = \prod_i (1-a_i^{-1})$ of the corresponding bundle. For a box $x \in \lambda$ we write $(x_1,x_2)$ for its (row, column) and $A_\lambda(x)$, $L_\lambda(x)$ for its arm and leg lengths.

The tangent space at the fixed point,
\begin{equation}\label{eq:normalbundle}
    \cN|_{\blambda} = \bigoplus_{i,j=1}^{N}\bigg[\bigoplus_{x\in\lambda_i}\bC_{\frac{s_i}{s_j} q_1^{-L_{\lambda_j}(x)} q_2^{A_{\lambda_i}(x)+1}} \oplus \bigoplus_{x\in\lambda_j}\bC_{\frac{s_i}{s_j}q_1^{L_{\lambda_i}(x)+1}q_2^{-A_{\lambda_j}(x)}}\bigg] \, ,
\end{equation}
gives the vector multiplet contribution
\begin{equation}\label{eq:zvecUN}
    z_{\rm{vec}} = \frac{1}{\wedge\cN^\vee} = \prod_{i,j=1}^N \bigg[ \prod_{x\in\lambda_i}\Big(1-\tfrac{s_j}{s_i}\,q_1^{L_{\lambda_j}(x)}q_2^{-A_{\lambda_i}(x)-1}\Big) \prod_{x\in\lambda_j}\Big(1-\tfrac{s_j}{s_i}\,q_1^{-L_{\lambda_i}(x)-1}q_2^{A_{\lambda_j}(x)}\Big) \bigg]^{-1} .
\end{equation}
Each fundamental hypermultiplet of fugacity $\mu_f$ contributes the matter bundle $\cE_f = \bigoplus_{j}\bigoplus_{x\in\lambda_j}\bC_{\mu_f s_j q_1^{x_1-1/2}q_2^{x_2-1/2}}$, i.e.
\begin{equation}\label{eq:zhyperUN}
    z_{{\rm{hyper}},f} = \wedge\cE_f^\vee = \prod_{j=1}^N \prod_{x\in\lambda_j}\Big(1-\tfrac{1}{\mu_f s_j}\,q_1^{1/2-x_1}q_2^{1/2-x_2}\Big) \, ,
\end{equation}
and a Chern--Simons term at level $k$ contributes the $k$-th power of the (dual) universal bundle,
\begin{equation}\label{eq:zCSUN}
    z_{CS} = \prod_{j=1}^N \prod_{x\in\lambda_j}\big(s_j^{-1}\,q_1^{1-x_1}q_2^{1-x_2}\big)^{k} \, .
\end{equation}
The full matter numerator over denominator is
\begin{equation}\label{instanton counting from hypers}
    \frac{\wedge\cE^\vee}{\wedge\cN^\vee}\bigg|_{\blambda} = z_{\rm{vec}}\prod_{f=1}^{N_f} z_{{\rm{hyper}},f} \, .
\end{equation}
At the unrefined point $q_1 = q_2^{-1} = q$ the box products \eqref{eq:zvecUN}--\eqref{eq:zCSUN} collapse onto the Nekrasov factors \eqref{eq:NFactor_finite_product} and the framing monomial $T_\lambda$, reproducing the vector, hypermultiplet and Chern--Simons contributions given below, see \eqref{eq:zvec}, \eqref{eq:zhyper} and \eqref{eq:zCS}.

The matter bundle $\cE = \oplus_i \cL_i$ enters \eqref{instanton counting from hypers} through its K-theoretic Euler class
\begin{equation}\label{vecotr bundle from instanton counting}
    \wedge\cE^\vee = \prod_i (1-\cL_i^{-1}) = \prod_i (1-e^{-x_i}) \, , \qquad \operatorname{ch}\cL_i = e^{x_i} \, ,
\end{equation}
which yields the numerator factors $1-x$ of \eqref{instanton counting from hypers}. K-theoretic instanton counting is the Witten index of the $\cN=4$ SQM with target $\cE \to \cM_{{\rm inst},k}$, and \eqref{vecotr bundle from instanton counting}, starting at $1$, assigns the vacuum zero charge under the flavour fugacity $e^{x_i}$. There is however an ambiguity on the charge of the vacuum, and we can equivalently make a more symmetric choice.\footnote{This is the standard $\det^{1/2}$ ambiguity of the K-theoretic index, seen already for a single complex fermion $\psi$ on $S^1$ (periodic spin structure) of $U(1)$ charge $1$: its zero mode $\{\psi_0,\bar\psi_0\}=1$ gives two ground states $\{|0\rangle,\bar\psi_0|0\rangle\}$, and the refined vacuum index depends on the charge of $|0\rangle$,
\begin{equation}\label{eq:fermionindex}
    \operatorname{tr}_{\cH_{\rm vac}}(-1)^F e^{a J} = \begin{cases}\ 1-e^{-a} & J_{|0\rangle}=0 \, ,\\[2pt]\ e^{a/2}-e^{-a/2} & J_{|0\rangle}=\tfrac12 \, ,\end{cases}
\end{equation}
reproducing \eqref{vecotr bundle from instanton counting} and \eqref{twisted vecotr bundle} respectively. The symmetric assignment $J_{|0\rangle}=\tfrac12$ makes the flavour symmetry manifest and defines the symmetrized contributions used throughout. }
The symmetric choice gives the vacuum charge $\tfrac12\sum_i x_i$, i.e.\ twists by $\det\cE^{1/2}$:
\begin{equation}\label{twisted vecotr bundle}
    \wedge\cE^\vee \ \longrightarrow\ \det\cE^{1/2} \otimes \,\wedge\cE^\vee = e^{\frac12\sum_i x_i}\prod_i(1-e^{-x_i}) = \prod_i\big(e^{x_i/2}-e^{-x_i/2}\big) \, .
\end{equation}
Every factor $1-A$ is thereby replaced by $A^{-1/2}-A^{1/2}$, i.e.\ each Nekrasov factor $\cN_{\lambda_j \emptyset}$ by the modified factor $\tilde{\cN}_{\lambda_j\emptyset}$ of \eqref{eq:NFactor_sym}. Mathematically, in the computation the twist replaces the Todd genus by the roof genus. In this paper, we always use the symmetric convention: 
\begin{equation}\label{eq:ZinstUNsym}
    Z_{\rm{inst}} = \sum_{\blambda} \fq^{\,|\blambda|}\, \frac{ \det\cE^{1/2}\, \otimes \wedge\cE^\vee}{\det\cN^{1/2}\, \otimes \wedge\cN^\vee}\bigg|_{\blambda}\, z_{CS} \, , \qquad |\blambda| = \sum_{j=1}^N |\lambda_j| \, ,
\end{equation}

\subsection{Instanton partition function contributions}

The outcome of the previous discussion is summarized as follows. At the unrefined point $q_1 = 1/q_2 = q$, the box products \eqref{eq:zvecUN}--\eqref{eq:zCSUN} of Appendix~\ref{subsec: instanton partition function} collapse onto the Nekrasov factors \eqref{eq:NFactor_finite_product} and the framing monomial $T_\lambda$. The vector, hypermultiplet and Chern--Simons contributions of a fixed point $\blambda = (\lambda_1 , \dots , \lambda_N)$ are
\begin{align}
    z_{\rm{vec}}(\mathbf{s} , q , \blambda ) &= \prod_{i,j=1}^N \widetilde{\mathcal{N}}^{-1}_{\lambda_i \lambda_j}\Big( \tfrac{s_j}{s_i} , q \Big) \, , \label{eq:zvec} \\
    z_{\rm{hyper}}(\mathbf{s} , \mu , q , \blambda ) &= \prod_{j=1}^N \widetilde{\mathcal{N}}_{\lambda_j \emptyset}\Big( \tfrac{1}{\mu s_j} , q \Big) \, , \label{eq:zhyper} \\
    z_{CS}^{(k)}(\mathbf{s} , q , \blambda) &= \prod_{j=1}^N  s_j^{-k |\lambda_j|} q^{\frac{k}{2} \kappa_{\lambda_j} }  \, , \label{eq:zCS}
\end{align}
from the tangent space \eqref{eq:normalbundle}, the matter bundle of fugacity $\mu$, and the level-$k$ universal bundle respectively.

Finally, for gauge group $SU(2)$ the perturbative contribution is
\begin{equation}\label{eq:ZpertSU2}
  Z_{\rm{pert}} = \operatorname{PExp}\left[\frac{q}{(1-q)^2} \left(2s^2 - \sum_{j=1}^{N_f} s ( \mu_j +  \mu_j^{-1})\right)\right] \, .
\end{equation}

\section{GV invariants from parallel branes and complex structure variations}
\label{app:GVcomplexStructureVariation}

In the paper, we see the topological string partition function from parallel branes play an important role in relating string theory degree of freedom (M-theory on generalized toric $CY3$ $X$) and field theory degree of freedom ($5d$ $\cN=1$ theory $\mathcal{T}_X$). For cases considered in this paper, this is argued in the main text both from $CPT$ symmetry from field theory perspective and local mirror symmetry from string perspective. In this appendix, we take a closer look on the $GV$ invariants of the parallel branes, which is modelled by the strip toric diagram.

\paragraph{Strip toric diagrams and parallel legs. } 
Let $Y$ be the local toric CY threefold defined by 
\begin{equation}
     \raisebox{-.5\height}{\scalebox{.7}{\begin{tikzpicture}
  \draw[mid arrow] (-2,-1.5)--(-1,-1);
  \draw[mid arrow] (-1,-1)--(0,0) node[midway, above left]{$Q_1$};
  \draw[mid arrow] (0,0)--(0,1) node[midway, right]{$\dots$};
  \draw[mid arrow] (0,1)--(-1,2) node[midway, left]{$Q_{N-1}$};
  \draw[mid arrow] (-1,2)--(-2,2.5);
  \draw[mid arrow] (-1,-1)--(2,-1);
  \draw[mid arrow] (0,0)--(2,0);
  \draw[mid arrow] (0,1)--(2,1);
  \draw[mid arrow] (-1,2)--(2,2);
\end{tikzpicture}}} 
\label{eq:strip0}
\end{equation}
This geometry is the local model for a stack of $N$ parallel external legs in a general toric Calabi--Yau threefold $X$. We denote by $C_a$, $a=1,\dots,N-1$, the compact curves corresponding to the $N-1$ internal edges. Geometrically, $Y$ is a one-parameter family of $A_{N-1}$ surfaces, and one might expect its GW invariants to vanish identically, since the ordinary GW invariants of the hyperk\"ahler surface $A_{N-1}$ vanish in every nonzero curve class.\footnote{The resolved $A_{N-1}$ surface is, as a real  manifold, the affine variety $xy=z^N+\alpha_2 z^{N-2}+\cdots+\alpha_N\subset\bC^3$ for a generic choice of complex structure (i.e.\ of the $\alpha_i$); affine varieties contain no compact curves, and since Gromov--Witten invariants do not depend on the choice of complex structure, they vanish in every nonzero class of $A_{N-1}$. Physically: the hyperk\"ahler structure gives $(4,4)$ worldsheet supersymmetry, which forbids nontrivial worldsheet instanton corrections.} This is not the case: $Y$ varies the complex structure of the fiber over the base, and, as we explain below, a holomorphic curve appears over each of finitely many isolated points of the base.\footnote{See \cite{bryan2007rootsystemsquantumcohomology} for the mathematical details at genus $0$; the reduced virtual fundamental classes and equivariant localization techniques used there should generalize directly to higher genus.}

The relevant complex structures of the $A_{N-1}$ fiber are described by the periods of its holomorphic symplectic form, using the standard identification $H^2(A_{N-1};\bC) \simeq \bC^{N-1} \simeq \mathfrak{h}_{\bC}$, the Cartan subalgebra of $\mathfrak{sl}_N$. The compact primitive classes $[C]\in H_2(A_{N-1};\bZ) \simeq \mathfrak{h}^\ast_{\bC}$ are identified with the roots of $\mathfrak{sl}_N$. The root $\alpha_{a,b}$ corresponds to the curve class $[C_{\alpha_{a,b}}]=[C_a]+[C_{a+1}]+\cdots+[C_b]$ with Kähler monomial $Q_a Q_{a+1}\cdots Q_b$.  Given a holomorphic symplectic form $\Omega_{A_{N-1}}$, the class $[C]$ is represented by a holomorphic curve only if
\begin{equation}
     \int_C \Omega_{A_{N-1}} = 0 \, ;
\end{equation}
each root therefore defines a hyperplane in the deformation space $\mathfrak h_{\bC}$. After choosing a positive chamber, the effective root classes are
\[
    \Delta^+_{A_{N-1}}
    =
    \{ \alpha_{a,b}=\alpha_a+\alpha_{a+1}+\cdots+\alpha_b
    \mid 1\leq a\leq b\leq N-1\},
\]
where $\alpha_a$ are the simple roots. 

Let
\begin{equation}
    f:\bC \longrightarrow H^2(A_{N-1};\bC)\simeq \mathfrak{h}_{\bC}
\end{equation}
be a generic affine-linear map. For each positive root $\alpha\in \Delta^+_{A_{N-1}}$, let $    H_\alpha = \{\xi\in \mathfrak{h}_{\bC}\mid \langle \alpha,\xi\rangle=0\} $ be the corresponding root hyperplane. For generic $f$, the inverse image $f^{-1}(H_\alpha)$ consists of a single point $t_\alpha\in \bC$, and these points are mutually distinct. Denote by $\Omega_{A_{N-1},t}$ the holomorphic symplectic form of the $A_{N-1}$ fiber with complex structure specified by $f(t)$. The holomorphic three-form on $Y$ is then
\begin{equation}
    \label{top holomorphic form for a strip}
    \Omega_Y = \mathrm{d} t \wedge \Omega_{A_{N-1},t}.
\end{equation}

The irreducible compact holomorphic curves in $Y$ are therefore labelled by the positive roots $\alpha\in \Delta^+_{A_{N-1}}$, with $C_\alpha$ sitting in the fiber over the isolated point $t_\alpha$. Since $C_\alpha$ exists only at this single point of the base and does not deform, $H^0(N_{C_\alpha/Y})=0$. By Grothendieck's splitting theorem, $N_{C_\alpha/Y}\simeq \cO_{\bP^1}(a)\oplus \cO_{\bP^1}(b)$ for some integers $a,b$. Adjunction, with $K_Y$ trivial and $K_{\bP^1}=\cO(-2)$, gives $a+b=-2$, while $H^0(N_{C_\alpha/Y})=0$ forces $a,b<0$. Hence each such curve has the local geometry of the resolved conifold: 
\begin{equation}
    N_{C_\alpha/Y} \simeq \cO_{\bP^1}(-1)\oplus \cO_{\bP^1}(-1) \, . 
    \label{eq:normalBundleRoots}
\end{equation}

As a result, the $GV$ invariants of $A_{N-1}$ strip toric diagram are the same as $n$ copies of conifolds, $n$ is the number of positive roots of $A_{N-1}$ Lie algebra, up to a overall $\pm$ sign. Since around the neighborhood of every resolved conifold of $A_{N-1}$ strip toric $CY$ is the same as the resolved conifold of $A_1$ strip toric $CY$, and the $GV$ invariant of $A_1$ strip toric $CY$ is the same as resolved conifold up to a $\pm$ sign.  To fix the sign ambiguity, we only need to compute the genus $0$, degree $1$ Gromov Witten\footnote{Here we are computing the Gromov Witten invariant, not the $GV$ invariant. Thanks to \eqref{GW=GV}, fix sign ambiguity of Gromov Witten is the same as fixing sign ambiguity of $GV$ invariants. } invariant of $A_1$ strip toric $CY$ $Y_{A_1}$. 
\paragraph{Fixing the sign ambiguity} First, theorem $1$ of \cite{maulik2013gromov} tells us:
\begin{equation}
   \int_{ [\cM_{0,0}(Y_{A_1},1)]^{vir}}1 = \int_{[\cM_{0,0}(A_1,1)]^{red}}1
\end{equation}
where the left-hand side the reduced virtual fundamental class. The reduced virtual fundamental class will be needed to have non trivial enumerative information of holomorphic symplectic manifold, e.g. the $A_1$ space here, see \cite{Lerche:2023wkj} for some other examples. Then Lemma 2.1 in \cite{maulik2009gromov} gives us:
\begin{equation}
    \int_{[\cM_{0,0}(A_1,1)]^{red}}1 = \chi_{\bP^1}(\cO(-2))\int_{\cM_{0,0}(\bP^1,1)}1 =\chi_{\bP^1}(\cO(-2)) =-1.
\end{equation}
Hence we have 
\begin{equation}
     \int_{ [\cM_{0,0}(Y_{A_1},1)]^{vir}}1 = -1 \, . 
\end{equation}
While it is known that the degree $1$ genus $0$ Gromov--Witten invariant of the resolved conifold is $1$.  Hence we see the Gromov Witten invariant and $GV$ invariant of $A_{1}$ strip toric $CY$ is the same as conifold with a $-$ sign. The same discussion can be generalized to $A_{N-1}$ strip toric $CY$.

The $-1$ sign relative to conifold can also be seen from topological vertex calculation, where the relative $-$ sign comes from canonical framing in $CS$ theory via open/closed string duality\cite{Aganagic:2003db,Aganagic:2002qg}.  The conifold partition function was computed in \eqref{eq:conifold}. For \eqref{eq:strip0}, equation \eqref{eq:strip} gives 
\begin{equation}
    \label{topological string amplitude for a strip}
    Z_{\text{top}} [Y] =\prod_{\alpha\in \Delta^+_{A_{N-1}}}
    \operatorname{PExp} \left(\frac{qQ_\alpha}{(1-q)^2}\right) \, . 
\end{equation}
Equivalently, in the GV expansion of Appendix \ref{app:notations},
\begin{equation}
    \GV_{0,[C_\alpha]}(Y)=-1,
    \qquad
    \GV_{r,[C_\alpha]}(Y)=0
    \quad \text{for } r>0,
    \qquad
    \alpha\in \Delta^+_{A_{N-1}} \, . 
\end{equation}
Again, we see the relative $-$ sign compare with conifold.

Thus the contribution associated with a stack of parallel external legs is a product of inverse conifold factors, one for each positive root of the $A_{N-1}$ root system determined by the compact curves of the strip. This is the factor denoted by $Z_{\parallel}$ in \eqref{eq:main}. The perturbative terms and constant-map contributions are not included in the reduced expression above; they belong to the normalization-dependent part $F_{\mathrm{pert}}$ discussed in Appendix \ref{app:notations}.

\section{Del Pezzo and generalized del Pezzo surfaces} 
\label{app:delPezzo}

\paragraph{Definitions.}
\begin{itemize}
    \item A \emph{del Pezzo} surface is a smooth complex projective surface $S$ with ample anticanonical class $-K_S$. These are $\bP^2$, $\bP^1\times\bP^1$, and the blow-ups $dP_n=\mathrm{Bl}_n\bP^2$ at $n\leq 8$ points of $\bP^2$ in general position; the \emph{degree} is $K_S^2=9-n$. 
    \item A \emph{generalized} (or weak) del Pezzo surface $S$ \cite{derenthal2008nef} is a smooth complex projective surface $S$ with \emph{nef and big} anticanonical class $-K_S$. The degree is $K_S^2$. For $K_S^2\leq7$ it is still a blow-up of $\bP^2$ at $n=9-K_S^2$ points in \emph{almost} general position (allowing infinitely near points, or several on a line or conic); at degree $8$ one has in addition $\bP^1\times\bP^1$ and $\mathbb F_2$.
\end{itemize}
The surface $dP_9$ can be defined as the blowup of $\mathbb{P}^2$ at the nine intersection points of two cubic curves. The two cubic curves provide a map $dP_9 \rightarrow \mathbb{P}^1$, which defines $dP_9$ as an elliptically fibered surface. Despite the notation, it is not a del Pezzo surface. 

\paragraph{Negative curves.}
For an irreducible curve $C$ on a smooth surface $S$, adjunction reads $C^2+K_S\cdot C=2g(C)-2$. When $-K_S$ is nef, $-K_S\cdot C\geq 0$ forces $C^2\geq 2g-2\geq -2$, with $C^2=-2$ only if $g=0$ and $-K\cdot C=0$. Hence:
\begin{itemize}
    \item del Pezzo ($-K_S$ ample): $-K_S\cdot C>0$ for every $C$, so $C^2\geq -1$; the only negative curves are the $(-1)$-curves (lines).
    \item generalized del Pezzo ($-K_S$ nef): additionally there may be $(-2)$-curves, characterized by $C^2=-2$, $-K_S\cdot C=0$. These are smooth rational and are exactly the curves contracted by the anticanonical morphism $S\to\overline{S}$, which develops the corresponding du Val (ADE) singularities.
\end{itemize}

\paragraph{Homology and the $E_{n \geq 3}$ lattice.}
A (generalized) del Pezzo $S$ of degree $d \leq 6$ ($n \geq 3$) is rational with $h^{1,0}=h^{2,0}=0$, so
\begin{equation}
    H_2(S;\bZ)\;\cong\;H^2(S;\bZ)\;\cong\;\operatorname{Pic}(S)\;\cong\;\bZ^{1,n},
\end{equation}
the odd unimodular lattice with basis $H,e_1,\dots,e_n$ ($H$ is the pullback of the line class, $e_i$ the exceptional classes corresponding to the blown up points) and intersection form
\begin{equation}
    H^2=1,\qquad e_i\cdot e_j=-\delta_{ij},\qquad H\cdot e_i=0 \, . 
\end{equation} 
In this basis, 
\begin{equation}
    K_S=-3H+\sum_{i=1}^{n}e_i.
\end{equation}
The orthogonal complement $K^\perp=\Gamma_{E_n}\subset H_2(S;\bZ)$ is the negative-definite $E_n$ root lattice, with simple roots
\begin{equation}
    \alpha_0=H-e_1-e_2-e_3,\qquad \alpha_i=e_i-e_{i+1}\quad(1\leq i\leq n-1),
\end{equation}
and Dynkin diagram $E_n$. Roots are the classes $\alpha$ with $\alpha^2=-2$, $\alpha\cdot K=0$; lines are the classes $\ell$ with $\ell^2=\ell\cdot K=-1$. The Weyl group $W(E_n)$ acts by lattice automorphisms fixing $K$; this is the symmetry that organizes the GV invariants of $dP_n$ into Weyl orbits in Section \ref{sec:localMirrorSymmetry}. For a generalized del Pezzo surface, the classes of the $(-2)$-curves span a \emph{sub}-root-system $R\subseteq E_n$, of ADE type equal to the du Val singularities of $\overline{S}$.

\paragraph{Local Calabi-Yau and curve classes.}
The threefold of interest is the total space $X=K_S=\operatorname{Tot}(K_S\to S)$ of the canonical bundle, a non-compact Calabi-Yau. The zero section $S\hookrightarrow X$ induces an isomorphism on compact $2$-cycles,
\begin{equation}
    H_2(X;\bZ)\;\cong\;H_2(S;\bZ)\;\cong\;\bZ^{1,n},
\end{equation}
so the Gopakumar-Vafa and Gromov-Witten invariants are labelled by $\beta\in\operatorname{Pic}(S)$, with anticanonical degree $d(\beta)=-K_S\cdot\beta$. The cone of effective curves $\NEbar(S)$ is generated by the $(-1)$- and $(-2)$-curves. By the above, the $(-2)$-curves are precisely the irreducible curves of degree $d(\beta)=0$: in the brane web these are the compact curves stretched between parallel external legs, whose contribution decouples from the $5d$ SCFT. A genuine del Pezzo is Fano, has no such class, and correspondingly $Z_{\text{top}}(K_{dP_n})$ carries no parallel-brane factor.

\paragraph{Toric generalized del Pezzo.}
A smooth complete toric surface $S$ defined by its fan $\Sigma$ is a generalized del Pezzo iff $-K_S=\sum_i D_i$ is nef and big, equivalently iff its ray generators $v_1,\dots,v_r$ (cyclically ordered) are precisely the boundary lattice points of a \emph{reflexive} polygon $P=\operatorname{conv}(v_i)$, with $\mathbf 0$ the unique interior point, consecutive pairs forming a $\bZ$-basis \cite{cox2024toric}. The wall relation is 
\begin{equation}
    v_{i-1}+v_{i+1} + D_i^2 \, v_i = 0 \, . 
\end{equation}
Hence 
\begin{equation}
    D_i^2=-2\;\Longleftrightarrow\;v_{i-1}+v_{i+1}=2v_i\;\Longleftrightarrow\;v_i\ \text{is interior to an edge of }P.
\end{equation}
Thus the $(-2)$-curves are the non-vertex boundary points of $P$, dual to pairs of external parallel legs of the web. The degree is
\begin{equation}
     K_S^2=12-\#(\partial P\cap N) 
\end{equation}
with $N$ the usual toric lattice. 
For a reflexive polygon $\#(\partial P\cap N)+\#(\partial P^\ast\cap M)=12$, and any polygon has at least three boundary points, so $K_S^2 \geq 3$: \emph{there is no toric generalized del Pezzo of degree $<3$ (i.e. $n>6$)}, with no genericity assumption needed. 

The toric generalized del Pezzos of degrees $3\leq K_S^2 \leq 5$, in their maximally degenerate (largest-$R$) realizations, are listed in Table~\ref{tab:toricGdP}; each is obtained directly from its reflexive polygon, and the number of $(-2)$-curves matches the degree-$0$ (parallel-brane) GV invariants reported in Section \ref{sec:topologicalVertexEn}. Explicitly: $GdP_5$ is the resolution of the $4$-nodal degree-$4$ del Pezzo (a $(2,2)$ complete intersection in $\bP^4$, not a quartic surface); $GdP_6$ is the resolution of the toric cubic $xyz=w^3$, whose three edges each carry two consecutive $(-2)$-curves, giving $3A_2$. The full list of $16$ toric generalized del Pezzos (one per reflexive polygon) can be found in Table \ref{tab:GTPs_for_GdPs}.

\begin{table}[t]
    \centering
    \begin{tabular}{c|c|c|c|c|c}
        \hline
        $d$ & $n$ & $E_n=K^\perp$ & $R$ & $\#(-2)$ & Reflexive polygon \\ \hline
        $5$ & $4$ & $A_4$ & $2A_1$ & $2$ & \raisebox{-.5\height}{ \scalebox{.8}{  
\begin{tikzpicture}[x=.5cm,y=.5cm] 
\draw[gridline] (0,-0.5)--(0,2.5); 
\draw[gridline] (1,-0.5)--(1,2.5); 
\draw[gridline] (2,-0.5)--(2,2.5); 
\draw[gridline] (-0.5,0)--(2.5,0); 
\draw[gridline] (-0.5,1)--(2.5,1); 
\draw[gridline] (-0.5,2)--(2.5,2); 
\draw[ligne] (0,1)--(0,0); 
\draw[ligne] (0,0)--(1,0); 
\draw[ligne] (1,0)--(2,0); 
\draw[ligne] (2,0)--(2,1); 
\draw[ligne] (2,1)--(2,2); 
\draw[ligne] (2,2)--(1,2); 
\draw[ligne] (1,2)--(0,1); 
\node[bd] at (0,0) {}; 
\node[bd] at (1,0) {}; 
\node[bd] at (2,0) {}; 
\node[bd] at (2,1) {}; 
\node[bd] at (2,2) {}; 
\node[bd] at (1,2) {}; 
\node[bd] at (0,1) {}; 
\end{tikzpicture}}} \\
        $4$ & $5$ & $D_5$ & $4A_1$ & $4$ & \raisebox{-.5\height}{ \scalebox{.8}{  
\begin{tikzpicture}[x=.5cm,y=.5cm] 
\draw[gridline] (0,-0.5)--(0,2.5); 
\draw[gridline] (1,-0.5)--(1,2.5); 
\draw[gridline] (2,-0.5)--(2,2.5); 
\draw[gridline] (-0.5,0)--(2.5,0); 
\draw[gridline] (-0.5,1)--(2.5,1); 
\draw[gridline] (-0.5,2)--(2.5,2); 
\draw[ligne] (0,1)--(0,0); 
\draw[ligne] (0,0)--(1,0); 
\draw[ligne] (1,0)--(2,0); 
\draw[ligne] (2,0)--(2,1); 
\draw[ligne] (2,1)--(2,2); 
\draw[ligne] (2,2)--(1,2); 
\draw[ligne] (1,2)--(0,2); 
\draw[ligne] (0,2)--(0,1); 
\node[bd] at (0,0) {}; 
\node[bd] at (1,0) {}; 
\node[bd] at (2,0) {}; 
\node[bd] at (2,1) {}; 
\node[bd] at (2,2) {}; 
\node[bd] at (1,2) {}; 
\node[bd] at (0,2) {}; 
\node[bd] at (0,1) {}; 
\end{tikzpicture}}} \\
        $3$ & $6$ & $E_6$ & $3A_2$ & $6$ &   \raisebox{-.5\height}{\scalebox{.8}{  
\begin{tikzpicture}[x=.5cm,y=.5cm] 
\draw[gridline] (0,-0.5)--(0,3.5); 
\draw[gridline] (1,-0.5)--(1,3.5); 
\draw[gridline] (2,-0.5)--(2,3.5); 
\draw[gridline] (3,-0.5)--(3,3.5); 
\draw[gridline] (-0.5,0)--(3.5,0); 
\draw[gridline] (-0.5,1)--(3.5,1); 
\draw[gridline] (-0.5,2)--(3.5,2); 
\draw[gridline] (-0.5,3)--(3.5,3); 
\draw[ligne] (0,1)--(0,0); 
\draw[ligne] (0,0)--(1,0); 
\draw[ligne] (1,0)--(2,0); 
\draw[ligne] (2,0)--(3,0); 
\draw[ligne] (3,0)--(2,1); 
\draw[ligne] (2,1)--(1,2); 
\draw[ligne] (1,2)--(0,3); 
\draw[ligne] (0,3)--(0,2); 
\draw[ligne] (0,2)--(0,1); 
\node[bd] at (0,0) {}; 
\node[bd] at (1,0) {}; 
\node[bd] at (2,0) {}; 
\node[bd] at (3,0) {}; 
\node[bd] at (2,1) {}; 
\node[bd] at (1,2) {}; 
\node[bd] at (0,3) {}; 
\node[bd] at (0,2) {}; 
\node[bd] at (0,1) {}; 
\end{tikzpicture}}} \\
        \hline
    \end{tabular}
    \caption{Maximally degenerate toric generalized del Pezzo surfaces of degree $3\leq d\leq 5$. The $(-2)$-curve system $R$ is a sub-root-system of $E_n$; contracting these curves yields the corresponding du Val singularities, and removing their ($d(\beta)=0$) contributions from $Z_{\text{top}}(K_{GdP_n})$ reproduces $Z_{E_n\,\mathrm{SCFT}}=Z_{\text{top}}(K_{dP_n})$.}
    \label{tab:toricGdP}
\end{table}

\begin{table}[t]
    \centering
    \begin{tabular}{c|c|c|c}
       $n$ & Toric $dP_n$ & Toric $GdP_n$ & GTP $GdP_n$ \\ \hline
       0 & \raisebox{-.5\height}{ \scalebox{.8}{  \begin{tikzpicture}[x=.5cm,y=.5cm]  
       \foreach \n in {0,1,2} {\draw[gridline] (-.5,\n) -- (2.5,\n); \draw[gridline] (\n,-.5) -- (\n,2.5);};
       \draw[ligne] (1,0) -- (2,2) -- (0,1) -- (1,0);
       \node[bd] at (1,0) {};
       \node[bd] at (0,1) {};
       \node[bd] at (2,2) {};
       \end{tikzpicture}}}& & \\\hline
        1 &\raisebox{-.5\height}{ \scalebox{.8}{  
\begin{tikzpicture}[x=.5cm,y=.5cm] 
\draw[gridline] (0,-0.5)--(0,2.5); 
\draw[gridline] (1,-0.5)--(1,2.5); 
\draw[gridline] (2,-0.5)--(2,2.5); 
\draw[gridline] (-0.5,0)--(2.5,0); 
\draw[gridline] (-0.5,1)--(2.5,1); 
\draw[gridline] (-0.5,2)--(2.5,2); 
\draw[ligne] (0,1)--(1,0); 
\draw[ligne] (1,0)--(2,1); 
\draw[ligne] (2,1)--(1,2); 
\draw[ligne] (1,2)--(0,1); 
\node[bd] at (1,0) {}; 
\node[bd] at (2,1) {}; 
\node[bd] at (1,2) {}; 
\node[bd] at (0,1) {}; 
\end{tikzpicture}}},\raisebox{-.5\height}{ \scalebox{.8}{  
\begin{tikzpicture}[x=.5cm,y=.5cm] 
\draw[gridline] (0,-0.5)--(0,2.5); 
\draw[gridline] (1,-0.5)--(1,2.5); 
\draw[gridline] (2,-0.5)--(2,2.5); 
\draw[gridline] (-0.5,0)--(2.5,0); 
\draw[gridline] (-0.5,1)--(2.5,1); 
\draw[gridline] (-0.5,2)--(2.5,2); 
\draw[ligne] (0,1)--(1,0); 
\draw[ligne] (1,0)--(2,2); 
\draw[ligne] (2,2)--(1,2); 
\draw[ligne] (1,2)--(0,1); 
\node[bd] at (1,0) {}; 
\node[bd] at (2,2) {}; 
\node[bd] at (1,2) {}; 
\node[bd] at (0,1) {}; 
\end{tikzpicture}}} &\raisebox{-.5\height}{ \scalebox{.8}{  
\begin{tikzpicture}[x=.5cm,y=.5cm] 
\draw[gridline] (0,-0.5)--(0,2.5); 
\draw[gridline] (1,-0.5)--(1,2.5); 
\draw[gridline] (2,-0.5)--(2,2.5); 
\draw[gridline] (-0.5,0)--(2.5,0); 
\draw[gridline] (-0.5,1)--(2.5,1); 
\draw[gridline] (-0.5,2)--(2.5,2); 
\draw[ligne] (0,2)--(1,0); 
\draw[ligne] (1,0)--(2,2); 
\draw[ligne] (2,2)--(1,2); 
\draw[ligne] (1,2)--(0,2); 
\node[bd] at (1,0) {}; 
\node[bd] at (2,2) {}; 
\node[bd] at (1,2) {}; 
\node[bd] at (0,2) {}; 
\end{tikzpicture}}} & \raisebox{-.5\height}{ \scalebox{.8}{  
\begin{tikzpicture}[x=.5cm,y=.5cm] 
\draw[gridline] (0,-0.5)--(0,3.5); 
\draw[gridline] (1,-0.5)--(1,3.5); 
\draw[gridline] (2,-0.5)--(2,3.5); 
\draw[gridline] (-0.5,0)--(2.5,0); 
\draw[gridline] (-0.5,1)--(2.5,1); 
\draw[gridline] (-0.5,2)--(2.5,2); 
\draw[gridline] (-0.5,3)--(2.5,3); 
\draw[ligne] (0,2)--(1,0); 
\draw[ligne] (1,0)--(2,3); 
\draw[ligne] (0,2) -- (0,3) -- (2,3);
\draw[ligne] (2,3) -- (1,2)--(0,2); 
\node[bd] at (1,0) {};
\node[bd] at (0,3) {};
\node[wd] at (1,3) {};
\node[bd] at (2,3) {}; 
\node[bd] at (0,2) {}; 
\end{tikzpicture}}}, $ \,\ldots$\\\hline
        2 & \raisebox{-.5\height}{ \scalebox{.8}{  
\begin{tikzpicture}[x=.5cm,y=.5cm] 
\draw[gridline] (0,-0.5)--(0,2.5); 
\draw[gridline] (1,-0.5)--(1,2.5); 
\draw[gridline] (2,-0.5)--(2,2.5); 
\draw[gridline] (-0.5,0)--(2.5,0); 
\draw[gridline] (-0.5,1)--(2.5,1); 
\draw[gridline] (-0.5,2)--(2.5,2); 
\draw[ligne] (0,1)--(1,0); 
\draw[ligne] (1,0)--(2,1); 
\draw[ligne] (2,1)--(2,2); 
\draw[ligne] (2,2)--(1,2); 
\draw[ligne] (1,2)--(0,1); 
\node[bd] at (1,0) {}; 
\node[bd] at (2,1) {}; 
\node[bd] at (2,2) {}; 
\node[bd] at (1,2) {}; 
\node[bd] at (0,1) {}; 
\end{tikzpicture}}}&\raisebox{-.5\height}{ \scalebox{.8}{  
\begin{tikzpicture}[x=.5cm,y=.5cm] 
\draw[gridline] (0,-0.5)--(0,2.5); 
\draw[gridline] (1,-0.5)--(1,2.5); 
\draw[gridline] (2,-0.5)--(2,2.5); 
\draw[gridline] (-0.5,0)--(2.5,0); 
\draw[gridline] (-0.5,1)--(2.5,1); 
\draw[gridline] (-0.5,2)--(2.5,2); 
\draw[ligne] (0,2)--(1,0); 
\draw[ligne] (1,0)--(2,1); 
\draw[ligne] (2,1)--(2,2); 
\draw[ligne] (2,2)--(1,2); 
\draw[ligne] (1,2)--(0,2); 
\node[bd] at (1,0) {}; 
\node[bd] at (2,1) {}; 
\node[bd] at (2,2) {}; 
\node[bd] at (1,2) {}; 
\node[bd] at (0,2) {}; 
\end{tikzpicture}}} & \raisebox{-.5\height}{ \scalebox{.8}{  
\begin{tikzpicture}[x=.5cm,y=.5cm] 
\draw[gridline] (0,-0.5)--(0,3.5); 
\draw[gridline] (1,-0.5)--(1,3.5); 
\draw[gridline] (2,-0.5)--(2,3.5); 
\draw[gridline] (-0.5,0)--(2.5,0); 
\draw[gridline] (-0.5,1)--(2.5,1); 
\draw[gridline] (-0.5,2)--(2.5,2); 
\draw[gridline] (-0.5,3)--(2.5,3); 
\draw[ligne] (0,2)-- (0,1) -- (1,0); 
\draw[ligne] (1,0)--(2,3); 
\draw[ligne] (0,2) -- (0,3) -- (2,3);
\draw[ligne] (2,3) -- (1,2)--(0,2); 
\node[bd] at (1,0) {};
\node[bd] at (0,3) {};
\node[wd] at (1,3) {};
\node[bd] at (2,3) {}; 
\node[bd] at (0,2) {}; 
\node[bd] at (0,1) {};
\end{tikzpicture}}},\raisebox{-.5\height}{ \scalebox{.8}{  
\begin{tikzpicture}[x=.5cm,y=.5cm] 
\draw[gridline] (0,-0.5)--(0,3.5); 
\draw[gridline] (1,-0.5)--(1,3.5); 
\draw[gridline] (2,-0.5)--(2,3.5); 
\draw[gridline] (-0.5,0)--(2.5,0); 
\draw[gridline] (-0.5,1)--(2.5,1); 
\draw[gridline] (-0.5,2)--(2.5,2); 
\draw[gridline] (-0.5,3)--(2.5,3); 
\draw[ligne] (0,2)--(1,0); 
\draw[ligne] (1,0)-- (2,2) -- (2,3); 
\draw[ligne] (0,2) -- (0,3) -- (2,3);
\draw[ligne] (2,3) -- (1,2)--(0,2); 
\node[bd] at (1,0) {};
\node[bd] at (0,3) {};
\node[wd] at (1,3) {};
\node[bd] at (2,3) {}; 
\node[bd] at (0,2) {}; 
\node[bd] at (2,2) {};
\end{tikzpicture}}}, $ \,\ldots$ \\\hline
        3 & \raisebox{-.5\height}{ \scalebox{.8}{  
\begin{tikzpicture}[x=.5cm,y=.5cm] 
\draw[gridline] (0,-0.5)--(0,2.5); 
\draw[gridline] (1,-0.5)--(1,2.5); 
\draw[gridline] (2,-0.5)--(2,2.5); 
\draw[gridline] (-0.5,0)--(2.5,0); 
\draw[gridline] (-0.5,1)--(2.5,1); 
\draw[gridline] (-0.5,2)--(2.5,2); 
\draw[ligne] (0,1)--(0,0); 
\draw[ligne] (0,0)--(1,0); 
\draw[ligne] (1,0)--(2,1); 
\draw[ligne] (2,1)--(2,2); 
\draw[ligne] (2,2)--(1,2); 
\draw[ligne] (1,2)--(0,1); 
\node[bd] at (0,0) {}; 
\node[bd] at (1,0) {}; 
\node[bd] at (2,1) {}; 
\node[bd] at (2,2) {}; 
\node[bd] at (1,2) {}; 
\node[bd] at (0,1) {}; 
\end{tikzpicture}}} & \raisebox{-.5\height}{ \scalebox{.8}{  
\begin{tikzpicture}[x=.5cm,y=.5cm] 
\draw[gridline] (0,-0.5)--(0,2.5); 
\draw[gridline] (1,-0.5)--(1,2.5); 
\draw[gridline] (2,-0.5)--(2,2.5); 
\draw[gridline] (-0.5,0)--(2.5,0); 
\draw[gridline] (-0.5,1)--(2.5,1); 
\draw[gridline] (-0.5,2)--(2.5,2); 
\draw[ligne] (0,2)--(0,1); 
\draw[ligne] (0,1)--(1,0); 
\draw[ligne] (1,0)--(2,1); 
\draw[ligne] (2,1)--(2,2); 
\draw[ligne] (2,2)--(1,2); 
\draw[ligne] (1,2)--(0,2); 
\node[bd] at (0,2) {}; 
\node[bd] at (1,0) {}; 
\node[bd] at (2,1) {}; 
\node[bd] at (2,2) {}; 
\node[bd] at (1,2) {}; 
\node[bd] at (0,1) {}; 
\end{tikzpicture}}},\raisebox{-.5\height}{ \scalebox{.8}{  
\begin{tikzpicture}[x=.5cm,y=.5cm] 
\draw[gridline] (0,-0.5)--(0,2.5); 
\draw[gridline] (1,-0.5)--(1,2.5); 
\draw[gridline] (2,-0.5)--(2,2.5); 
\draw[gridline] (-0.5,0)--(2.5,0); 
\draw[gridline] (-0.5,1)--(2.5,1); 
\draw[gridline] (-0.5,2)--(2.5,2); 
\draw[ligne] (0,2)--(0,1); 
\draw[ligne] (0,1)--(2,0); 
\draw[ligne] (2,0)--(2,1); 
\draw[ligne] (2,1)--(2,2); 
\draw[ligne] (2,2)--(1,2); 
\draw[ligne] (1,2)--(0,2); 
\node[bd] at (0,2) {}; 
\node[bd] at (2,0) {}; 
\node[bd] at (2,1) {}; 
\node[bd] at (2,2) {}; 
\node[bd] at (1,2) {}; 
\node[bd] at (0,1) {}; 
\end{tikzpicture}}},\raisebox{-.5\height}{ \scalebox{.8}{  
\begin{tikzpicture}[x=.5cm,y=.5cm] 
\draw[gridline] (0,-1.5)--(0,2.5); 
\draw[gridline] (1,-1.5)--(1,2.5); 
\draw[gridline] (2,-1.5)--(2,2.5); 
\draw[gridline] (-0.5,-1)--(2.5,-1); 
\draw[gridline] (-0.5,0)--(2.5,0); 
\draw[gridline] (-0.5,1)--(2.5,1); 
\draw[gridline] (-0.5,2)--(2.5,2); 
\draw[ligne] (0,2)-- (2,-1) -- (2,0); 
\draw[ligne] (2,0)--(2,1); 
\draw[ligne] (2,1)--(2,2); 
\draw[ligne] (2,2)--(1,2); 
\draw[ligne] (1,2)--(0,2); 
\node[bd] at (2,0) {}; 
\node[bd] at (2,1) {}; 
\node[bd] at (2,2) {}; 
\node[bd] at (1,2) {}; 
\node[bd] at (0,2) {}; 
\node[bd] at (2,-1) {};
\end{tikzpicture}}}&\raisebox{-.5\height}{ \scalebox{.8}{  
\begin{tikzpicture}[x=.5cm,y=.5cm] 
\draw[gridline] (0,-0.5)--(0,3.5); 
\draw[gridline] (1,-0.5)--(1,3.5); 
\draw[gridline] (2,-0.5)--(2,3.5); 
\draw[gridline] (-0.5,0)--(2.5,0); 
\draw[gridline] (-0.5,1)--(2.5,1); 
\draw[gridline] (-0.5,2)--(2.5,2); 
\draw[gridline] (-0.5,3)--(2.5,3); 
\draw[ligne] (0,2)-- (0,1) -- (1,0); 
\draw[ligne] (1,0)-- (2,2) -- (2,3); 
\draw[ligne] (0,2) -- (0,3) -- (2,3);
\draw[ligne] (2,3) -- (1,2)--(0,2); 
\node[bd] at (1,0) {};
\node[bd] at (0,3) {};
\node[wd] at (1,3) {};
\node[bd] at (2,3) {}; 
\node[bd] at (0,2) {}; 
\node[bd] at (2,2) {};
\node[bd] at (0,1) {};
\end{tikzpicture}}},\raisebox{-.5\height}{ \scalebox{.8}{  
\begin{tikzpicture}[x=.5cm,y=.5cm] 
\draw[gridline] (0,-0.5)--(0,3.5); 
\draw[gridline] (1,-0.5)--(1,3.5); 
\draw[gridline] (2,-0.5)--(2,3.5); 
\draw[gridline] (-0.5,0)--(2.5,0); 
\draw[gridline] (-0.5,1)--(2.5,1); 
\draw[gridline] (-0.5,2)--(2.5,2); 
\draw[gridline] (-0.5,3)--(2.5,3); 
\draw[ligne] (0,2)-- (0,1) -- (0,0) -- (1,0); 
\draw[ligne] (1,0)-- (2,3); 
\draw[ligne] (0,2) -- (0,3) -- (2,3);
\draw[ligne] (2,3) -- (1,2)--(0,2); 
\node[bd] at (1,0) {};
\node[bd] at (0,3) {};
\node[wd] at (1,3) {};
\node[bd] at (2,3) {}; 
\node[bd] at (0,2) {}; 
\node[bd] at (0,0) {};
\node[bd] at (0,1) {};
\end{tikzpicture}}}, $ \,\ldots$ \\\hline
        4 & &\raisebox{-.5\height}{ \scalebox{.8}{  
\begin{tikzpicture}[x=.5cm,y=.5cm] 
\draw[gridline] (0,-0.5)--(0,2.5); 
\draw[gridline] (1,-0.5)--(1,2.5); 
\draw[gridline] (2,-0.5)--(2,2.5); 
\draw[gridline] (-0.5,0)--(2.5,0); 
\draw[gridline] (-0.5,1)--(2.5,1); 
\draw[gridline] (-0.5,2)--(2.5,2); 
\draw[ligne] (0,2)--(0,1); 
\draw[ligne] (0,1)--(1,0); 
\draw[ligne] (1,0)--(2,0); 
\draw[ligne] (2,0)--(2,1); 
\draw[ligne] (2,1)--(2,2); 
\draw[ligne] (2,2)--(1,2); 
\draw[ligne] (1,2)--(0,2); 
\node[bd] at (0,1) {}; 
\node[bd] at (1,0) {}; 
\node[bd] at (2,0) {}; 
\node[bd] at (2,1) {}; 
\node[bd] at (2,2) {}; 
\node[bd] at (1,2) {}; 
\node[bd] at (0,2) {}; 
\end{tikzpicture}}},\raisebox{-.5\height}{ \scalebox{.8}{  
\begin{tikzpicture}[x=.5cm,y=.5cm] 
\draw[gridline] (0,-1.5)--(0,2.5); 
\draw[gridline] (1,-1.5)--(1,2.5); 
\draw[gridline] (2,-1.5)--(2,2.5); 
\draw[gridline] (-0.5,-1)--(2.5,-1); 
\draw[gridline] (-0.5,0)--(2.5,0); 
\draw[gridline] (-0.5,1)--(2.5,1); 
\draw[gridline] (-0.5,2)--(2.5,2); 
\draw[ligne] (0,1)--(1,0); 
\draw[ligne] (1,0)-- (2,-1) -- (2,0); 
\draw[ligne] (2,0)--(2,1); 
\draw[ligne] (2,1)--(2,2); 
\draw[ligne] (2,2)--(1,2); 
\draw[ligne] (1,2)--(0,1); 
\node[bd] at (0,1) {}; 
\node[bd] at (1,0) {}; 
\node[bd] at (2,1) {}; 
\node[bd] at (2,2) {}; 
\node[bd] at (1,2) {}; 
\node[bd] at (2,0) {}; 
\node[bd] at (2,-1) {};
\end{tikzpicture}}}&  \raisebox{-.5\height}{ \scalebox{.8}{  
\begin{tikzpicture}[x=.5cm,y=.5cm] 
\draw[gridline] (0,-0.5)--(0,3.5); 
\draw[gridline] (1,-0.5)--(1,3.5); 
\draw[gridline] (2,-0.5)--(2,3.5); 
\draw[gridline] (-0.5,0)--(2.5,0); 
\draw[gridline] (-0.5,1)--(2.5,1); 
\draw[gridline] (-0.5,2)--(2.5,2); 
\draw[gridline] (-0.5,3)--(2.5,3); 
\draw[ligne] (0,1)--(0,0); 
\draw[ligne] (0,0)--(2,1); 
\draw[ligne] (2,1)--(2,2); 
\draw[ligne] (2,2)--(2,3); 
\draw[ligne] (2,3)--(1,3); 
\draw[ligne] (1,3)--(0,3); 
\draw[ligne] (0,3)--(0,2); 
\draw[ligne] (0,2)--(0,1); 
\draw[ligne] (0,2) -- (1,2) -- (2,3);
\node[bd] at (0,0) {}; 
\node[bd] at (2,1) {}; 
\node[bd] at (2,2) {}; 
\node[bd] at (2,3) {}; 
\node[wd] at (1,3) {}; 
\node[bd] at (0,3) {}; 
\node[bd] at (0,2) {}; 
\node[bd] at (0,1) {}; 
\end{tikzpicture}}}, \raisebox{-.5\height}{ \scalebox{.8}{  
\begin{tikzpicture}[x=.5cm,y=.5cm] 
\draw[gridline] (0,-0.5)--(0,3.5); 
\draw[gridline] (1,-0.5)--(1,3.5); 
\draw[gridline] (2,-0.5)--(2,3.5); 
\draw[gridline] (-0.5,0)--(2.5,0); 
\draw[gridline] (-0.5,1)--(2.5,1); 
\draw[gridline] (-0.5,2)--(2.5,2); 
\draw[gridline] (-0.5,3)--(2.5,3); 
\draw[ligne] (0,1)--(1,0); 
\draw[ligne] (1,0)--(2,1); 
\draw[ligne] (2,1)--(2,2); 
\draw[ligne] (2,2)--(2,3); 
\draw[ligne] (2,3)--(1,3); 
\draw[ligne] (1,3)--(0,3); 
\draw[ligne] (0,3)--(0,2); 
\draw[ligne] (0,2)--(0,1); 
\draw[ligne] (0,2) -- (1,2) -- (2,3);
\node[bd] at (1,0) {}; 
\node[bd] at (2,1) {}; 
\node[bd] at (2,2) {}; 
\node[bd] at (2,3) {}; 
\node[wd] at (1,3) {}; 
\node[bd] at (0,3) {}; 
\node[bd] at (0,2) {}; 
\node[bd] at (0,1) {}; 
\end{tikzpicture}}}, 
\raisebox{-.5\height}{ \scalebox{.8}{  
\begin{tikzpicture}[x=.5cm,y=.5cm] 
\draw[gridline] (0,-1.5)--(0,3.5); 
\draw[gridline] (1,-1.5)--(1,3.5); 
\draw[gridline] (2,-1.5)--(2,3.5); 
\draw[gridline] (-0.5,-1)--(2.5,-1); 
\draw[gridline] (-0.5,0)--(2.5,0); 
\draw[gridline] (-0.5,1)--(2.5,1); 
\draw[gridline] (-0.5,2)--(2.5,2); 
\draw[gridline] (-0.5,3)--(2.5,3); 
\draw[ligne] (0,1)--(0,-1); 
\draw[ligne] (0,-1)--(2,2); 
\draw[ligne] (2,2)--(2,3); 
\draw[ligne] (2,3)--(1,3); 
\draw[ligne] (1,3)--(0,3); 
\draw[ligne] (0,3)--(0,2); 
\draw[ligne] (0,2)--(0,1); 
\draw[ligne] (0,2) -- (1,2) -- (2,3);
\node[bd] at (0,0) {}; 
\node[bd] at (0,-1) {}; 
\node[bd] at (2,2) {}; 
\node[bd] at (2,3) {}; 
\node[wd] at (1,3) {}; 
\node[bd] at (0,3) {}; 
\node[bd] at (0,2) {}; 
\node[bd] at (0,1) {}; 
\end{tikzpicture}}},$ \,\ldots$\\\hline
        5 & &\raisebox{-.5\height}{ \scalebox{.8}{  
\begin{tikzpicture}[x=.5cm,y=.5cm] 
\draw[gridline] (0,-0.5)--(0,2.5); 
\draw[gridline] (1,-0.5)--(1,2.5); 
\draw[gridline] (2,-0.5)--(2,2.5); 
\draw[gridline] (-0.5,0)--(2.5,0); 
\draw[gridline] (-0.5,1)--(2.5,1); 
\draw[gridline] (-0.5,2)--(2.5,2); 
\draw[ligne] (0,1)--(0,0); 
\draw[ligne] (0,0)--(1,0); 
\draw[ligne] (1,0)--(2,0); 
\draw[ligne] (2,0)--(2,1); 
\draw[ligne] (2,1)--(2,2); 
\draw[ligne] (2,2)--(1,2); 
\draw[ligne] (1,2)--(0,2); 
\draw[ligne] (0,2)--(0,1); 
\node[bd] at (0,0) {}; 
\node[bd] at (1,0) {}; 
\node[bd] at (2,0) {}; 
\node[bd] at (2,1) {}; 
\node[bd] at (2,2) {}; 
\node[bd] at (1,2) {}; 
\node[bd] at (0,2) {}; 
\node[bd] at (0,1) {}; 
\end{tikzpicture}}},\raisebox{-.5\height}{ \scalebox{.8}{  \begin{tikzpicture}[x=.5cm,y=.5cm] 
\draw[gridline] (0,-1.5)--(0,2.5); 
\draw[gridline] (1,-1.5)--(1,2.5); 
\draw[gridline] (2,-1.5)--(2,2.5); 
\draw[gridline] (-0.5,-1)--(2.5,-1); 
\draw[gridline] (-0.5,0)--(2.5,0); 
\draw[gridline] (-0.5,1)--(2.5,1); 
\draw[gridline] (-0.5,2)--(2.5,2); 
\draw[ligne] (0,2)--(0,1); 
\draw[ligne] (0,1)--(1,0); 
\draw[ligne] (1,0)-- (2,-1) -- (2,0); 
\draw[ligne] (2,0)--(2,1); 
\draw[ligne] (2,1)--(2,2); 
\draw[ligne] (2,2)--(1,2); 
\draw[ligne] (1,2)--(0,2); 
\node[bd] at (0,1) {}; 
\node[bd] at (1,0) {}; 
\node[bd] at (2,0) {}; 
\node[bd] at (2,1) {}; 
\node[bd] at (2,2) {}; 
\node[bd] at (1,2) {}; 
\node[bd] at (0,2) {}; 
\node[bd] at (2,-1) {};
\end{tikzpicture}}},\raisebox{-.5\height}{ \scalebox{.8}{  \begin{tikzpicture}[x=.5cm,y=.5cm] 
\draw[gridline] (0,-1.5)--(0,3.5); 
\draw[gridline] (1,-1.5)--(1,3.5); 
\draw[gridline] (2,-1.5)--(2,3.5); 
\draw[gridline] (-0.5,-1)--(2.5,-1); 
\draw[gridline] (-0.5,0)--(2.5,0); 
\draw[gridline] (-0.5,1)--(2.5,1); 
\draw[gridline] (-0.5,2)--(2.5,2); 
\draw[gridline] (-0.5,3)--(2.5,3); 
\draw[ligne] (0,1)--(1,0); 
\draw[ligne] (1,0)-- (2,-1) -- (2,0); 
\draw[ligne] (2,0)--(2,1); 
\draw[ligne] (0,1) -- (2,3) -- (2,1);
\node[bd] at (0,1) {}; 
\node[bd] at (1,0) {}; 
\node[bd] at (2,0) {}; 
\node[bd] at (2,1) {}; 
\node[bd] at (2,2) {}; 
\node[bd] at (1,2) {}; 
\node[bd] at (2,3) {}; 
\node[bd] at (2,-1) {};
\end{tikzpicture}}} & \raisebox{-.5\height}{ \scalebox{.8}{  
\begin{tikzpicture}[x=.5cm,y=.5cm] 
\draw[gridline] (0,-0.5)--(0,3.5); 
\draw[gridline] (1,-0.5)--(1,3.5); 
\draw[gridline] (2,-0.5)--(2,3.5); 
\draw[gridline] (-0.5,0)--(2.5,0); 
\draw[gridline] (-0.5,1)--(2.5,1); 
\draw[gridline] (-0.5,2)--(2.5,2); 
\draw[gridline] (-0.5,3)--(2.5,3); 
\draw[ligne] (0,1)--(0,0); 
\draw[ligne] (0,0)--(1,0); 
\draw[ligne] (1,0)--(2,1); 
\draw[ligne] (2,1)--(2,2); 
\draw[ligne] (2,2)--(2,3); 
\draw[ligne] (2,3)--(1,3); 
\draw[ligne] (1,3)--(0,3); 
\draw[ligne] (0,3)--(0,2); 
\draw[ligne] (0,2)--(0,1); 
\draw[ligne] (0,2) -- (1,2) -- (2,3);
\node[bd] at (0,0) {}; 
\node[bd] at (1,0) {}; 
\node[bd] at (2,1) {}; 
\node[bd] at (2,2) {}; 
\node[bd] at (2,3) {}; 
\node[wd] at (1,3) {}; 
\node[bd] at (0,3) {}; 
\node[bd] at (0,2) {}; 
\node[bd] at (0,1) {}; 
\end{tikzpicture}}}, \raisebox{-.5\height}{ \scalebox{.8}{  
\begin{tikzpicture}[x=.5cm,y=.5cm]  
\foreach \n in {-1,...,3} {\draw[gridline] (\n,-1.5) -- (\n,3.5); \draw[gridline] (-1.5,\n) -- (3.5,\n);};
\draw[ligne] (-1,0) -- (-1,3) -- (3,-1) -- (1,-1) -- (-1,0);
\draw[ligne] (-1,3) -- (0,1) -- (2,-1);
\draw[ligne] (1,0) -- (1,1);
\foreach \n in {1,...,3} {\node[bd] at (-1,\n) {};
\node[bd] at (\n,-1) {};};
\node[bd] at (-1,0) {};
\node[bd] at (1,1) {};
\node[wd] at (2,0) {};
\node[wd] at (0,2) {};
\end{tikzpicture}}},$ \,\ldots$ \\\hline
        6 & &\raisebox{-.5\height}{\scalebox{.8}{  
\begin{tikzpicture}[x=.5cm,y=.5cm] 
\draw[gridline] (0,-0.5)--(0,3.5); 
\draw[gridline] (1,-0.5)--(1,3.5); 
\draw[gridline] (2,-0.5)--(2,3.5); 
\draw[gridline] (3,-0.5)--(3,3.5); 
\draw[gridline] (-0.5,0)--(3.5,0); 
\draw[gridline] (-0.5,1)--(3.5,1); 
\draw[gridline] (-0.5,2)--(3.5,2); 
\draw[gridline] (-0.5,3)--(3.5,3); 
\draw[ligne] (0,1)--(0,0); 
\draw[ligne] (0,0)--(1,0); 
\draw[ligne] (1,0)--(2,0); 
\draw[ligne] (2,0)--(3,0); 
\draw[ligne] (3,0)--(2,1); 
\draw[ligne] (2,1)--(1,2); 
\draw[ligne] (1,2)--(0,3); 
\draw[ligne] (0,3)--(0,2); 
\draw[ligne] (0,2)--(0,1); 
\node[bd] at (0,0) {}; 
\node[bd] at (1,0) {}; 
\node[bd] at (2,0) {}; 
\node[bd] at (3,0) {}; 
\node[bd] at (2,1) {}; 
\node[bd] at (1,2) {}; 
\node[bd] at (0,3) {}; 
\node[bd] at (0,2) {}; 
\node[bd] at (0,1) {}; 
\end{tikzpicture}}} & \raisebox{-.5\height}{ \scalebox{.8}{  
\begin{tikzpicture}[x=.5cm,y=.5cm] 
\draw[gridline] (0,-0.5)--(0,3.5); 
\draw[gridline] (1,-0.5)--(1,3.5); 
\draw[gridline] (2,-0.5)--(2,3.5); 
\draw[gridline] (-0.5,0)--(2.5,0); 
\draw[gridline] (-0.5,1)--(2.5,1); 
\draw[gridline] (-0.5,2)--(2.5,2); 
\draw[gridline] (-0.5,3)--(2.5,3); 
\draw[ligne] (0,1)--(0,0); 
\draw[ligne] (0,0)--(1,0); 
\draw[ligne] (1,0)--(2,0); 
\draw[ligne] (2,0)--(2,1); 
\draw[ligne] (2,1)--(2,2); 
\draw[ligne] (2,2)--(2,3); 
\draw[ligne] (2,3)--(1,3); 
\draw[ligne] (1,3)--(0,3); 
\draw[ligne] (0,3)--(0,2); 
\draw[ligne] (0,2)--(0,1); 
\draw[ligne] (0,2) -- (1,2) -- (2,3);
\node[bd] at (0,0) {}; 
\node[bd] at (1,0) {}; 
\node[bd] at (2,0) {}; 
\node[bd] at (2,1) {}; 
\node[bd] at (2,2) {}; 
\node[bd] at (2,3) {}; 
\node[wd] at (1,3) {}; 
\node[bd] at (0,3) {}; 
\node[bd] at (0,2) {}; 
\node[bd] at (0,1) {}; 
\end{tikzpicture}}},\raisebox{-.5\height}{ \scalebox{.8}{  
\begin{tikzpicture}[x=.5cm,y=.5cm] 
\draw[gridline] (0,-1.5)--(0,3.5); 
\draw[gridline] (1,-1.5)--(1,3.5); 
\draw[gridline] (2,-1.5)--(2,3.5); 
\draw[gridline] (-0.5,-1)--(2.5,-1); 
\draw[gridline] (-0.5,0)--(2.5,0); 
\draw[gridline] (-0.5,1)--(2.5,1); 
\draw[gridline] (-0.5,2)--(2.5,2); 
\draw[gridline] (-0.5,3)--(2.5,3); 
\draw[ligne] (0,1)--(0,0); 
\draw[ligne] (0,0)--(0,-1) -- (2,1); 
\draw[ligne] (2,1)--(2,2); 
\draw[ligne] (2,2)--(2,3); 
\draw[ligne] (2,3)--(1,3); 
\draw[ligne] (1,3)--(0,3); 
\draw[ligne] (0,3)--(0,2); 
\draw[ligne] (0,2)--(0,1); 
\draw[ligne] (0,2) -- (1,2) -- (2,3);
\node[bd] at (0,0) {}; 
\node[bd] at (1,0) {}; 
\node[bd] at (0,-1) {}; 
\node[bd] at (2,1) {}; 
\node[bd] at (2,2) {}; 
\node[bd] at (2,3) {}; 
\node[wd] at (1,3) {}; 
\node[bd] at (0,3) {}; 
\node[bd] at (0,2) {}; 
\node[bd] at (0,1) {}; 
\end{tikzpicture}}}, \raisebox{-.5\height}{ \scalebox{.8}{  
\begin{tikzpicture}[x=.5cm,y=.5cm]  
\foreach \n in {-1,...,3} {\draw[gridline] (\n,-1.5) -- (\n,3.5); \draw[gridline] (-1.5,\n) -- (3.5,\n);};
\draw[ligne] (-1,0) -- (-1,3) -- (3,-1) -- (0,-1) -- (-1,0);
\draw[ligne] (-1,3) -- (0,1) -- (2,-1);
\draw[ligne] (1,0) -- (1,1);
\foreach \n in {0,...,3} {\node[bd] at (-1,\n) {};\node[bd] at (\n,-1) {};};
\node[bd] at (1,1) {};
\node[wd] at (2,0) {};
\node[wd] at (0,2) {};
\end{tikzpicture}}},$ \,\ldots$\\\hline
        7 & & &\raisebox{-.5\height}{ \scalebox{.8}{  
\begin{tikzpicture}[x=.5cm,y=.5cm] 
\draw[gridline] (0,-1.5)--(0,3.5); 
\draw[gridline] (1,-1.5)--(1,3.5); 
\draw[gridline] (2,-1.5)--(2,3.5); 
\draw[gridline] (-0.5,-1)--(2.5,-1); 
\draw[gridline] (-0.5,0)--(2.5,0); 
\draw[gridline] (-0.5,1)--(2.5,1); 
\draw[gridline] (-0.5,2)--(2.5,2); 
\draw[gridline] (-0.5,3)--(2.5,3); 
\draw[ligne] (0,1) -- (0,0) -- (0,-1) -- (2,-1) -- (2,0);
\draw[ligne] (0,-1) -- (1,0) -- (2,0);
\draw[ligne] (1,0)--(2,0); 
\draw[ligne] (2,0)--(2,1); 
\draw[ligne] (2,1)--(2,2); 
\draw[ligne] (2,2)--(2,3); 
\draw[ligne] (2,3)--(1,3); 
\draw[ligne] (1,3)--(0,3); 
\draw[ligne] (0,3)--(0,2); 
\draw[ligne] (0,2)--(0,1); 
\draw[ligne] (0,2) -- (1,2) -- (2,3);
\node[bd] at (0,0) {}; 
\node[bd] at (0,-1) {};
\node[bd] at (2,-1) {};
\node[wd] at (1,-1) {}; 
\node[bd] at (2,0) {}; 
\node[bd] at (2,1) {}; 
\node[bd] at (2,2) {}; 
\node[bd] at (2,3) {}; 
\node[wd] at (1,3) {}; 
\node[bd] at (0,3) {}; 
\node[bd] at (0,2) {}; 
\node[bd] at (0,1) {}; 
\end{tikzpicture}}},\raisebox{-.5\height}{ \scalebox{.8}{  
\begin{tikzpicture}[x=.5cm,y=.5cm]  
\foreach \n in {-1,...,3} {\draw[gridline] (\n,-1.5) -- (\n,3.5); \draw[gridline] (-1.5,\n) -- (3.5,\n);};
\draw[ligne] (-1,-1) -- (-1,3) -- (3,-1) -- (-1,-1);
\draw[ligne] (-1,3) -- (0,1) -- (2,-1);
\draw[ligne] (1,0) -- (1,1);
\foreach \n in {-1,...,3} {\node[bd] at (-1,\n) {};\node[bd] at (\n,-1) {};};
\node[bd] at (1,1) {};
\node[wd] at (2,0) {};
\node[wd] at (0,2) {};
\end{tikzpicture}}},\raisebox{-.5\height}{ \scalebox{.8}{  
\begin{tikzpicture}[x=.5cm,y=.5cm]  
\foreach \n in {-1,...,3} {\draw[gridline] (\n,-1.5) -- (\n,3.5); \draw[gridline] (-1.5,\n) -- (3.5,\n);};
\draw[ligne] (-1,-1) -- (-1,3) -- (3,-1) -- (-1,-1);
\draw[ligne] (-1,2) -- (0,1) -- (1,1) -- (1,0) -- (2,-1);
\foreach \n in {-1,...,3} {\node[bd] at (-1,\n) {};\node[bd] at (\n,-1) {};};
\node[bd] at (1,1) {};
\node[wd] at (2,0) {};
\node[wd] at (0,2) {};
\end{tikzpicture}}}, $ \,\ldots$ \\\hline
        8 & & & \raisebox{-.5\height}{ \scalebox{.8}{  \begin{tikzpicture}[x=.5cm,y=.5cm]  
        \foreach \n in {-1,...,5} {\draw[gridline] (-1.5,\n) -- (5.5,\n); \draw[gridline] (\n,-1.5) -- (\n, 5.5);};
        \draw[ligne] (-1,0) -- (-1,4) -- (0,2) -- (2,0) -- (1,0) -- (1,-1) -- (0,-0) -- (-1,0);
        \draw[ligne] (-1,4) -- (-1,5) -- (5,-1) -- (-1,-1) -- (-1,0);
        \draw[ligne] (-1,4) -- (3,0) -- (5,-1);
        \draw[ligne] (1,1) -- (1,2) -- (2,2);
        \draw[ligne] (3,-1) -- (2,0) -- (3,0);
        \foreach \n in  {-1,...,5} \node[bd] at (-1,\n) {};
        \node[bd] at (2,2) {};
        \node[bd] at (1,-1) {};
        \node[bd] at (3,1) {};
        \node[bd] at (5,-1) {};
        \node[wd] at (0,3) {};
        \node[wd] at (2,1) {};
        \foreach \n in {0,2,4} \node[wd] at (\n,-1) {};
        \foreach \n in {0,1,3,4} \node[wd] at (\n,4-\n) {};
        \end{tikzpicture}}},\raisebox{-.5\height}{ \scalebox{.8}{  \begin{tikzpicture}[x=.5cm,y=.5cm] 
        \foreach \n in {0,...,6} {\draw[gridline] (-.5,\n) -- (3.5,\n); \node[bd] at (0,\n) {};};
        \foreach \n in {0,...,3}\draw[gridline] (\n,-.5) -- (\n,6.5);
        \draw[ligne] (0,2) -- (1,2) -- (2,1) -- (2,3) -- (3,4) -- (2,4) -- (2,5) -- (1,4) -- (0,4)--(0,2);
        \draw[ligne] (0,2) -- (0,0) -- (3,0) -- (3,6) -- (0,6) -- (0,4);
        \draw[ligne] (0,5) -- (2,5) -- (3,6);
        \draw[ligne] (2,2) -- (3,2);
        \draw[ligne] (0,1) -- (2,1) -- (3,0);
        \foreach \n in {0,2,4,6} \node[bd] at (3,\n) {};
        \foreach \n in {1,3,5} \node[wd] at (3,\n) {};
        \node[wd] at (1,0) {};
        \node[wd] at (2,0) {};
        \node[wd] at (1,6) {};
        \node[wd] at (2,6) {};
        \node[wd] at (1,1) {};
        \node[wd] at (1,5) {};
        \end{tikzpicture}}},$ \,\ldots$ 
    \end{tabular}
    \caption{(G)TPs for local del Pezzo and generalized del Pezzo surfaces.}
    \label{tab:GTPs_for_GdPs}
\end{table}

\afterpage{
    \clearpage
    \thispagestyle{empty}
    \begin{landscape}
    \begin{table}
        \hspace{-3.8cm}
        \begin{tabular}{c||c|c|c|c|c|c}
         $d_f \backslash d_b $ & 0&1&2&3&4&5 \\ \hline \hline
        0   &&-2&&&&\\ \hline
        1   & -2 &-4&-6&-8&-10&-12 \\ \hline
        2   &&-6&-32, 9& -110, 68, -12& -288, 300, -116, 15&-644, 988, -628, 176, -18 \\ \hline
        3   &&-8&-110, 68, -12&-756, 1016, -580, 156, -16&-3556, 7792, -8042, 4680, -1560, 276, -20& -13072, 41376, -64624, 60840, ...\\ \hline
        4 && -10 & -288, 300, -116, 15& -3556, 7792, -8042, 4680, -1560, 276, -20& -27264, 95313, -167936, 184056, ...& -153324, 760764, -1964440, ...  \\ \hline 
        5 && -12 & -644, 988, -628, 176, -18 &-13072, 41376, -64624, 60840, -36408, ... & -153324, 760764, -1964440, 3288688, ...& -1252040, 8695048, -32242268, ... \\ \hline
        6 && -14 & -1280, 2698, -2488, 1130, -248, 21& -40338, 172124, -371980, 501440,  ... & -690400, 4552692, -15913228, ... & -7877210, 71859628, ... 
    \end{tabular}
    \caption{Gopakumar-Vafa invariants in low degree, for the local $\mathbb F_0$ surface (with the natural degree/grading). \emph{How to read the table}: in degree $\beta = (3,2)$, the non-vanishing invariants are $\text{GV}_{0,\beta} = -110$, $\text{GV}_{1,\beta} = 68$ and $\text{GV}_{2,\beta} = -12$.}
    \label{tab:GV_F0}
    \end{table}
    \begin{table}
        \hspace{-3cm}
        \begin{tabular}{c||c|c|c|c|c|c|c|c}
        $d_f \backslash d_b $ & 0&1&2&3&4&5&6&7 \\ \hline \hline
       0    & &{\color{blue} \textbf{-1}} &&&&&& \\  \hline
       1    & -2 & -2 &&&&&&\\ \hline
       2    & & -4 &&&&&& \\ \hline
       3    & & -6 & -6 &&&&&\\ \hline
       4    & & -8 & -32, 9 & -8 &&& \\ \hline
       5    & & -10 & -110, 68, -12 & -110, 68, -12 & -10 &&& \\ \hline
       6    & & -12 & -288, 300, -116, 15 & -756, 1016, -580, 156, -16& -288, 300, -116, 15& -12& \\ \hline
       7    & & -14 & -644, 988, -628, 176, -18& -3556, 7792, -8042, 4680, -1560, 276, -20&  -3556, 7792, -8042, 4680, -1560, 276, -20& -644, 988, -628, 176, -18& -14&
    \end{tabular}
    \caption{Gopakumar-Vafa invariants in low degree, for the local $\mathbb F_2$ surface. \emph{How to read the table}: in degree $\beta = (4,2)$, the non-vanishing invariants are $\text{GV}_{0,\beta} = -32$ and $\text{GV}_{1,\beta} = 9$. In blue, we single out the contribution from the two parallel semi-infinite 5-branes. Table 11 of \cite{chiang1999localmirrorsymmetrycalculations} records $-1/2$ for this class; we find $-1$, consistent with the inverse conifold contribution of the two parallel legs discussed in Section \ref{subsec: Mori cone F0 F2}. }
    \label{tab:GV_F2}
    \end{table}
    \end{landscape}
}

\afterpage{
    \clearpage
    \thispagestyle{empty}
    \begin{landscape}
    \begin{table}
        \hspace*{-3.5cm}\begin{tabular}{c||c|c|c|c|c}
             $d_f \backslash d_b $ & 0 & 1 & 2 & 3 & 4  \\\hline\hline
            0 &  & 1 &  &  &    \\\hline
            1 & -2 & 3 &  &  &    \\\hline
            2 &  & 5 & -6 &  &    \\\hline
            3 &  & 7 & -32, 9 & 27, -10 &   \\\hline
            4 & & 9 & -110, 68, -12 & 286, -288, 108, -14 & -192, 231, -102, 15  \\\hline
            5 & & 11 & -288, 300, -116, 15 & 1651, -2938, 2353, -992, 212, -18 &
             -3038, 6984, -7506, 4519, -1542, 276, -20 \\\hline
            6 & & 13 & -644, 988, -628, 176, -18 & 6885, -18470, 23910, -18118, 8368, -2308, 348, -22 &
             -25216, 90131, -161760, 179995, -131972, 64688, -20914, 4266, -496, 25 
        \end{tabular}\vspace{.2cm}
        \hspace*{-3.5cm}\begin{tabular}{c||c|c}
            $d_f \backslash d_b $ &  5 & 6 \\\hline\hline
            0 &    &  \\\hline
            1 &   &  \\\hline
            2 &    &  \\\hline
            3 &    &  \\\hline
            4  & & \\\hline
            5 & 1695, -4452, 5430, -3672, 1386, -270, 21 & \\\hline
            6 & 35870, -152622, 329544, -447502, 407661, -254310, 108440, -30948, 5634, -590, 27 & -17064, 80948, -194022, 290853, -290400, 196857, -90390, 27538, -5310, 585, -28 
        \end{tabular}
    \caption{Gopakumar-Vafa invariants in low degree, for the local $\mathbb F_1$ surface (with the natural degree/grading). \emph{How to read the table}: in degree $\beta = (4,2)$, the non-vanishing invariants are $\text{GV}_{0,\beta} = -110$, $\text{GV}_{1,\beta} = 68$ and $\text{GV}_{2,\beta} = -12$.}
    \label{tab:GV_F1}
    \end{table}
    \end{landscape} 
}

\bibliographystyle{JHEP}
\bibliography{bibli}

 \end{document}